\documentclass{aa}
\usepackage[utf8]{inputenc}
\usepackage{natbib}
\usepackage[flushleft]{threeparttable}
\usepackage{pdflscape}
\usepackage{tikz}
\usepackage{graphicx}
\usetikzlibrary{positioning}
\usepackage[labelfont=bf]{caption}
\usepackage{subcaption}
\bibpunct{(}{)}{;}{a}{}{,} 

\title{Empirical mass-loss rates and clumping properties of Galactic early-type O supergiants}
\author{C. Hawcroft\inst{\ref{inst1}}\and H. Sana\inst{\ref{inst1}} \and L. Mahy\inst{\ref{inst1}, \ref{inst2}} \and J.O. Sundqvist\inst{\ref{inst1}} \and M. Abdul-Masih\inst{\ref{inst1}} \and J.C. Bouret\inst{\ref{inst4}} \and S. A. Brands\inst{\ref{inst3}} \and A. de Koter\inst{\ref{inst3}, \ref{inst1}} \and F. A. Driessen\inst{\ref{inst1}} \and J. Puls\inst{\ref{inst5}}}

\institute{Institute of Astronomy, KU Leuven, Celestijnenlaan 200D, 3001, Leuven, Belgium \\ email: calum.hawcroft@kuleuven.be\label{inst1} \and Royal Observatory of Belgium, Avenue Circulaire 3, B-1180 Brussels, Belgium\label{inst2} \and Astronomical Institute Anton Pannekoek, Amsterdam University, Science Park 904, 1098 XH Amsterdam, The Netherlands\label{inst3} \and Aix Marseille Univ, CNRS, CNES, LAM, Marseille, France\label{inst4} \and LMU M{\"u}nchen, Universit{\"a}tssternwarte, Scheinerstr. 1, 81679 M{\"u}nchen, Germany\label{inst5}}

\date{August 18, 2021}

\abstract{}
{We investigate the impact of optically thick clumping on spectroscopic stellar wind diagnostics in O supergiants and constrain wind parameters associated with porosity in velocity space. This is the first time the effects of optically thick clumping have been investigated for a sample of massive hot stars, using models which include a full optically thick clumping description.
}{We re-analyse existing spectroscopic observations of a sample of eight O supergiants previously analysed with the non-local-thermodynamic-equilibrium (NLTE) atmosphere code CMFGEN. Using a genetic algorithm wrapper around the NLTE atmosphere code FASTWIND we obtain simultaneous fits to optical and ultraviolet spectra and determine photospheric properties, chemical surface abundances and wind properties. 
}{We provide empirical constraints on a number of wind parameters including the clumping factors, mass-loss rates and terminal wind velocities. Additionally, we establish the first systematic empirical constraints on velocity filling factors and interclump densities. These are parameters that describe clump distribution in velocity-space and density of the interclump medium in physical-space, respectively. We observe a mass-loss rate reduction of a factor of 3.6 compared to theoretical predictions from \citet{Vink2000}, and mass-loss rates within a factor 1.4 of theoretical predictions from \citet{Bjorklund2020}. 
}{We confirm that including optically thick clumping allows simultaneous fitting of optical recombination lines and ultraviolet resonance lines, including the unsaturated ultraviolet phosphorus lines (\ion{P}{V} $\lambda\lambda$1118-1128), without reducing the phosphorus abundance. We find that, on average, half of the wind velocity field is covered by dense clumps. We also find that these clumps are 25 times denser than the average wind, and that the interclump medium is 3-10 times less dense than the mean wind. The former result agrees well with theoretical predictions, the latter suggests that lateral filling-in of radially compressed gas might be critical for setting the scale of the rarefied interclump matter.}

\keywords{stars: early-type -- stars: massive -- stars: mass-loss -- stars: atmospheres -- stars: winds, outflows -- stars: individual: HD 16691, HD 66811, HD 190429A, HD 15570, HD 14947, HD 210839, HD 163758, HD 192639} 

\begin{document}

\maketitle

\section{Introduction}

Massive stars ($M_{\rm{ini}} > 8M_{\odot}$) have a profound influence on their surroundings, whether chemically through their ejecta enriched with heavy elements or physically with ionising radiation, momentum, and kinetic energy (\citealp{Chiosi1986}, \citealp{Matteucci2007a}, \citealp{Geen2020}). Their strong outflows shape the interstellar medium and can play an integral role in triggering star formation (\citealp{MacLow2004}, \citealp{Chen2006}, \citealp{Elmegreen2011}, \citealp{Kennicutt2012}). Massive stars are also proxies of the first generations of stars and progenitors of a range of exotic phenomena from supernovae to black holes (\citealp{Haiman1996}, \citealp{Zaroubi2012}). The breadth of the influence of massive stars in astronomy is difficult to overstate and an understanding of their evolution is imperative. In this context we focus on one of the most critical factors in massive star evolution, the mass loss through stellar winds \citep{Langer2012}.

The source of mass loss in these stars is their strong line-driven winds (see e.g. \citealp{Puls2008a}), which are inherently unstable due to the line deshadowing instability (LDI; \citealp{MacGregor1979a}, \citealp{Carlberg1980}, \citealp{Owocki1984}). Full time-dependent, radiation-hydrodynamic wind models show the LDI leading to the development of a highly structured wind morphology consisting of out-flowing over-dense clumps, where wind opacity can increase, separated by gaps of tenuous material, which allows more photons to escape (see e.g. \citealp{Owocki1988}; \citealp{Feldmeier1995}; \citealp{Dessart2003}; \citealp{Sundqvist2013}; \citealp{Sundqvist2018a}; \citealp{Driessen2019}).

Most modern spectroscopic models of hot stars approximately include the effects of the wind instability by assuming that all the wind mass is concentrated in optically thin clumps, occupying a fraction of the total volume of the wind (e.g. CMFGEN, \citealp{Hillier1998} or PoWR, \citealp{Grafener2002}). But if the clumps start to become optically thick there will be additional light leakage effects, due to the Doppler shift, through porous channels in velocity space. In particular, there can be significant light leakage for spectral lines formed in the wind (e.g. \citealp{Oskinova2007}, \citealp{Sundqvist2009}, \citealp{Surlan2013}). This effect is unaccounted for in such a volume filling factor approach. In this paper we employ the unified, non-local-thermodynamic-equilibrium (NLTE), steady-state stellar atmosphere and wind modelling code FASTWIND (\citealp{Santolaya-Rey1997a}, and follow-up papers as specified in Sect. 3), where the effect of clumps of arbitrary optical thickness has been fully implemented \citep{Sundqvist2018}. Due to the flexibility and speed of FASTWIND, we are able to examine effects of velocity porosity (and a potentially non-void `interclump medium') upon strategic diagnostics in a systematic way that has not been possible before.

To model the inherent physical processes of line transitions, stellar atmosphere models solve the stellar wind physics in a NLTE atmosphere computation and the production of the synthetic spectrum is handled by the formal solution of the radiative transfer equation. This is one of the key differences between the wind porosity included in FASTWIND and other tools used in studies such as \citet{Oskinova2007} and \citet{Surlan2013}. For example, these other methods look only at what is the impact of the wind opacity and porosity on spectral lines in the formal integral. We are able to more realistically account for these effects with FASTWIND, which includes wind opacity and porosity in the computation of the stellar and wind structure, for example, how the modified radiation field affects temperature and ionisation balances throughout the stellar atmosphere and wind. 

To derive good empirical constraints on the various parameters describing the wind, spectra that include both recombination and resonance wind lines are required, each of which probe different aspects of stellar wind physics. This highlights the importance of both optical and ultraviolet coverage. In the optical, various lines produced by recombination processes are observed, while in the ultraviolet, lines are often produced by resonance scattering. Rates of recombination processes are proportional to the density squared, implying that the strength of a spectral line resulting from a recombination process is shaped strongly by the assumed overdensities of the wind clumps in the inner wind region. On the other hand, resonance scattering is proportional to density, meaning these diagnostics generally react differently to a clumped and potentially porous wind. As a result, lines arising through each process are affected uniquely by the different wind parameters.

Currently, mass loss is one of the most significant sources of uncertainty in the evolution of massive stars, exemplified by the well-established discrepancy between theoretical predictions of mass loss and those derived from observations (see e.g. \citealp{Crowther2002}, \citealp{Hillier2003}, \citealp{Bouret2004a}, \citealp{Fullerton2005}, \citealp{Puls2006} \citealp{Mokiem2007}, \citealp{Tramper2014}, \citealp{Bestenlehner2014}). Another current problem is that no simultaneous fits to optical and ultraviolet spectra, including the \ion{P}{V} $\lambda\lambda$1118-1128 line, can be achieved without considering the wind structure (e.g. \citealp{Oskinova2007}, \citealp{Sundqvist2009}, \citealp{Sundqvist2011a}, \citealp{Surlan2013}). These problems are strong motivations for an empirical study of the wind structure and mass loss of massive stars. Here, we focus on the end of the main sequence, with a sample of stars close to the terminal-age main sequence (TAMS). At this stage, the future evolution of the star is highly sensitive to the mass loss; it can help define the post-main-sequence evolution and the stars' eventual end fate. This sample also represents typical rotation rates of O-type stars; coupled with the evolutionary status, this provides a good opportunity to study CNO enrichment on the main sequence, as detailed in \citet{Bouret2012}. We may also obtain clues of inhomogeneities from other physical processes (rotation, stellar pulsations, etc.) by comparing theoretical and observational wind structure properties. 

The layout of the paper is as follows. Section 2 describes the stellar sample and observations used. Section 3 covers our modelling software, procedure, and assumptions. Sections 4 and 5 give a generalised overview of stellar and wind diagnostics, respectively. In Sect. 6 we present the results of our analysis, including stellar and wind parameters. Section 7 is a discussion of our findings, especially wind parameters. Finally our conclusions are outlined in Sect. 8. Appendices A and B include an in-depth discussion of the analysis of each object and figures of final best fits obtained, respectively. 

\begin{table}
\centering
\caption{Overview of the instruments used to obtain spectra for this sample.}
\label{tbl:Instruments}
\vspace{2mm}

\begin{tabular}{cccc}
\hline
\hline

Regime & Instrument	& Wavelength [\AA] & Resolving Power \\
\hline
FUV	&	IUE SWP	&	1187-2000	&	10\,000	\\
FUV	&	HST FUSE	&	905-1187	&	20\,000	\\
FUV &   Copernicus  &   907-1443    &   20\,000  \\
FUV &   Copernicus  &   1426-1650    &   15\,000  \\
UV &   Copernicus  &   1408-3196    &   7\,500  \\
Optical	&	ELODIE	&	3895-6815	&	42\,000	\\
Optical	&	NARVAL	&	3700-10050	&	65\,000	\\
Optical	&	FEROS	&	3800-8800	&	48\,000	\\
\hline
\\              
\end{tabular}
\begin{tablenotes}
\item{}
\end{tablenotes}
\end{table}

\section{Sample and Observations}

The sample in the present paper is the full sample of eight O supergiants adopted from Bouret et al. (2012, hereafter BHL12). Almost all of these objects are spectral standards for their respective class (\citealp{Sota2011}) and their classification has not changed in follow-up studies. As this is a pilot study of wind porosity, the adopted sample presents the advantage of minimising the influence of binarity on the spectral analysis. While the majority of massive stars are found in binary systems (see e.g. \citealp{Sana2012a}), if we are to include the effects of optically thick clumping for the first time, in the spectral fitting of a representative sample, it is best to ensure there are as few other processes as possible impacting the winds of these stars.

Table \ref{tbl:Instruments} shows an overview of the instruments utilised to collect the data analysed in this study while Table \ref{tbl:Sample} shows an overview of the sample used in this study. All objects have IUE and FUSE spectra. BHL12 used observations in the far-ultraviolet (FUV) obtained by Copernicus for HD66811 ($\zeta$ Puppis) and so these observations are also used here (further discussion on these observations can be found in  \citealp{Morton1977}, \citealp{Pauldrach1994}). Northern objects have spectra from ELODIE, except $\lambda$ Cep, which was observed with NARVAL. The southern objects with FEROS spectra are HD163758 and HD66811. For further details on the observations and data reduction we refer to BHL12. It is important to note that for this analysis we use the spectra as they are presented in BHL12, the normalisation is not adjusted.

\begin{table}
\centering
\caption{Overview of the sample used in this work.}
\label{tbl:Sample}
\vspace{2mm}

\begin{tabular}{cccc}
\hline
\hline

ID & Sp. Type	& log$(\frac{L}{L_{\odot}})$ & S/N \cr 
 & & [dex] & FUV-UV-Opt \\
\hline
HD 16691	&  O4 If	&	5.94	&	22-9-110	\\
HD 66811	&	O4 I(n)fp	&	5.91	&	183-155-404	\\
HD 190429A &   O4 If  &   5.96    &   25-41-130  \\
HD 15570 &   O4 If  &   5.94    &   8-36-190  \\
HD 14947 &   O4.5 If  &   5.83    &   22-9-110  \\
HD 210839	&	O6 I(n)fp	&	5.80	&	20-14-126	\\
HD 163758	&	O6.5 If	&	5.76	&	17-10-256	\\
HD 192639	&	O7.5 Iabf	&	5.68	&	21-13-188	\\
\hline
\\              
\end{tabular}
\begin{tablenotes}
\item{}
\end{tablenotes}
\end{table}

\section{Methods}

Our analysis is performed with synthetic spectra produced by the stellar atmosphere and wind modelling code FASTWIND (\citealp{Santolaya-Rey1997a}; \citealp{Puls2005}; \citealp{RiveroGonzalez2011}; \citealp{Carneiro2016}, specifically the version presented in \citealp{Sundqvist2018}). FASTWIND solves NLTE number density rate equations, assuming a steady-state spherically symmetric envelope of material encompassing the photosphere and wind of the star, where wind-inhomogeneities (clumping) are treated in a statistical way. FASTWIND accounts for thousands of spectral lines from various metals, using detailed model atoms (H, He, C, N, O, Si \& P), with full co-moving frame radiation transfer for the elements mentioned here. A simplified approach is applied for other elements to account for the effects of line-blanketing while minimising runtime (see \citealp{Puls2005}). 

As detailed in \cite{Santolaya-Rey1997a}, FASTWIND uses a unified approach to compute the complete atmosphere of the star by connecting a hydrostatic photosphere to an outflowing wind at a transition point defined at 10\% of the sound speed. The wind follows a $\beta$-law velocity (see e.g. Eq. 2 in \citealp{Sundqvist2018}), while the density structure is derived from the equation of continuity for a pre-defined mass-loss rate, $\dot{M}$. The user inputs include a number of standard photospheric parameters: effective temperature $T_{\rm{eff}}$ (defined at Rosseland optical depth $\tau = 2/3$), surface gravity log $g$, stellar radius $R$ ($R_{\rm{eff}}$), and chemical abundances. If the outflows are assumed to be homogeneous, we define: $\dot{M}$, $\beta$, and the terminal wind speed $v_{\infty}$. For clumped winds a number of extra inputs are required, these are detailed in the following section (Sect. 3.1), where the number of parameters is larger for optically thick clumping than for optically thin clumping.

A microturbulent velocity ($v_{\rm{turb}}$) is included with a depth-dependent prescription. During the NLTE computation, microturbulence is a fixed constant, in our case $v_{\rm{mic}}$      $= 15\, \rm{km}\, \rm{s^{-1}}$. In the formal integral we select the maximum value between $v_{\rm{turb}}$  $= 15\, \rm{km}\, \rm{s^{-1}}$ and $v_{\rm{turb}}$ $= 0.1v$, where $v$ is the wind speed. This essentially acts as a fixed value of microturbulence at the photosphere, which then increases linearly with wind speed in the outer, wind-dominated atmosphere. BHL12 also implement a similar depth-dependent microturbulence, of the form $v_{\rm{turb}} = v_{\rm{min}} + (v_{\rm{max}} - v_{\rm{min}})v(r)/v_{\infty}$, where $v_{\rm{min}}$ $= 15\, \rm{km}\, \rm{s^{-1}}$ and $v_{\rm{max}}$ is of the order of $0.1v_{\infty}$. This inclusion of supersonic turbulence is physically motivated as a way to account for velocity dispersion in the wind. Such velocity dispersion is predicted by LDI simulations and observed as saturation over larger wavelength regions than predicted with subsonic turbulence in UV P-Cygni troughs (\citealp{Hamann1980}, \citealp{Lucy1982}, \citealp{Puls1993}, \citealp{Sundqvist2011a}). In Sect. 3.3 we perform a robustness check to investigate what effect changing the minimum microturbulence has on the spectrum, but we do not change our microturbulence description from what is defined in this paragraph in any of our fits.

One significant difference between our study and BHL12 is that we do not include the effects of shock-generated X-ray emission. The impact of the inclusion of X-rays in our models is discussed further in Sect. 3.4. In any spectral fitting with optically thin clumping we therefore take the same assumptions as BHL12, with the exception of X-rays. Another key difference is the fitting method employed; BHL12 optimised models by eye while we use a genetic fitting algorithm, this is discussed in Sect. 3.2.

\subsection{Wind clumping formalism}

The standard clumping formalism is to include only the effects of optically thin clumps, parameterised by a generalised clump volume filling factor $f_\mathrm{vol}$ and onset velocity $v_\mathrm{cl}$. The optically thin clumping formalisation is detailed in, for example, BHL12. The volume filling factor $f_\mathrm{vol}$ generally acts as the inverse of the clumping factor $f_\mathrm{cl}$, discussed below, and becomes exactly 1/$f_\mathrm{vol}$ in the limit of optically thin clumping and a void interclump medium. 

The adopted version of FASTWIND (FW v10.3, \citealp{Sundqvist2018}) includes the leakage of light, where the clumping is parameterised with five main factors: i) a clumping factor $f_\mathrm{cl}$, ii) a velocity filling factor $f_\mathrm{vel}$, iii) the interclump density factor $f_\mathrm{ic}$, iv) the clumping onset velocity $v_\mathrm{cl}$ and v) the velocity at which maximum clumping is reached $v_\mathrm{cl, max}$. For the assumption of optically thin clumping only $f_\mathrm{cl}$, $v_\mathrm{cl}$, and $v_\mathrm{cl, max}$ are required \footnote{Formally, one more input clumping parameter is required by FASTWIND, namely the porosity length $h$. However, since $h$ may only influence continuum absorption and emission, it is not relevant for the current project where diagnostic lines are considered. As such, throughout this work we simply set this parameter to the standard $h/R_\mathrm{eff} = v/ v_\infty$, as recommended by Sundqvist \& Puls (2018).}. The wind can be thought of as a two-component medium consisting of a mixture of dense clumps and tenuous interclump material. The material in the wind is compressed into clumps, described through the clumping factor, $f_\mathrm{cl}$. The discreet clumps occupy a fraction of the total volume, this volume is the aforementioned clump volume filling factor $f_\mathrm{vol} (\approx 1/f_\mathrm{cl}$ in the case of a small $f_\mathrm{ic}$). These factors are defined in relation to the mass density $\rho$ of the mean wind as 

\begin{equation}
f_\mathrm{cl} = \frac{\langle\rho{}^{2}\rangle}{\langle\rho{}\rangle^{2}} = \frac{f_\mathrm{vol}\rho{}_\mathrm{cl}^{2} + (1 - f_\mathrm{vol})\rho_\mathrm{ic}^{2}}{(f_\mathrm{vol}\rho_\mathrm{cl} + (1 - f_\mathrm{vol})\rho_\mathrm{ic})^{2}} \geq 1.
\end{equation}  

\noindent Here the average mass density conserves mass relative to a smooth medium with the same mass-loss rate. The density of the material in the clumps is $\rho_\mathrm{cl}$, and $\rho_\mathrm{ic}$ is the density of the tenuous interclump medium. The impact of $f_\mathrm{cl}$ on observed line profiles is discussed further in Sect. 5.1. The velocity filling factor can be thought of as an equivalent scaling in velocity space

\begin{equation}
f_\mathrm{vel} = \frac{\delta{v}}{\delta{v} + \Delta{v}} \leq 1,
\end{equation}

\noindent where $\delta{v}$ is the velocity span in a clump and $\Delta{v}$ is the velocity separation between clumps. The impact of this parameter on observed line profiles is discussed further in Sect. 5.3. The interclump density factor

\begin{equation}
f_\mathrm{ic} = \frac{\rho_\mathrm{ic}}{\langle\rho\rangle}
\end{equation}

\noindent defines the density of the tenuous interclump material.

An effective opacity formalism is used to approximate this two-component medium. This effective opacity $\chi_\mathrm{eff}$ is given as

\begin{equation}
\chi_\mathrm{eff} = \langle\chi\rangle\frac{1 + \tau_\mathrm{cl}f_\mathrm{ic}}{1 + \tau_\mathrm{cl}},
\end{equation}

\noindent where $\tau_\mathrm{cl}$ is the clump optical depth, which takes different forms for continuum and line absorption (see eqs. 9 and 11 in \citealp{Sundqvist2018}). The interclump density $f_\mathrm{ic}$  is defined in Eq. 3. The mean opacity $\langle\chi\rangle$ is defined as $\langle\chi\rangle = \sigma{}n/f_\mathrm{cl}$, where $\sigma$ and $n$ are the atomic cross-section and occupation number density (corrected for stimulated emission) of a fiducial clump, respectively. This mean opacity is computed first, using the NLTE occupation numbers of fiducial clumps with densities $\rho_\mathrm{f} = \langle\rho\rangle f_\mathrm{cl} = \rho_\mathrm{sm} f_\mathrm{cl}$ where the `smooth' density $\rho_\mathrm{sm} = \dot{M}/4\pi{}v(r)r^{2}$ comes from the input mass-loss rate, assuming a $\beta$ velocity law (see eq. 2 in \citealp{Sundqvist2018}).

Using these general expressions one can relate back to the more commonly used case of optically thin clumps by considering appropriate limits: In the case $\tau_\mathrm{cl} \ll 1$, and assuming a void interclump medium, $f_\mathrm{ic} = 0$, one recovers $f_\mathrm{cl} = 1/f_\mathrm{vol}$. Moreover, Eq. 4 for the effective opacity then yields $\chi_\mathrm{eff} =\langle\chi\rangle$. In this limiting case, the description for the mean opacities is consistent with previous implementations of optically thin clumping (e.g. CMFGEN, \citealp{Hillier1998} or PoWR, \citealp{Grafener2002}). In the case of opacities that linearly depend on density:

\begin{equation}
\chi_\mathrm{eff} \propto \frac{\rho_\mathrm{f}}{f_\mathrm{cl}} \propto \frac{\langle\rho\rangle f_\mathrm{cl}}{f_\mathrm{cl}} \propto \langle\rho\rangle,
\end{equation}

\noindent and in the case of opacities that depend on the square of the density:

\begin{equation}
\chi_\mathrm{eff} \propto \frac{\rho_\mathrm{f}^{2}}{f_\mathrm{cl}} \propto \frac{\langle\rho\rangle^{2} f_\mathrm{cl}^{2}}{f_\mathrm{cl}} \propto \langle\rho\rangle^{2}f_\mathrm{cl}.
\end{equation}

The above parametrisation then aims to capture the overall effect of the time-dependent structure expected from the LDI while implementing a time-independent constant mass loss and a parameterised form for the clumping based on a set of input parameters. The time-dependent nature of the LDI is visually represented by figure 2 in \cite{Sundqvist2013} and figure 3 in \cite{Driessen2019}.

\subsection{Fitting procedure - Genetic algorithm}

To perform a global fit to each atmosphere of our sample of stars we utilised a genetic algorithm (GA) wrapper for FASTWIND, adapted from similar codes (\citealp{Mokiem2005}, \citealp{Tramper2014}) originally based on the pikaia framework \citep{Charbonneau1995}. This algorithm is designed to effectively explore a large parameter space and avoid local minima, which is ideally suited to attempt to simultaneously fit for all the stellar parameters that we are able to determine spectroscopically using FASTWIND.
The GA works analogously to genetic evolution and natural selection. We begin with an initial population randomly sampling the full parameter space, the `genes' of which are encoded with the input parameters for FASTWIND. The models are then ranked by a fitness metric $F$ defined as:

\begin{equation}
F = \Bigg(\sum_{i}^{N} \chi_{\rm{red,i}}^{2} \Bigg)^{-1},
\end{equation}

\noindent where $N$ is the number of spectral lines in the fit and $\chi_\mathrm{red,i}^{2}$ is the reduced chi-square of each line \citep{Mokiem2005}. We compute the $\chi^{2}$ for each individual line by comparing the modelled line profile to the observed spectrum. In this study we weigh all lines equally. 

Within a given generation, models with a higher fitness are more likely to be selected to pair with another model to then produce two offspring models. These offspring models comprise the next generation and the process is repeated iteratively until convergence to an optimised fitness. To ensure the parameter space is thoroughly explored we introduce random mutations to digits of the offspring genes. These mutations are applied at a variable rate. The idea being that after a number of generations the parameter space exploration will be refined and narrowed around the models with high fitness; at this point the mutation rate is increased to re-explore the parameter space. Error margins associated with the GA exploration are described in Sect. 6. 
The GA has been used to fit a multitude of stellar types and datasets (see e.g. \citealp{Mokiem2006}; \citealp{Mokiem2007}; \citealp{Tramper2011}; \citealp{Tramper2014}; \citealp{Ramirez-Agudelo2017a}; \citealp{Abdul-Masih2019}). The version we use is presented in \cite{Abdul-Masih2021}. 

The updates that we make to the GA for this study include allowing clumps to become optically thick (Sect. 3.1), a modified list of diagnostic lines (Table \ref{tbl:Linelist}) and a bolometric luminosity anchor (Sect. 4.1). We also tailor the population size and generations as necessary to converge on a best fit with suitable confidence intervals. For a population with 250 individuals we iterate for 200 to 300 generations, yielding a grand total of around 50\,000 to 75\,000 FASTWIND models per fit.

The key motivation of this work is the inclusion of optically thick clumping. To isolate the effects of a modified effective wind opacity and porosity in our analysis we split the fitting procedure into two stages. The initial fit strictly follows the prescription of optically thin clumps. In this case we aim to constrain the photospheric properties of these stars. The line list over which we define our global quality of fit can be seen in Table \ref{tbl:Linelist}. We do not include the \ion{P}{v} $\lambda\lambda$1118-1128 lines when computing models with optically thin clumping due to known issues in replicating this line with optically thin clumps (BHL12), which we also reproduced when fitting models assuming optically thin clumps. These issues are discussed in Sect. 7.5. For the second round of fitting we extend our wind prescription in the atmosphere modelling to allow for optically thick clumping. For clarity we present only the analysis with optically thick clumping in the main text and all discussion of fits with optically thin clumping with FASTWIND are included in Appendix A. 

The commonly used `by-eye' method for massive-star spectroscopic fitting relies on optimising models to fit a number of specific diagnostics to determine individual parameters, for example iterating the surface gravity of the model to fit the wings of Balmer lines. While we are not using such a direct method to constrain individual parameters, as the exploration of the parameter space is automated based on the fitness ranking of the models in a population, we ensure such diagnostic lines are included. The general line list used for most GA runs is shown in Table \ref{tbl:Linelist}. If, for an individual object, any line is omitted we discuss this on a case-by-case basis in Appendix A. 

If used blindly, such automated fitting is not guaranteed to ensure each parameter is constrained to its individual true value. The spectra must be manually prepared and fitting regions carefully selected. As mentioned previously we do not adjust the normalisation of these spectra from BHL12 but we do remove any blends (that are not included in the model) within our selected fitting regions from the spectra as these would negatively influence the fitness metric. Figures of all line regions used for fitting are provided in Appendix B and indicate the diagnostic lines used as well as the omitted spectral ranges. We also monitor the fitness line-by-line for each parameter to investigate the sensitivity of each line to individual parameters. Therefore, with careful manual consideration of the input observations and physical shortcomings of input models, the GA provides an automated method to converge on unique, reproduceable solutions with a well defined statistical foundation for evaluation. The overall agreement between GA-based fitting and a by-eye approach has been investigated by \citet{Markova2020} and was found to be very good.

\subsection{Robustness - minimum microturbulence}

When varying minimum microturbulence ($v_{\rm{turb}}$ as defined earlier in Sect. 3) in these stars the largest effect is on photospheric metal lines and weak helium lines: an increase in $v_{\rm{turb}}$ will increase the strength of these lines \citep{Mcerlean1998}. The effect on strong resonance lines in the UV is marginal. Therefore, in these models the microturbulence stands only to impact abundance diagnostics. As abundance determination is not a main focus of this study we leave more extensive analysis of the microturbulence prescription to future studies. 

\subsection{Robustness - X-rays}

\begin{figure}[t!]
    \centering
    \includegraphics[scale=0.78]{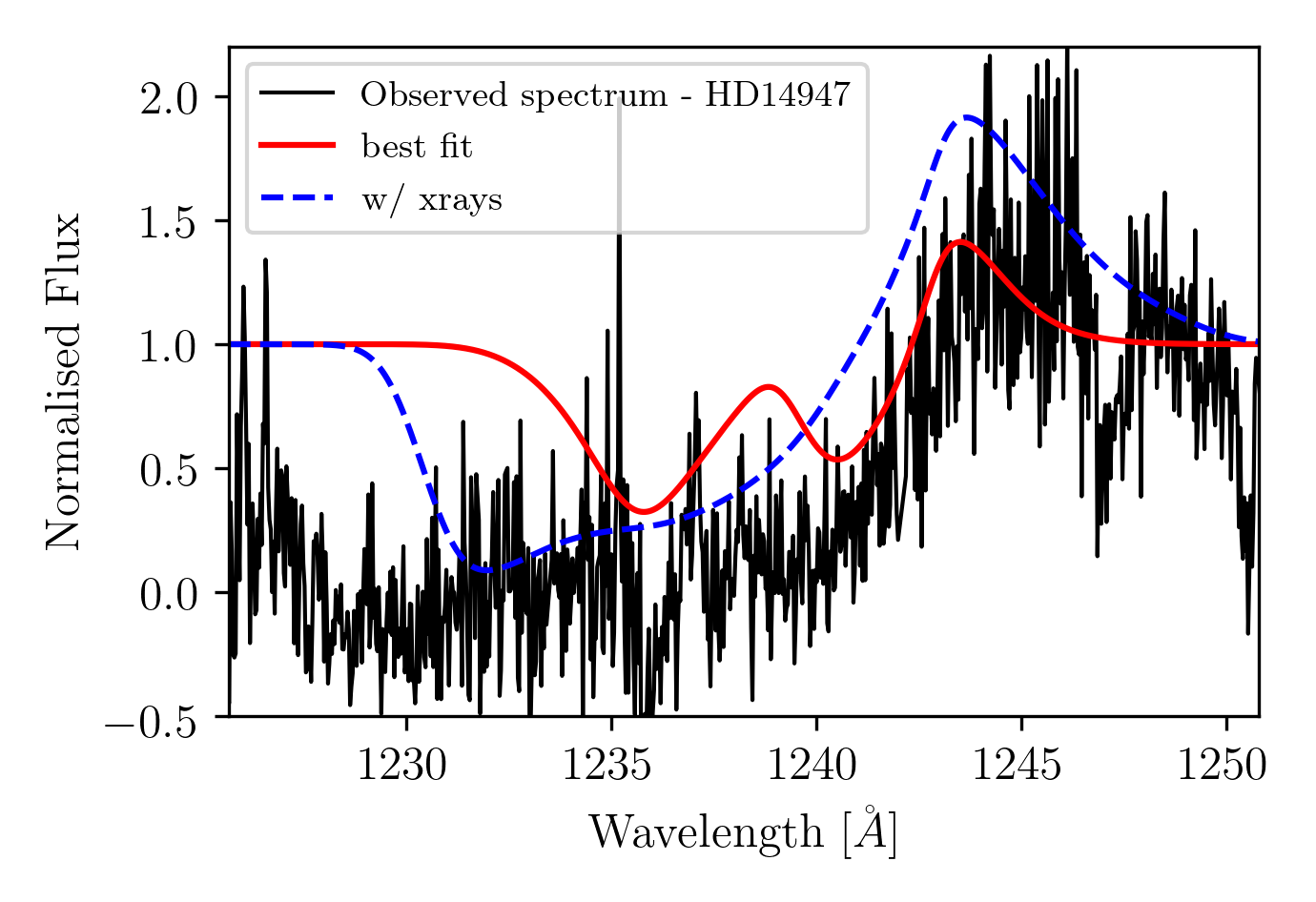}
    \caption{Effect of the inclusion of X-rays on the resonance line \ion{N}{v} $\lambda$1239-1243. GA best fit model without X-rays - red solid line, best fit with X-rays - blue dashed line. We note that \ion{N}{v} $\lambda$1239-1243 is contaminated with Lyman-$\alpha$ in the blue part of the spectral line, which causes the discrepancy in width of the absorption trough between the spectrum and model.} 
    \label{fig: Xray Diagnostics}
\end{figure}

The inclusion of X-rays, as currently parameterised in FASTWIND, does not have a significant effect  on any line profiles investigated in this study (see Table 3 for the list of line profiles). Therefore, we do not include X-rays and all best-fit parameters were obtained using models without X-rays. For this robustness check we manually run an individual FASTWIND model, shown in Fig. \ref{fig: Xray Diagnostics}.  In this model we include X-ray parameters from \citet{Abdul-Masih2019}, based on recommendations in \citet{Carneiro2016}. These recommendations are tailored for typical O stars, meaning for this robustness check we have not examined potential effects of higher X-ray luminosities than those normally assumed for O stars. On the other hand, for lines from a high ionisation stage such as \ion{N}{v} $\lambda$1239-1243, X-rays have a significant effect on the profile. Included in Fig. \ref{fig: Xray Diagnostics} is the high ionisation stage UV line \ion{N}{v} $\lambda$1239-1243. For this line, and others of high ionisation stage, X-rays are required to reproduce the line profile, these lines are excluded from this analysis.

\section{Stellar parameter diagnostics}

\vspace{2mm}

\begin{table}
\caption{List of diagnostic spectral lines used for fitting in the GA.}
\centering
\vspace{2mm}
\begin{tabular}{c@{\hskip 1.8in}c}
\hline
\hline
(F)UV & Optical \\
\hline
        
\ion{He}{ii} 1640 & H$\delta$\\
\ion{C}{iii} 1176 & H$\gamma$\\
\ion{C}{iii} 1620 & H$\beta$\\
\ion{C}{iv} 1548-1550 & H$\alpha$\\
\ion{N}{iii} 1750 & \ion{He}{i} 4026 \\
\ion{N}{iv} 1718 & \ion{He}{i} 4387 \\
\ion{O}{iv} 1340 & \ion{He}{i} 4471 \\ 
\ion{O}{v} 1371 & \ion{He}{i} 4200 \\
\ion{Si}{iv} 1393-1402 & \ion{He}{ii} 4541 \\
\ion{P}{v} 1118-1128* & \ion{He}{ii} 4686 \\
     & \ion{N}{iv} 4058 \\
     & \ion{O}{iii} 5592 \\
\hline
\\              
\end{tabular}
\begin{tablenotes}
\item{ *\ion{P}{v}$\lambda\lambda$1118-1128 is only included in the fits that allow optically thick clumping and this line is excluded when fitting with optically thin clumping.}
\end{tablenotes}
\label{tbl:Linelist}
\end{table}

\vspace{2mm}

\subsection{Radius}

The stellar radius is an important FASTWIND input. We adjust the model radius to balance with the varying temperature so as to maintain a fixed luminosity throughout the GA run. For this analysis we opted to use a bolometric luminosity anchor, taking luminosities calculated in BHL12 and adjusting such that in any model the radius followed $R = (L/(4\pi{}\sigma_{\rm{SB}}T_{\rm{eff}}^{4}))^{0.5}$, where $L$ is the bolometric luminosity, $\sigma_{\rm{SB}}$ is the Stefan-Boltzmann constant, and the effective temperature is a free parameter. 

\subsection{Effective temperature}

\begin{figure}[t!]
    \centering
    \includegraphics[scale=0.8]{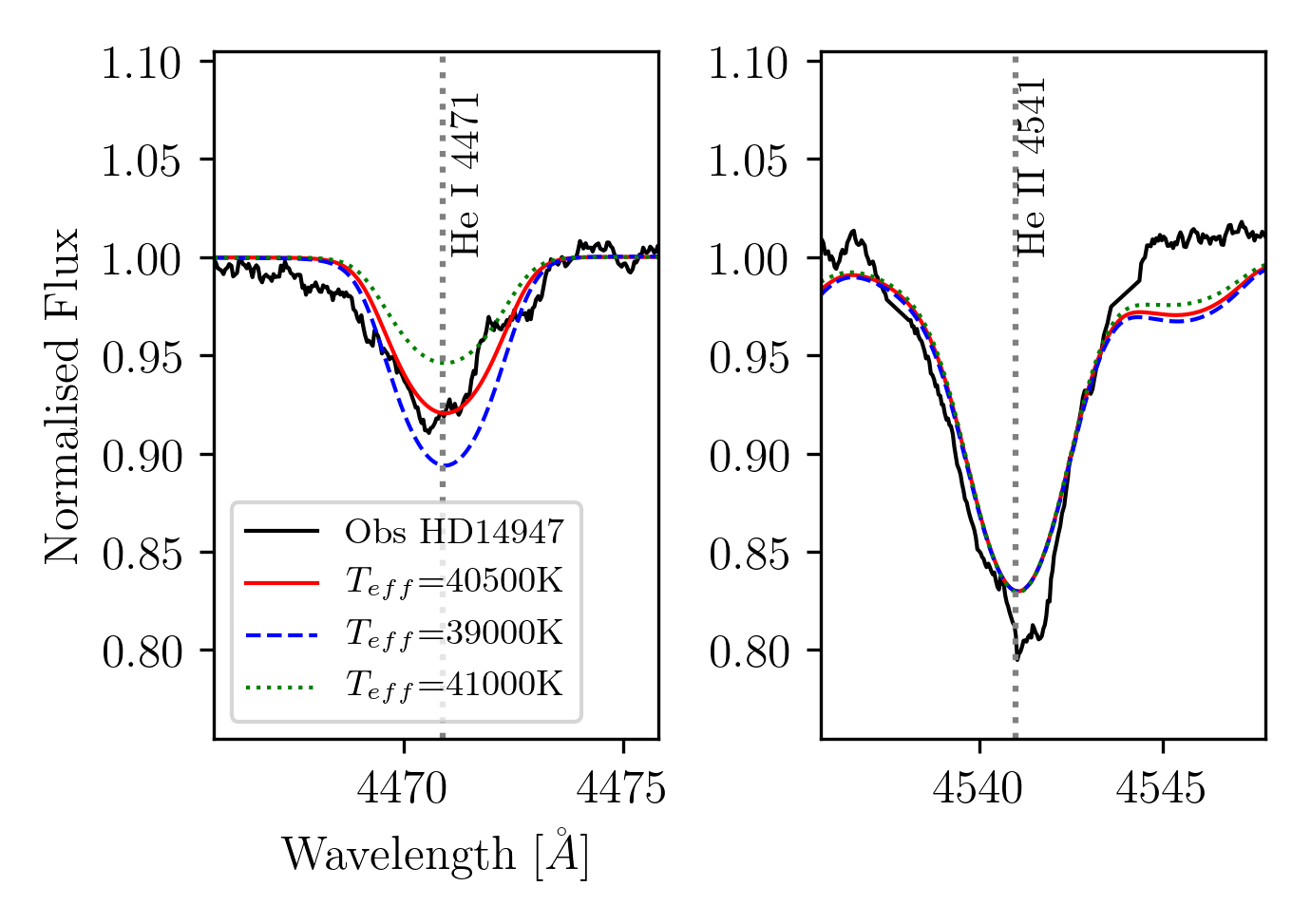}
    \caption{Effect of changing effective temperature on diagnostic lines \ion{He}{i} $\lambda$4471 and \ion{He}{ii} $\lambda$4541. GA best-fit model - red solid line, best fit with 1kK lower $T_\mathrm{eff}$ - blue dashed line, best fit with 1kK higher $T_\mathrm{eff}$ - green dotted line.} 
    \label{fig: Temperature Diagnostics}
\end{figure}

Effective temperature is often constrained by the ionisation balance of available elements. For example, if multiple helium absorption lines are present in the spectrum from different energy levels and ions, comparing their relative strengths (equivalent widths) can be a robust temperature estimate. In our $\chi^{2}$ based approach, we rely on the absolute strength of all diagnostics lines, including (ideally) He, O and N lines of different ionisation stages (Table 3) and the model sensitivity in the parameter regime. Figure \ref{fig: Temperature Diagnostics} shows the effect of changing $T_{\rm{eff}}$ on diagnostic helium lines.

\subsection{Surface gravity}

\begin{figure}[t!]
    \centering
    \includegraphics[scale=0.78]{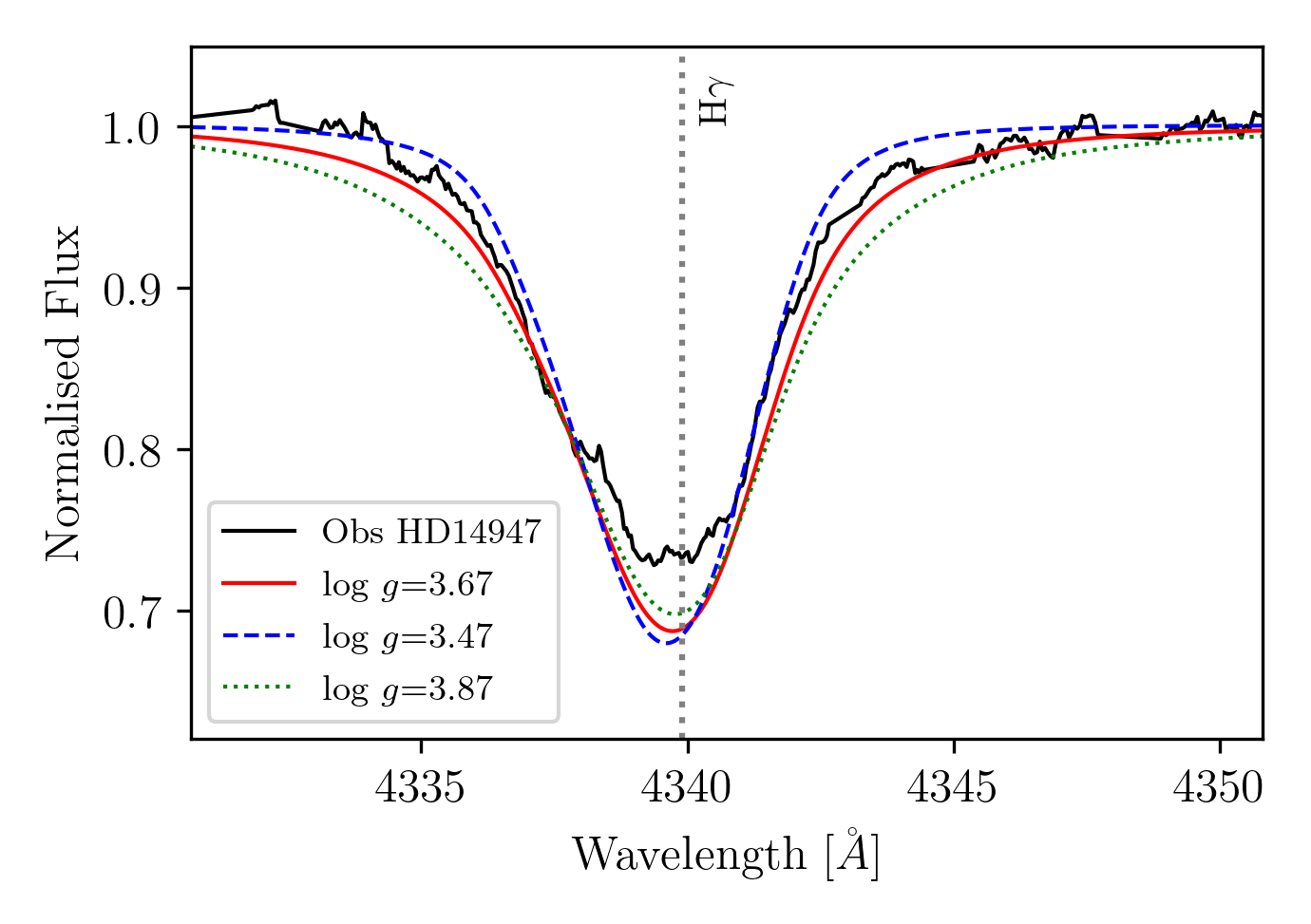}
    \caption{Effect of changing surface gravity on H$\gamma$. GA best-fit model - red solid line, best fit with 0.2 lower log $g$ - blue dashed line, best fit with 0.2 higher log $g$ - green dotted line.} 
    \label{fig: Surface Gravity Diagnostics}
\end{figure}

The surface gravity is typically constrained using the hydrogen Balmer series, which is affected by Stark broadening. This shapes the wings of these profiles and is strongly dependent on the electron density at the surface of the star. Figure \ref{fig: Surface Gravity Diagnostics} shows the effect of changing log $g$ on H$\gamma$. Mostly, Balmer transitions from higher levels (H$\delta$ and H$\gamma$) are uncontaminated by wind infilling emission and are ideal surface gravity diagnostics. H$\beta$ remains dominated by an absorption profile in most objects but the wind emission begins to fill in the line core, which can also affect the line wings, in objects with strong winds. 

The effective temperature also affects the electron pressure so there is some degeneracy between temperature and surface gravity. To help break this degeneracy our list of diagnostic lines (Table 3) includes temperature diagnostics that are not as sensitive to log $g$, as discussed in Sect. 4.2. The present GA based approach allows us to take these correlations into account while converging towards the optimal set of atmospheric and wind parameters.

\subsection{Surface abundances}

Determining surface abundances is not one of the main focuses of this study; therefore, we do not include an extensive line list. This means we do not rigorously examine the abundances, rather we include a subset of lines and compare with what is found in BHL12 where the surface abundances were a focal point. We include multiple lines from N, C and O in both the optical and UV (Table 3). We aim to include at least two different ionisation stages from each element, and for each ionisation stage a photospheric line profile relatively free from contamination from other processes like winds or blends. However, for Si we only include one ion so we will not discuss the Si abundance further in our analysis.

\subsection{Rotation and turbulence}

After the formal integral the synthetic spectra are broadened with an instrumental, rotational and macroturbulent profile. The rotational and macroturbulent velocities are free parameters in the GA, and metal absorption lines are good diagnostics for rotational and macroturbulent broadening. Figures 4 and 5 show the effect of changing rotation and macroturbulence on the diagnostic line \ion{O}{iii} $\lambda$5592, respectively. Most objects are included in \cite{Simon-Diaz2014}, who provided an in-depth look at the contributions from macroturbulence and rotation in the overall line broadening. Generally we find that our confidence intervals include the values determined in \cite{Simon-Diaz2014}, this is discussed on a case-by-case basis in Appendix A.

\begin{figure}[t!]
    \centering
    \includegraphics[scale=0.78]{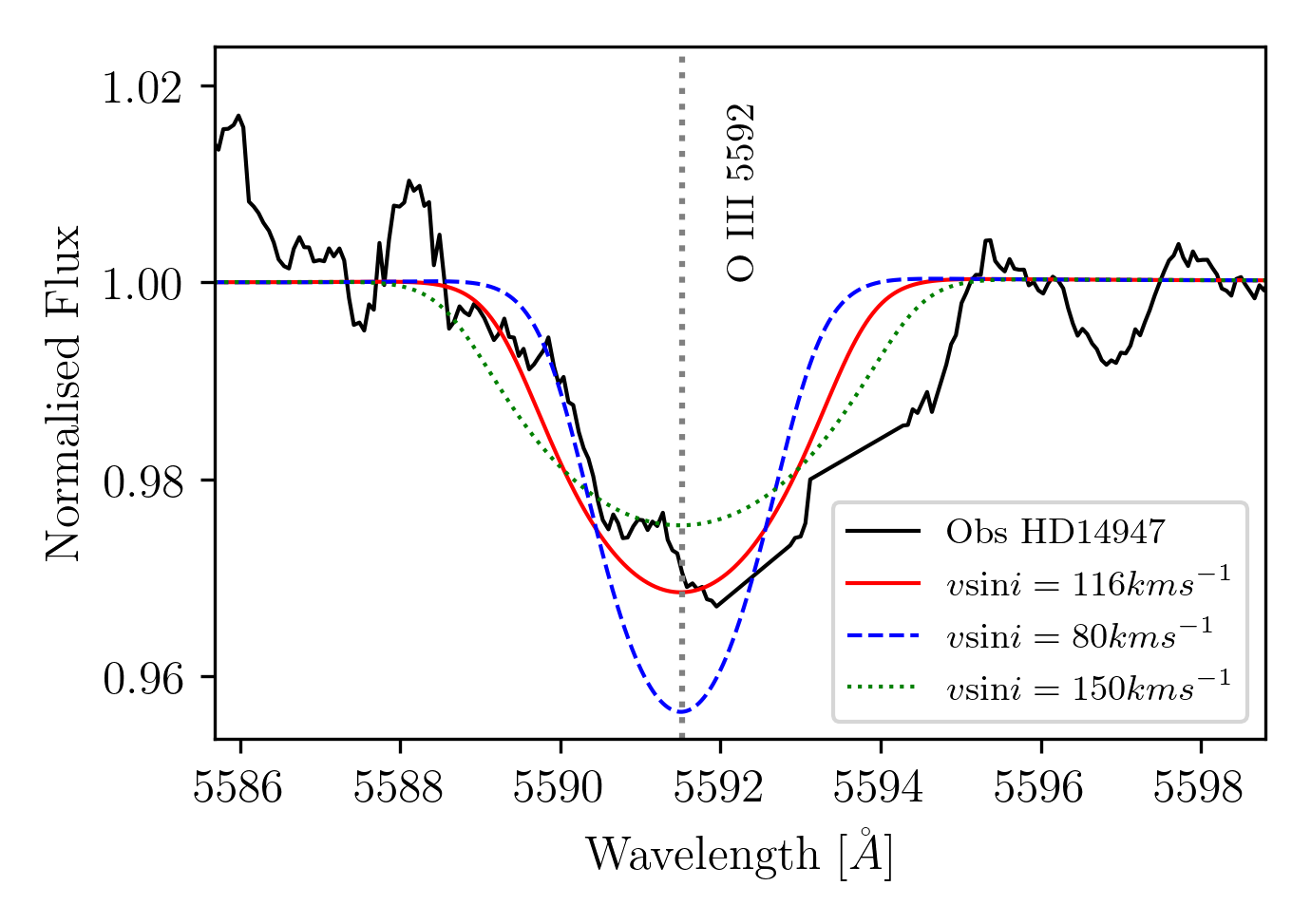}
    \caption{Effect of changing rotation on the broadening of \ion{O}{iii} $\lambda$5592. GA best-fit model - red solid line, best fit with lower rotation - blue dashed line, best fit with higher rotation - green dotted line.} 
    \label{fig: Rotation Diagnostics}
\end{figure}

\begin{figure}[t!]
    \centering
    \includegraphics[scale=0.78]{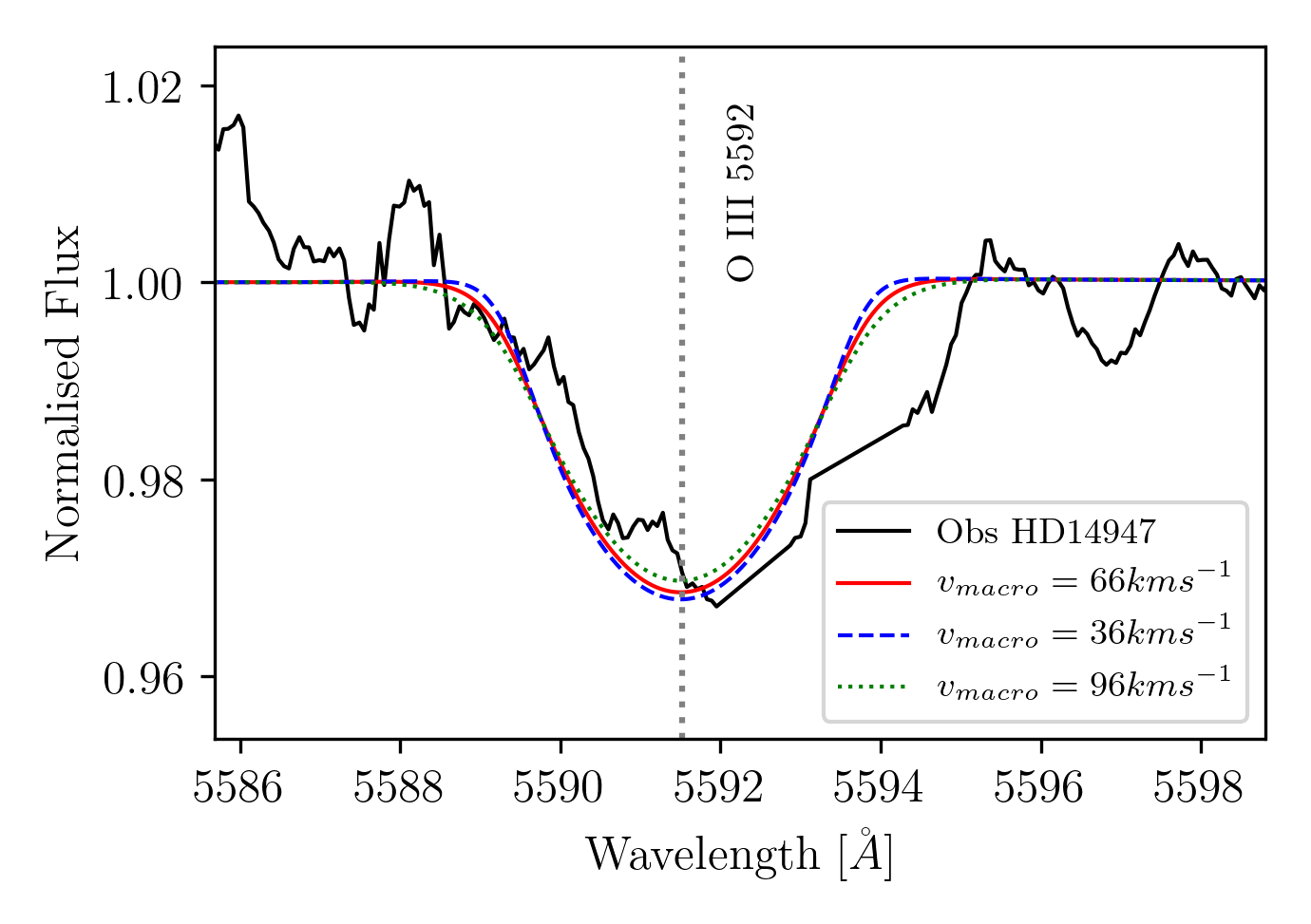}
    \caption{Effect of changing macroturbulent velocity on the broadening of \ion{O}{iii} $\lambda$5592. GA best-fit model - red solid line, best fit with lower macroturbulence - blue dashed line, best fit with higher macroturbulence - green dotted line. Note that the effect is small in this case but increases for lower $v\sin i$.} 
    \label{fig: Macroturbulence Diagnostics}
\end{figure}

\subsection{Radial velocities}

We find radial velocities for these stars by fitting a 1D Gaussian profile to optical helium lines in the observed spectra and by measuring the offset between the mean of the Gaussian and the line centre wavelength used in FASTWIND. This offset correction is then applied to all FASTWIND models produced by the GA before the fitness of the model is evaluated.

\section{Wind parameter diagnostics}

When fitting with optically thin clumping we allow the mass-loss rate, $\beta$, and terminal wind speed to vary and use a fixed clumping factor corresponding to those found in BHL12. 
When running the GA with optically thick clumping the number of free parameters is increased. We include velocity filling factor, interclump density factor, and the clumping onset velocity alongside all other parameters. We also test varying a subset of parameters while fixing others to the values found in BHL12. Generally this leads to less accurate fitting, either due to systematic differences between CMFGEN (used by BHL12) and FASTWIND or due to optimisation on a case-by-case basis resulting from the difference in fitting techniques.

\subsection{Mass-loss rate, clumping, and terminal wind velocity}

In the optical, the mass-loss rate and clumping factor are highly degenerate. Fortunately, they affect resonance lines and recombination lines differently. The recombination lines are particularly sensitive to the clumping factor as the resultant increase in recombination rates is proportional to density squared and thus the ionisation balance changes. Clumping also has an impact on the optical depth, affecting the radiative transfer. The resonance lines are less sensitive as this process is proportional to density. This means the mass-loss rate should be constrained more so by the absorption components of resonance lines while the clumping factor is constrained by the recombination lines.

\begin{figure}[t!]
    \centering
    \includegraphics[scale=0.78]{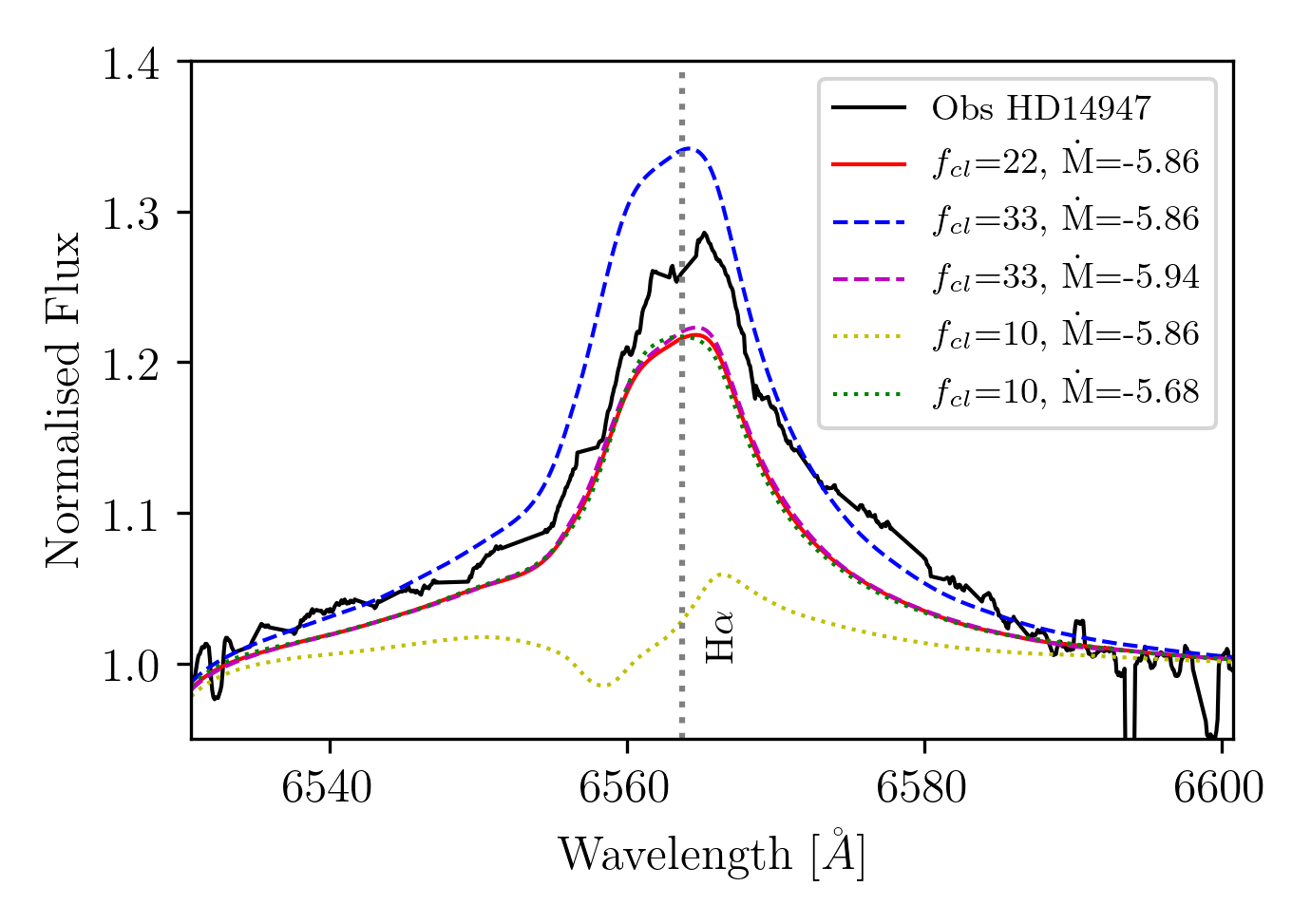}
    \includegraphics[scale=0.78]{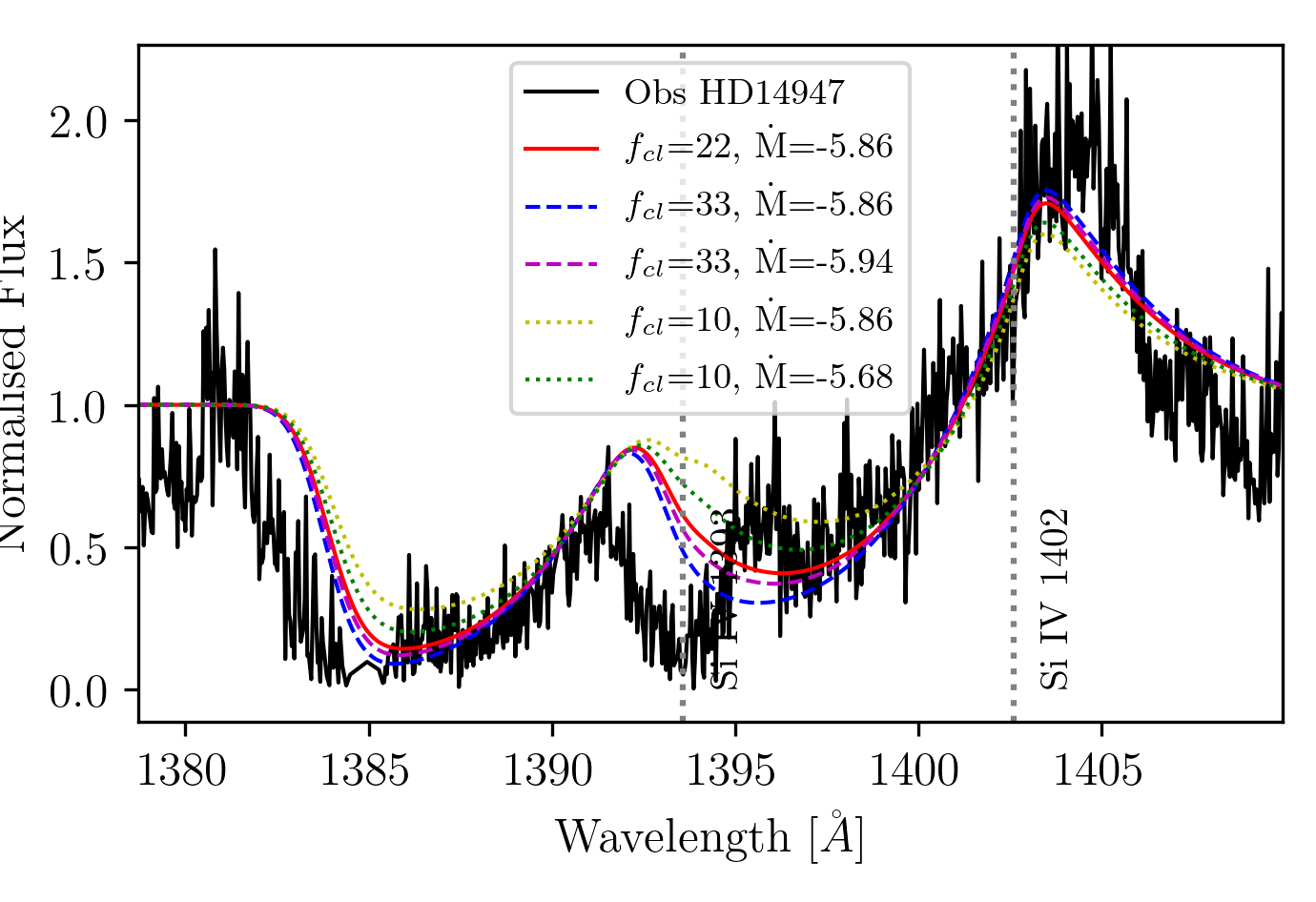}
    \caption{Effect of changing mass-loss rate and clumping factor on the H$\alpha$ recombination line and on the \ion{Si}{iv} $\lambda\lambda$1393-1402 resonance line, using the best-fit model including optically thick clumping. GA best-fit model - red solid line. Best fit with increased clumping and fixed mass-loss rate - blue dashed line. Best fit with reduced clumping and fixed mass-loss rate - yellow dotted line. Best fit with increased clumping and reduced mass-loss rate - magenta dashed line. Best fit with reduced clumping and increased mass-loss rate - green dotted line.} 
    \label{fig: mass-loss Diagnostics}
\end{figure}

To illustrate this, Fig. \ref{fig: mass-loss Diagnostics} shows FASTWIND models for the H$\alpha$ line with various combinations of clumping factor and mass-loss rate, applied to our best-fit model with optically thick clumping. In general mass-loss rates and clumping factors are degenerate in recombination lines; such that, if the best-fit model has a clumping factor of 20 and a mass-loss rate log($\dot{M}) = -5.8$, and we were to reduce the clumping factor to 10 and increase the mass-loss rate accordingly \footnote{For the case of spatially constant clumping, we conserve the quantity $\dot{M}*\sqrt{f_\mathrm{cl}}$ to maintain a constant line strength. We note that this relation is strictly true in the case of optically thin clumping. Allowing optically thick clumping may have some, as of yet unquantified, effect on this relation.}, we would produce a FASTWIND model that is statistically indistinguishable from our best fit within a reasonable error margin. This degeneracy is broken by the simultaneous fitting of resonance lines, which are less sensitive to the clumping factor (Fig. \ref{fig: mass-loss Diagnostics} bottom panel).

The terminal wind speed is usually determined by the extent of the bluest edge of the absorption in the saturated P-Cygni profiles of the UV resonance lines, of which \ion{C}{iv} $\lambda\lambda$1548-1550 is often a particularly good example \citep{Prinja1990}. This value can be determined almost independently from any other if one observes a saturated P-Cygni line such that the absorption flux is essentially zero up until the blue edge of the absorption trough. One should consider the microturbulence when fitting the blue edge of the P-Cygni line as we include a depth-dependent microturbulence $v_{\rm{turb}}$ defined in Sect. 3; however, the effect here is marginal, as discussed in Sect. 3.3.

\subsection{Beta wind parameter and clumping onset}

\begin{figure}[t!]
    \centering
    \includegraphics[scale=0.78]{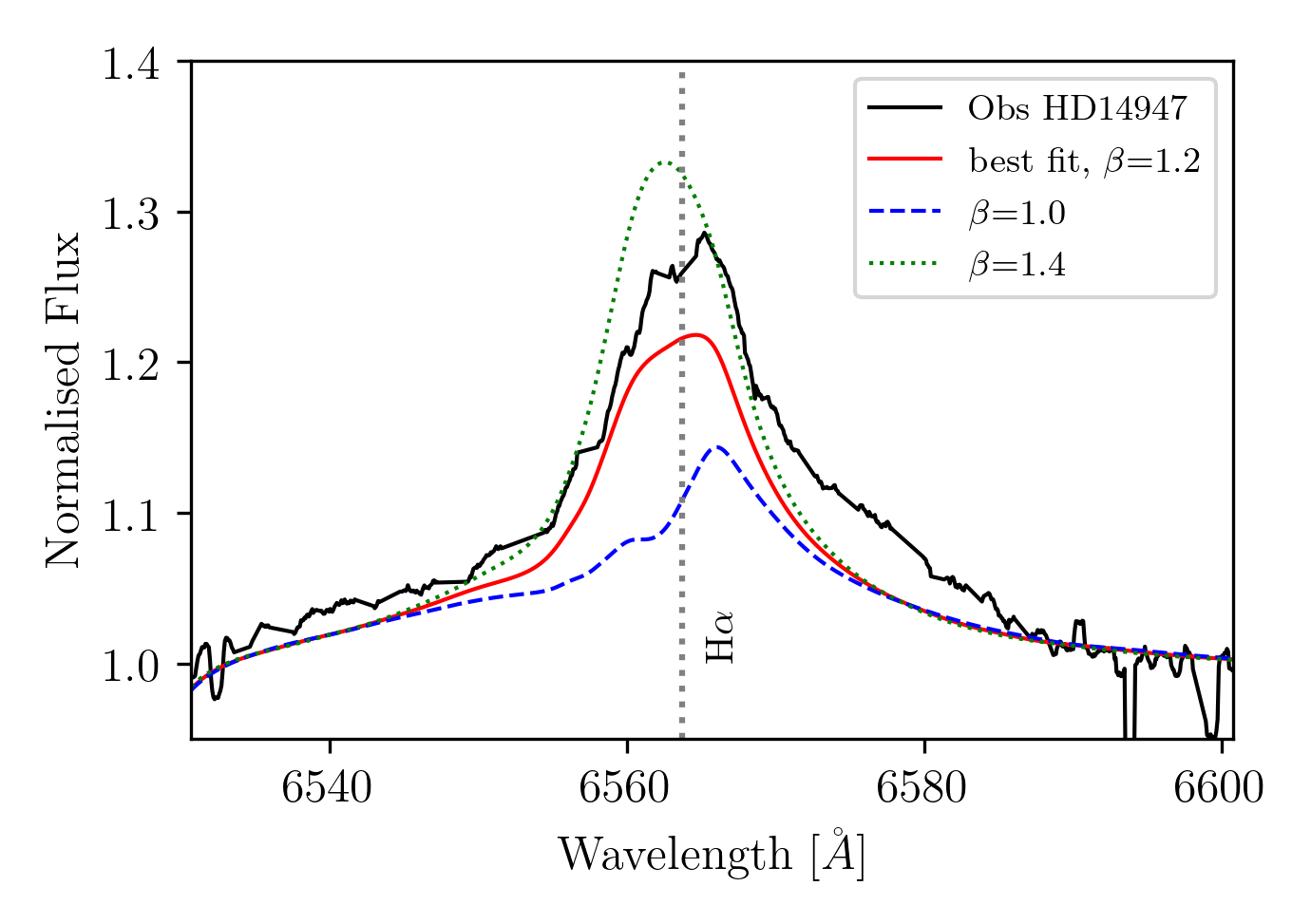}
    \caption{Effect of changing $\beta$ on a recombination diagnostic line H$\alpha$. GA best-fit model - red, best fit with decreased $\beta$ - blue, best fit with increased $\beta$ - green.}
    \label{fig: Beta Diagnostics}
\end{figure}

These wind parameters control the wind acceleration and the location where clumping starts in the wind. A low $\beta$ means the wind accelerates quickly thus casting an array of effects on line profiles, depending on which wind region dominates the line formation. In lines that are formed close to the stellar surface, for example H$\alpha$, a low $\beta$ causes a significant reduction in emission due to the reduction in wind density that results from the high velocity gradient close to the wind onset. The clumping onset velocity sets the point at which clumps develop in the wind: if the clumping onset velocity is high the clumps begin to develop further away from the star. Within our formalism the clumping factor then increases linearly to the maximum clumping factor. For this analysis, we fix the velocity at which the wind reaches maximum clumpiness to twice the onset velocity. Figure \ref{fig: Beta Diagnostics} shows the effect of varying $\beta$ on H$\alpha$. 

\subsection{Velocity filling factor and interclump density}

\begin{figure}[t!]
    \centering
    \includegraphics[scale=0.78]{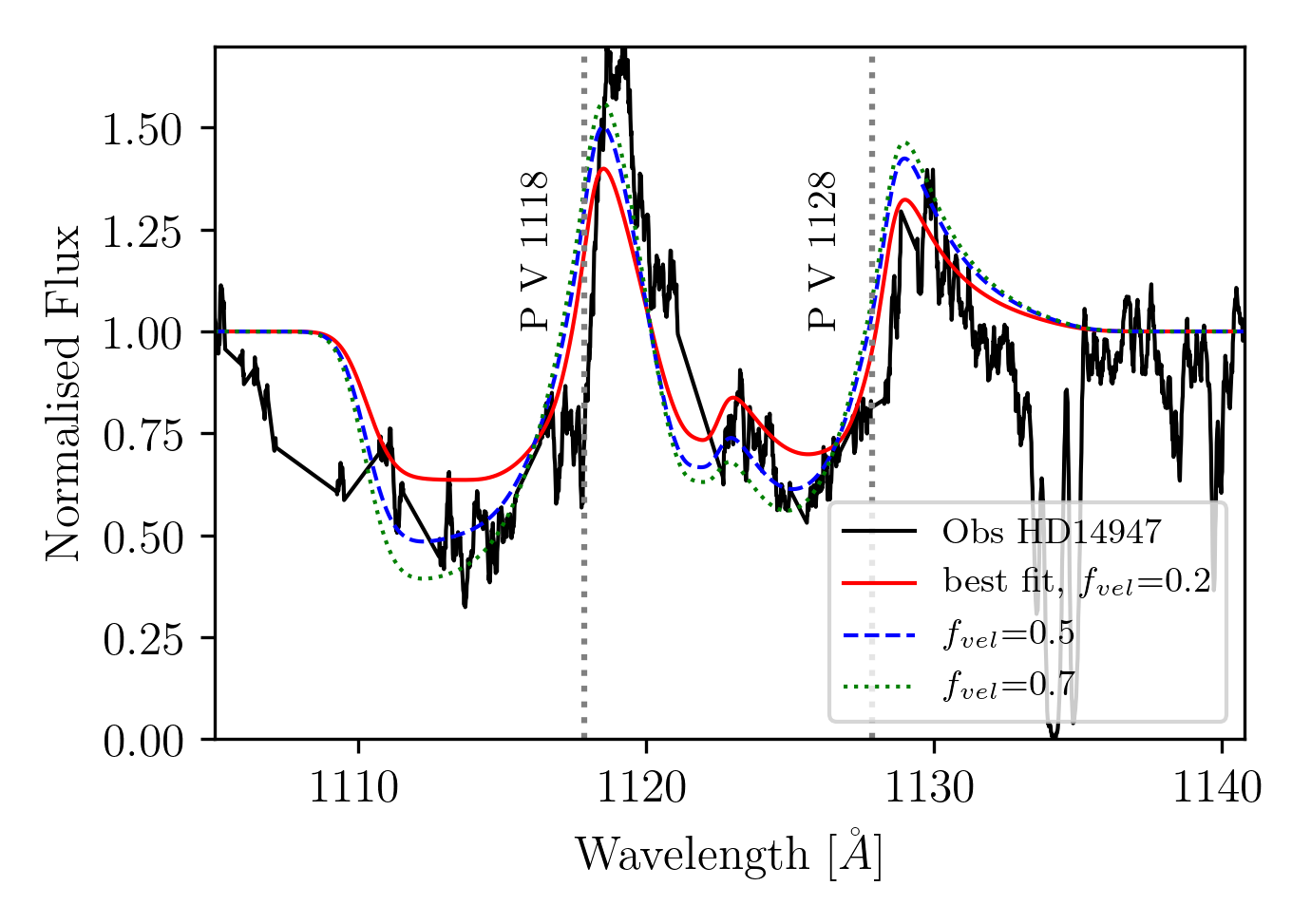}
    \caption{Effect of changing $f_\mathrm{vel}$ on \ion{P}{v} $\lambda\lambda$1118-1128 resonance lines. GA best-fit model - red solid line, best fit with 0.3 higher $f_\mathrm{vel}$ - blue dashed line, best fit with 0.5 higher $f_\mathrm{vel}$ - green dotted line.} 
    \label{fig: Velocity Filling Diagnostics}
\end{figure}

These factors act as light blockers, as they increase so does the probability of light interacting with material in velocity space and between clumps, respectively. These parameters are constrained using the P-Cygni troughs in UV resonance lines. As $f_\mathrm{vel}$ and $f_\mathrm{ic}$ are reduced the troughs become less saturated. This effect is shown in Fig. \ref{fig: Velocity Filling Diagnostics}.

\section{Results}

For each object, we present best-fit model optically thick clumping parameters and further photospheric and wind parameters as found by the GA in Tables \ref{tbl:Wind Parameters} and \ref{tbl:Parameters}, respectively. In Table \ref{tbl:Parameters} there are two rows per object; these correspond to the best fit found by BHL12, who use optically thin clumping, and by the GA, which allows for optically thick clumping. We also produce a number of different GA best fits using different clumping assumptions and fitting strategies. These are discussed in Appendix A and presented in Table \ref{tbl: Parameters Thin}. The best-fit model presented here is the model with the minimum global $\chi^{2} = \sum_{i=1}^{N} \chi_{\rm{i}}^{2} $, where $\chi_{\rm{i}}^{2}$ is the squared difference between model and data for the individual lines. The GA returns approximately 60\,000 models in order to explore the parameter space. An example of these model distributions is shown in Fig. \ref{fig: Chi2 distrib HD14947}. For an in-depth analysis of the fit quality per star see the Appendix A. 

For half the sample, our fits that allow for optically thick clumping agree with BHL12s effective temperature within 0.5kK. For HD210839 we find a temperature 1kK higher than BHL12. For the three remaining objects (HD16691, HD190429A and HD14947) we find clearly different solutions, with 1.5kK lower than BHL12, 3kK higher and 3.5kK higher. These stars are all O4.5 or earlier but these results are not pointing towards systematic differences in fitting, rather these are different fitting solutions than found in BHL12. These are discussed in the Appendix A. Our surface gravity values again present no systematic differences, although for the two stars of latest type we find significantly larger log $g$ values and higher $\beta$. Also, our surface gravity for HD16691 is high, but this also appears too high in the best fit, there are a number of difficulties in fitting this object discussed in Appendix A. 

The mass-loss rates are remarkably consistent with BHL12, with the largest difference being 0.1 dex, again in the two latest objects. In all objects we agree with BHL12 within 0.1 dex in mass-loss rate, despite including the effects of optically thick clumping and often finding significantly different clumping factors. We note that BHL12 had to reduce the phosphorus abundance in their models to find these mass-loss rates while we do not. We obtain a systematically higher broadening than BHL12 when considering the sum of rotation and macroturbulent broadening applied to each object. This can be explained by the implementation of macroturbulent broadening. In BHL12 an isotropic prescription is used while we implement a radial-tangential broadening. Three of the objects are included in \cite{Simon-Diaz2014}, the broadening values found by these authors using a Fourier method are consistent with those found from the GA, the largest discrepancies found in $v\sin i$ and $v_\mathrm{mac}$ are 5\% and 20\%, respectively. 

In terms of abundances, we find similar trends as BHL12, confirming the evolutionary status of these stars. The only clear difference is the difficulty in determining an oxygen abundance for $\zeta$ Puppis due to low signal-to-noise ratio (S/N) in the optical oxygen diagnostic. Abundances measurements are less consistent in the fits with optically thin clumping, due to a combination of mainly using diagnostic lines dominated by wind processes and an insufficient wind prescription in the fitting. This shows that if one attempts to determine surface abundances using wind line profiles the wind prescription has a significant effect. 

We find higher clumping onset velocities than BHL12, which may result from the different clumping laws used. This is discussed in detail in Sect. 7.4. For velocity filling, we find an average value throughout the sample of 0.44 $\pm$ 0.33 and an average interclump density of 0.13 $\pm$ 0.08. These findings are discussed in detail in Sect. 7.2.

The confidence intervals on the best-fit parameters are calculated as in \citet{Tramper2011}, \citet{Tramper2014}, \citet{Ramirez-Agudelo2017a} and \citet{Abdul-Masih2019}: all $\chi^{2}$ values are normalised such that the best fitting model has $\chi_{\rm{red}}^{2}=1$ and the error bounds include all models within the 95$\%$ confidence interval and are considered statistically equivalent. In this scheme, the exact confidence boundaries depend on the sampling around the peak. Another usual limitation of this approach is the assumption that the best-fit model is a good replication of the observed spectrum. 

To overcome the possibility of underestimated errors we run an orthogonal grid of models around the best fit, varying each free parameter independently, to complete any confidence interval coverage the GA may have missed. This method works well for parameters that have a fairly consistent effect on the global fitness and correlate little with other fitting parameters; for example, an abundance variation will have the same effect on the fitness of any relevant metal lines and other lines will be fairly unaffected, therefore giving a smooth fitness distribution around the global best fit. On the other hand, effective temperature has various effects on fitness depending on the line, an increase in temperature could well increase the fit to the Balmer series but reduce fitness to low ionisation helium and metal lines, thus the distribution of fitness of models around the best fit are less smooth. The errors on temperature quoted here are then extremely sensitive to the quality of fit, the line list used, and the weight of each line. We note also that these errors are statistical formal errors from the GA; the actual errors may be larger due to a number of effects. These effects include: Using a fixed luminosities, which affects the radii and thus mass-loss rates, potential shortcomings in the spectra such as normalisation, which has an impact on log $g$, and the limited line list used in the GA. In the case that errors remain unrealistically small we adopt a minimum error region, these are identifiable in Tables \ref{tbl:Wind Parameters} and \ref{tbl:Parameters} as any symmetric error values around the best-fit value.

\onecolumn

\begin{figure}[t!]
    \includegraphics[scale=0.43]{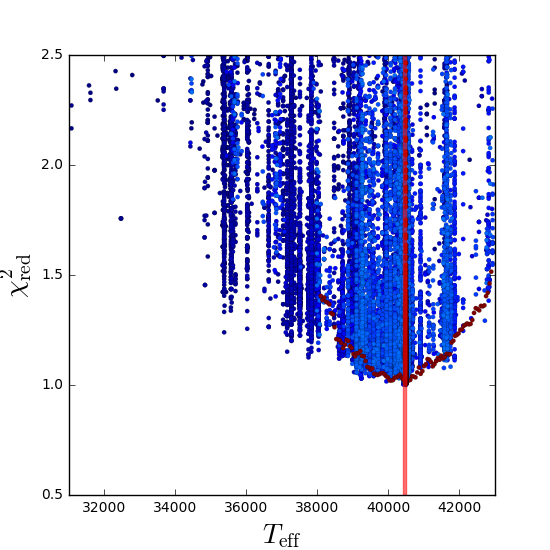}
    \includegraphics[scale=0.43]{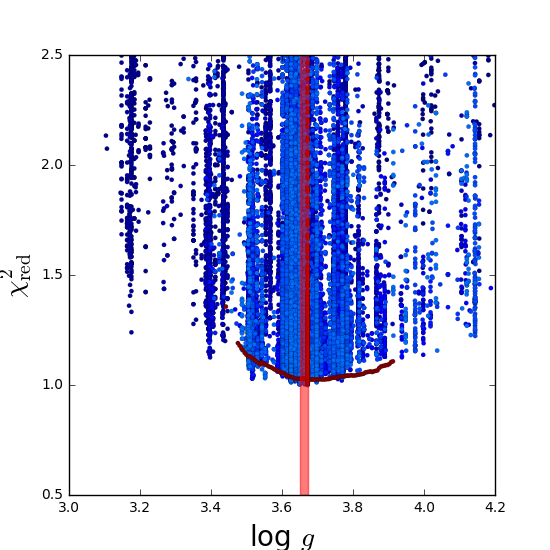}
    \includegraphics[scale=0.43]{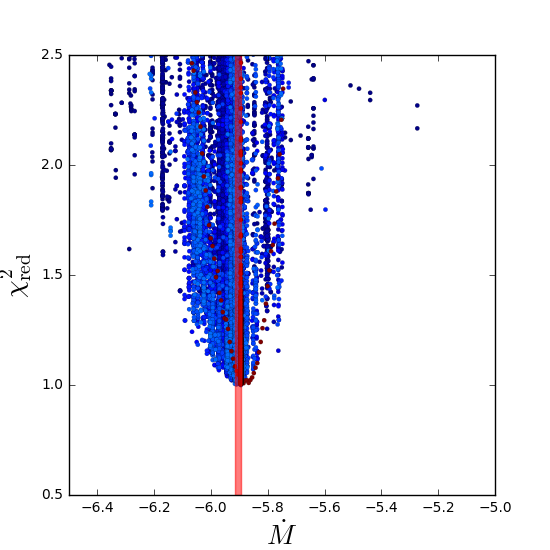}
    \includegraphics[scale=0.43]{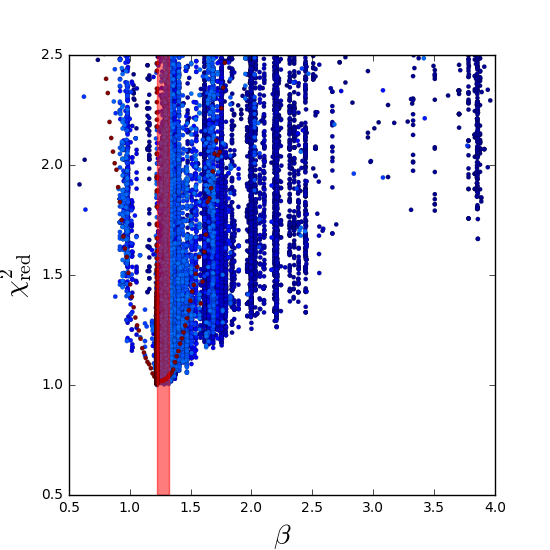}
    \includegraphics[scale=0.43]{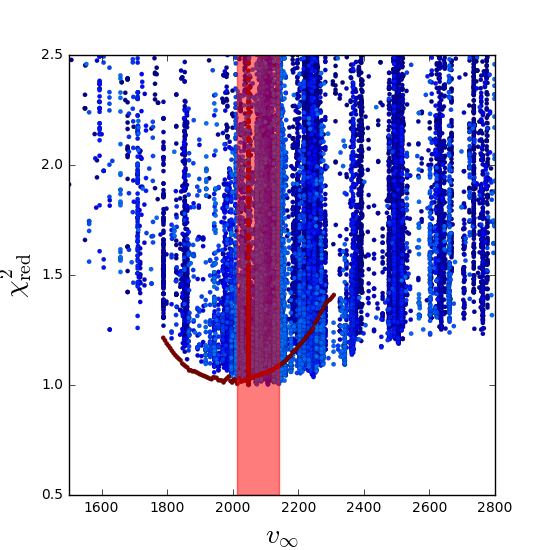}
    \includegraphics[scale=0.43]{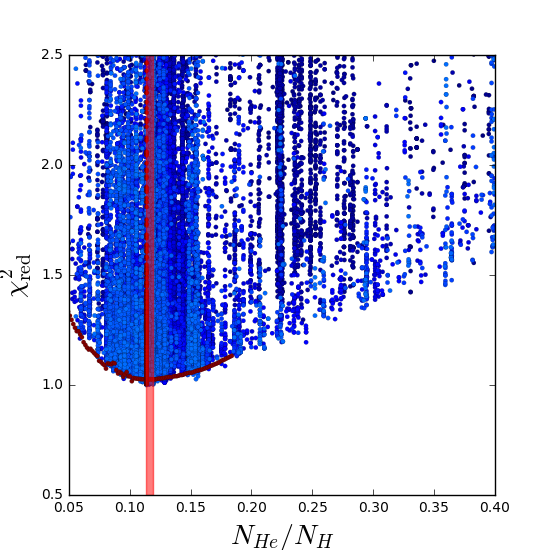}
    \includegraphics[scale=0.43]{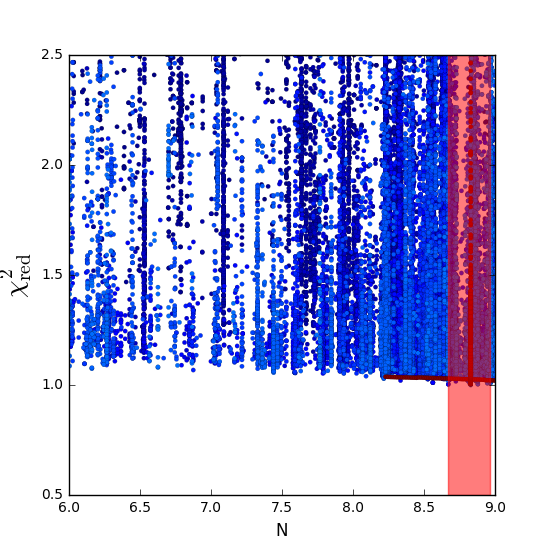}
    \includegraphics[scale=0.43]{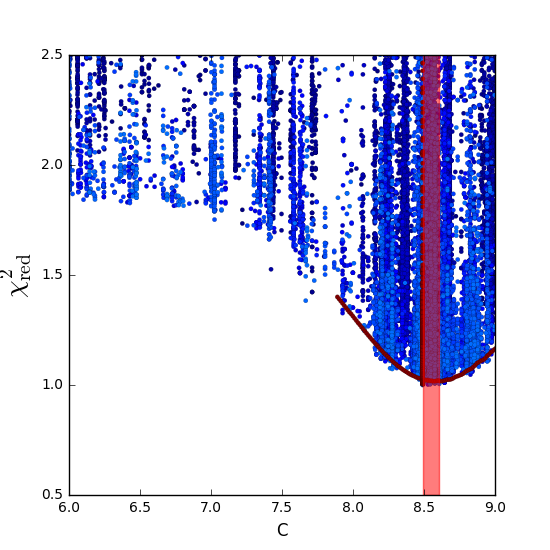}
    \includegraphics[scale=0.43]{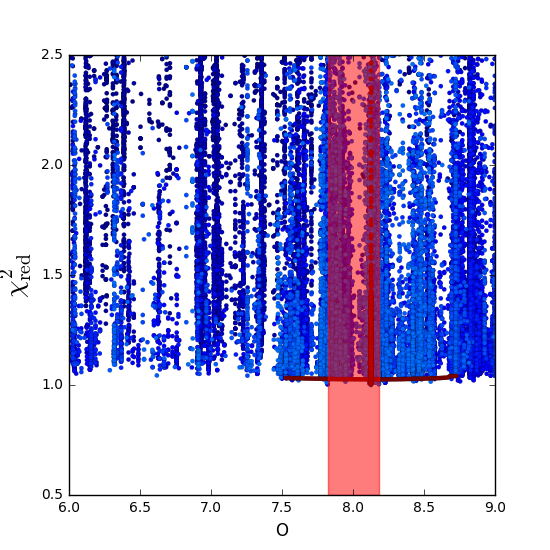}
    \caption{$\chi^{2}$ distribution per parameter for the stellar and wind parameters and chemical surface abundances of HD14947. The shaded region shows the confidence interval, marking out the upper and lower bounds surrounding the best fitting model for each parameter. The colour scheme evolves with the generations to give one a sense of the convergence of the algorithm, the darker blue points are the earliest generations and lighter blue points correspond to the latest models. The red points show the model grids run afterwards to constrain errors. This figure is continued on the next page.}
    \label{fig: Chi2 distrib HD14947}
\end{figure}

\begin{figure}[t!]
    \ContinuedFloat
    \includegraphics[scale=0.43]{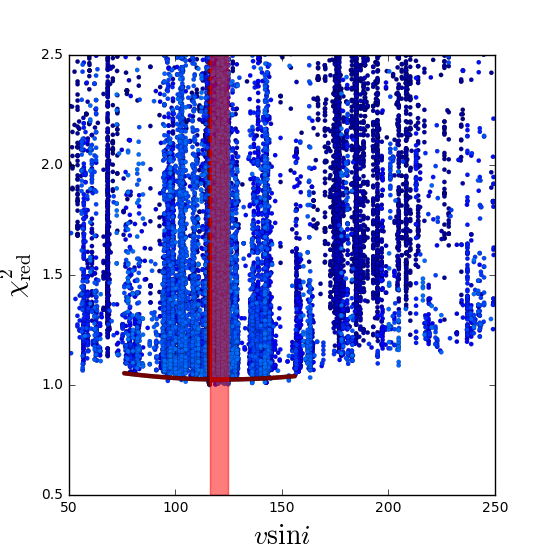}
    \includegraphics[scale=0.43]{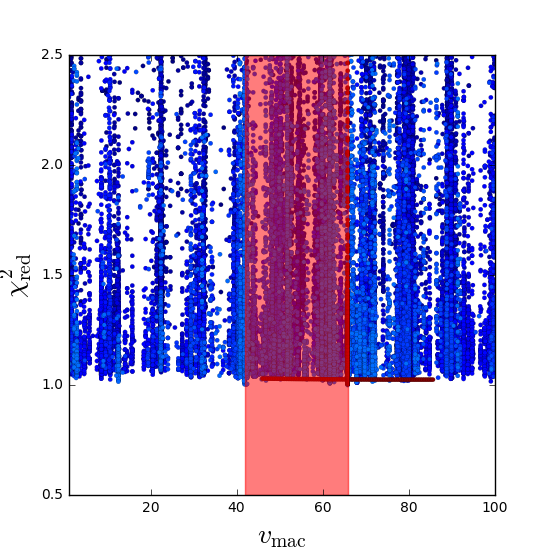}
    \includegraphics[scale=0.43]{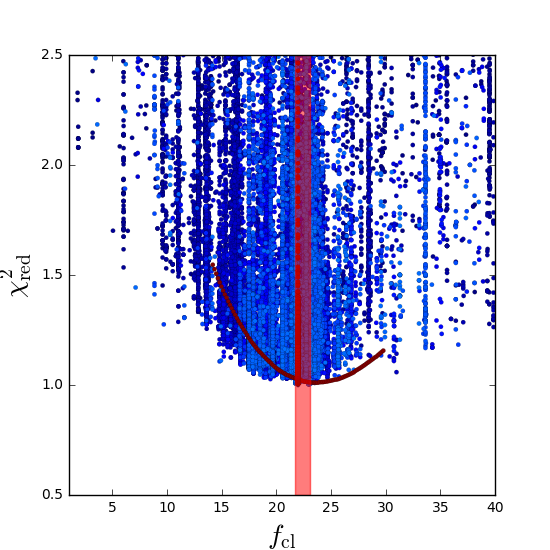}
    \includegraphics[scale=0.43]{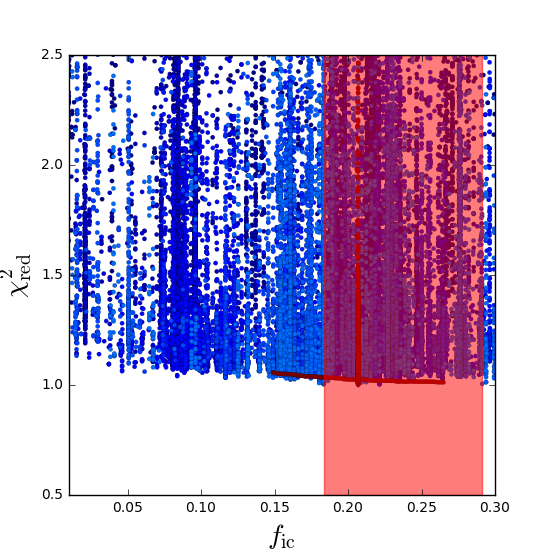}
    \includegraphics[scale=0.43]{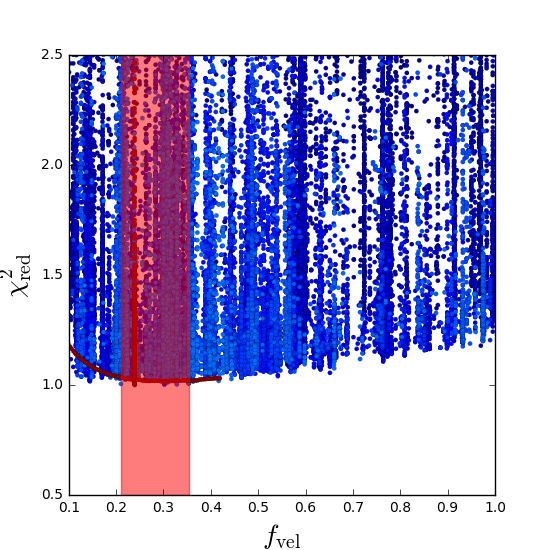}
    \includegraphics[scale=0.43]{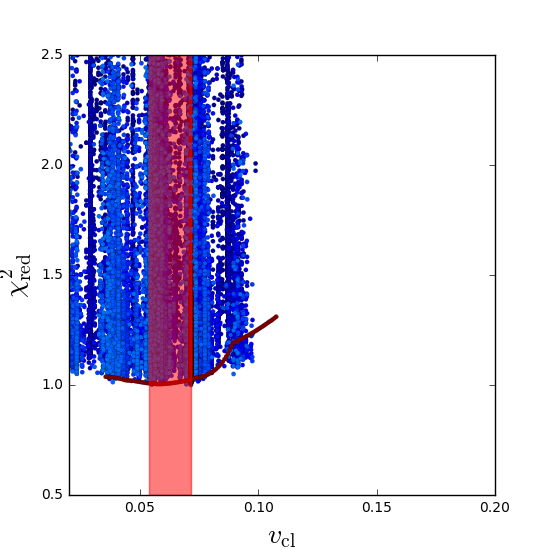}
    \caption{Continued.} 
\end{figure}

\vspace{2mm}

\begin{table}
    
\caption{Best fit wind parameters from GA fitting including optically thick clumping.}

\label{tbl:Wind Parameters}
\begin{tabular}{llllllll}
\hline
\hline

ID & $\dot{M}$ & $\beta$ & $f_\mathrm{cl}$ & $v_\mathrm{cl}$ & $v_\mathrm{cl}$ & $f_\mathrm{vel}$ & $f_\mathrm{ic}$ \cr
 & [$M_{\odot}yr^{-1}$] & & & [$v_{\infty}$] & [$R_{\rm{eff}}$] \\
\hline
HD16691	&	$-5.60	\pm 0.05$ 	&	$1.6	\pm 0.1 $	&	$20	_{-	2	}^{+	2 }$   &  $ 0.14 \pm 0.01$ & 1.10  & $ 0.99	_{-		0.06}^{+	0.05	}$ &   $0.14	_{-		0.06}^{+	0.05  }$ \\
HD66811 & $-5.70	\pm 0.05 $ & $0.8_{-		0.2}^{+	0.1	}$ & $	23	_{-	1	}^{+	8	}$ & $0.09 	_{-		0.1}^{+	0.2	}$& 1.12 & $0.20 	_{-	0.09	}^{+	0.05 }$ & $0.07 _{-	0.05	}^{+	0.06 }$  \\
HD190429A & $-5.69	\pm 0.05 $ & $0.9	\pm 0.1$ & $30 _{-	1	}^{+	3	}$ & $0.03	_{-		0.03}^{+	0.01	}$ & 1.04 & $0.46 	_{-	0.07	}^{+	0.05}$ & $ 0.06 \pm 0.05 $ \\
HD15570 & $-5.63	\pm 0.05 $ & $1.6 	\pm 0.1$ & $20_{-	1	}^{+	3	}$ & $0.07 \pm 0.02$ & 1.06 & $0.52 	_{-	0.05	}^{+	0.09}$ & $ 0.03 _{-	0.05	}^{+	0.06 }$ \\
HD14947 & $-5.89	\pm 0.05 $ & $1.2	\pm 0.1$  & $22 \pm 1$ & $0.07 	_{-		0.02}^{+	0.01	}$ & 1.06 & $0.24 	_{-	0.05	}^{+	0.11}$ & $0.21 _{-	0.05	}^{+	0.08 }$ \\
HD210839 & $-5.86	\pm 0.05 $ & $0.9	\pm 0.1$  & $23_{-	1	}^{+	2	}$& $0.12	_{-		0.03}^{+	0.01	}$ & 1.16 & $0.18 	_{-	0.09	}^{+	0.05}$ & $ 0.16 _{-	0.05	}^{+	0.06 }$ \\
HD163758 & $-5.81	\pm 0.05 $ & $2.3	\pm 0.1$  & $30	\pm 1 $ & $0.12 \pm 0.01$ & 1.06 & $0.84 _{-	0.18	}^{+	0.16	}$ & $0.12 \pm 0.06 $ \\
HD192639 & $-5.85	\pm 0.05 $ & $2.5	_{-	0.1	}^{+	0.3	}$& $32	\pm 2 $ & $0.14 \pm 0.01$ & 1.06 & $0.10 	_{-	0.07	}^{+	0.05}$ & $0.27 _{-	0.06	}^{+	0.05 }$ \\
\hline
\\              
\end{tabular}
\end{table}

\thispagestyle{plain}

\begin{landscape}

\thispagestyle{plain}

\begin{table}

\caption{Photospheric \& Wind Parameter best fits. The first row, for each object, shows the best-fit model parameters from BHL12, which uses optically thin clumping. The second row shows the GA run that allows for optically thick clumping.}

\vspace{2mm}

\label{tbl:Parameters}
\begin{tabular}{lllllllllllllll}
\hline
\hline
ID (HD) & $T_{\rm{eff}}$ & log $g$ & $R_\mathrm{eff}$ & $\dot{M}$ & $v_{\infty}$ & $\beta$ & $f_\mathrm{cl}$  & $v \sin i$ & $v_\mathrm{mac}$ & $N_{He}/N_{H}$ & $\epsilon$(C) & $\epsilon$(N) & $\epsilon$(O) \cr
Sp Type & [kK] & [cgs] & [$R_{\odot}$] & [$M_{\odot}yr^{-1}$] & [$\rm{km}\, \rm{s^{-1}}$] & &  & [$\rm{km}\, \rm{s^{-1}}$] & [$\rm{km}\, \rm{s^{-1}}$] & \\
\hline
16691	&	$41.0\pm1.0$ & $3.66\pm0.1$ & $18.4\pm0.9$ &$-5.52\pm0.03$& $2300\pm 100$ &	1.2	&	17&	135	&	37	&	0.15&	$6.5\pm0.2$ 	&$9.0\pm0.2$&	$7.8\pm0.3$	\\
O4 If	& $39.5 \pm 0.5$ & $	3.85_{-	0.05}^{+0.14}$& $19.9\pm0.5$&$	-5.60	\pm 0.05 $&$	2300	\pm 100 $&$	*1.6	\pm 0.1 $&$	20	_{-	2	}^{+	2	}$&$	188	_{-	25	}^{+	42	}$&$	68	_{-	21	}^{+	25	}$&$	0.13	\pm 0.02 $&$	6.6	_{-	0.1	}^{+	0.6	}$&$	9.0	\pm 0.1 $&$ 8.4	_{-	0.4	}^{+	0.3	}$ \\
&	&	&	&	&	&	&	&	&	&	&	&	\\																																																																
66811	&	$40.0 \pm 1.0$					&	$3.64 \pm 0.1$ & $18.7\pm0.9$					&$	-5.70_{-	0.05	}^{+	0.04	}$					&	$2300\pm 100$					&	0.9					&	20					&	210					&	90					&	0.16					&	$7.6\pm0.3$					&	$9.1\pm0.2$					&	$8.1\pm0.3$		\\
O4 I(n)fp	&$	40.0\pm 0.5 $&$	*3.52\pm 0.05 $&$18.6\pm0.5$&$-5.70\pm0.05$&$	2300	\pm 100 $&$	0.8	_{-	0.2	}^{+	0.1	}$&$	23	_{-	1	}^{+	8	}$&$	203	_{-	10	}^{+	40	}$&$	96	_{-	65	}^{+	10	}$&$	*0.25	_{-	0.03	}^{+	0.02	}$&$	*8.3	_{-	0.3	}^{+	0.1	}$&$	9.0	_{-	0.1	}^{+	0.2	}$&$	*6.9	_{-	0.7	}^{+	0.1	}$\\
&	&	&	&	&	&	&	&	&	&	&	&	\\																																																																	
190429A	&	$39.0 \pm 1.0$					&	$3.62 \pm 0.1$ &$20.8\pm1.1$					&	$-5.68\pm0.04$					&	$2300\pm 100$					&	1.0					&	25					&	150					&	57					&	0.15					&	$7.1\pm0.2$					&	$8.9\pm0.2$					&	$7.8\pm0.3$		\\
O4 If	&$	42.5	_{-	0.7	}^{+	0.5	}$&$	3.62	\pm 0.05 $&$17.4\pm0.4$&$	-5.69	\pm 0.05 $&$	2400	\pm 100 $&$	0.9	\pm 0.1 $&$	30	_{-	1	}^{+	3	}$&$	192	_{-	54	}^{+	10	}$&$	44	_{-	10	}^{+	56	}$&$	0.16	\pm 0.02 $&$	7.7	_{-	0.3	}^{+	0.1	}$&$	9.3	_{-	0.4	}^{+	0.1	}$&$	8.0	_{-	0.2	}^{+	0.4	}$\\
&	&	&	&	&	&	&	&	&	&	&	&	&	\\																																																																	
15570	&	$38.0 \pm 1.0$					&	$3.51 \pm 0.1$&$21.4\pm1.1$					&	$-5.66\pm0.04$					&	$2200\pm 100$					&	1.1					&	20					&	97					&	40					&	0.10					&	$7.5\pm0.2$					&	$8.6\pm0.2$					&	$8.3\pm0.2$		\\
O4 If	&$	38.0 \pm 0.5 $&$	3.45	_{-	0.06	}^{+	0.05	}$&$21.2\pm0.6$&$	-5.63	\pm 0.05 $&$	2700	\pm 100 $&$	*1.6	\pm 0.1 $&$	20	_{-	1	}^{+	3	}$&$	103	_{-	10	}^{+	11	}$&$	98	\pm 10 $&$	0.09	\pm 0.02 $&$	7.9	_{-	0.2	}^{+	0.2	}$&$	8.7	_{-	0.5	}^{+	0.5	}$&$	8.3	_{-	0.4	}^{+	0.1	}$\\
	&	&	&	&	&	&	&	&	&	&	&	&	&	\\																																																																	
14947	&	$37.0 \pm 1.0$					&	$3.52 \pm 0.1$ &$19.9\pm1.1$					&	$-5.85	_{-	0.07	}^{+	0.06	}$					&	$2300\pm 100$					&	1.3					&	33					&	130					&	36					&	0.12					&	$8.3\pm0.2$					&	$8.8\pm0.1$					&	$8.1\pm0.2$		\\
O4.5 If	&$	40.5 \pm 0.5 $&$	3.67	\pm 0.05 $&$16.6\pm0.4$&$	-5.89	\pm 0.05 $&$	2000	\pm 100 $&$	1.2	\pm 0.1 $&$	22	\pm 1 $&$	116	\pm 10 $&$	66	_{-	24	}^{+	10	}$&$	0.11	\pm 0.02 $&$	8.5	_{-	0.1	}^{+	0.1	}$&$	8.8	_{-	0.2	}^{+	0.1	}$&$	8.1	_{-	0.3	}^{+	0.1	}$\\
	&	&	&	&	&	&	&	&	&	&	&	&	&	\\																																																																	
210839	&	$36.0 \pm 1.0$					&	$3.54 \pm 0.1$ &$20.3\pm1.1$					&	$-5.85	_{-	0.07	}^{+	0.06	}$					&	$2100\pm 100$					&	1.0					&	20					&	210					&	80					&	0.12					&	$8.2\pm0.2$					&	$8.7\pm0.2$					&	$8.5\pm0.1$		\\
O6 I(n)fp	&$	37.0	\pm 0.5 $&$	3.47	_{-	0.09	}^{+	0.05	}$&$19.4\pm0.6$&$	-5.86	\pm 0.05 $&$	2100	_{-	128	}^{+	100	}$&$	0.9	\pm 0.1 $&$	23	_{-	1	}^{+	2	}$&$	214	_{-	10	}^{+	22	}$&$	62	_{-	12	}^{+	12	}$&$	0.17 \pm 0.02 $&$	8.0	\pm 0.1 $&$	8.9	_{-	0.6	}^{+	0.1	}$&$	8.7	_{-	0.6	}^{+	0.2	}$\\
	&	&	&	&	&	&	&	&	&	&	&	&	&	\\																																																																	
163758	&	$34.5 \pm 1.0$					&	$3.41 \pm 0.1$	&$21.1\pm1.2$				&	$-5.8	_{-	0.06	}^{+	0.05	}$					&	$2100\pm 100$					&	1.1					&	20					&	94					&	34					&	0.15					&	$8.6\pm0.2$					&	$8.8\pm0.2$					&	$8.4\pm0.2$		\\
O6.5 If	&$	34.0	\pm 0.5 $&$	3.45 \pm 0.05$&$21.5\pm0.7$&$	-5.81	\pm 0.05 $&$	2600	\pm 100 $&$	*2.3	\pm 0.1 $&$	30 \pm 1$&$	116	\pm 10 $&$	34	_{-	20	}^{+	10	}$&$	*0.28	\pm 0.02 $&$	7.9	\pm 0.1 $&$	9.3	\pm 0.1 $&$	8.6	\pm 0.1 $ \\
	&	&	&	&	&	&	&	&	&	&	&	&	&	\\																																																																	
192639	&	$33.5 \pm 1.0$	&	$3.42\pm0.1$ &$20.4\pm1.2$	&	$-5.92	_{-	0.08	}^{+	0.07	}$	 &	$1900\pm 100$	&	1.3	&	20	&	90	&	43	&	0.15	&	$8.2\pm0.2$	&	$8.8\pm0.2$	&	$8.6\pm0.2$	\\
O7.5 Iabf	&$	33.5	\pm 0.5 $&$	3.71	\pm 0.05 $&$20.7\pm0.6$&$	-5.85	\pm 0.05 $&$	*2700	\pm 100 $&$	*2.5	_{-	0.1	}^{+	0.3	}$&$	32	\pm 2 $&$	103	\pm 14 $&$	100	\pm 19 $&$	0.19	\pm 0.02 $&$	7.8	\pm 0.1 $&$	8.7	\pm 0.1 $&$	8.6	\pm 0.1 $\\
\hline
\\         
\end{tabular}
\begin{tablenotes}
\item{Asterisks denote dubious values, the origins of these values are discussed in Appendix A. The issue with a number of high $\beta$ values is discussed in Sect. 7.4.}
\end{tablenotes}
\end{table}

\end{landscape}

\twocolumn

\section{Discussion}

\subsection{Mass-loss rates and clumping factors}

It is now well established that neglecting clumping may lead to erroneous estimates of mass-loss rates. Up until now, mostly models including optically thin clumping have been used to constrain spectroscopic mass-loss rates. The effect of optically thick clumping on various line diagnostics has been studied but the resultant systematic effects on empirical mass-loss rates across a sample of early-type stars remains unclear. It is thought that the additional leakage of light through porous channels between clumps, allowed by optically thick clumping, may have a significant effect on the derived mass-loss rate. 

Overall our mass-loss rates are very similar to BHL12. The largest difference of 0.08 dex is only found in two stars while the average difference between the studies is 0.03 dex. For the present sample it seems that the mass-loss rates found with optically thick clumping and optically thin clumping do not diverge significantly. However, this is in the case of using a reduced phosphorus abundance to fit the unsaturated PV lines, in the rates determined using optically thin clumping. If a solar phosphorus abundance was used to determine the mass-loss rates with optically thin clumping we would see a systematic difference compared to the rates found with optically thick clumping. Instead, discrepancies are due more to goodness-of-fit on a case by case basis.

\begin{figure}[t!]
    \centering
    \includegraphics[scale=0.25]{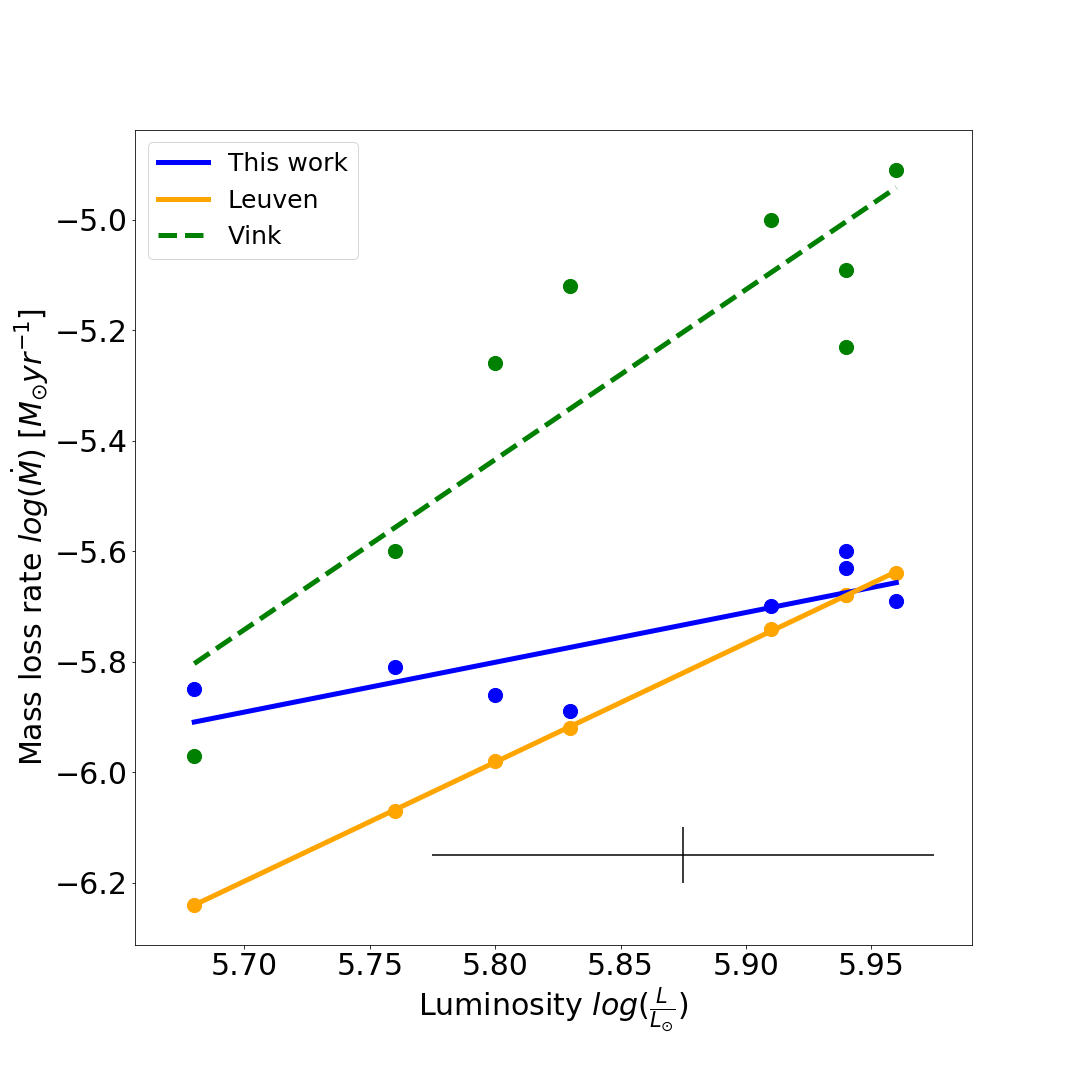}
    \includegraphics[scale=0.25]{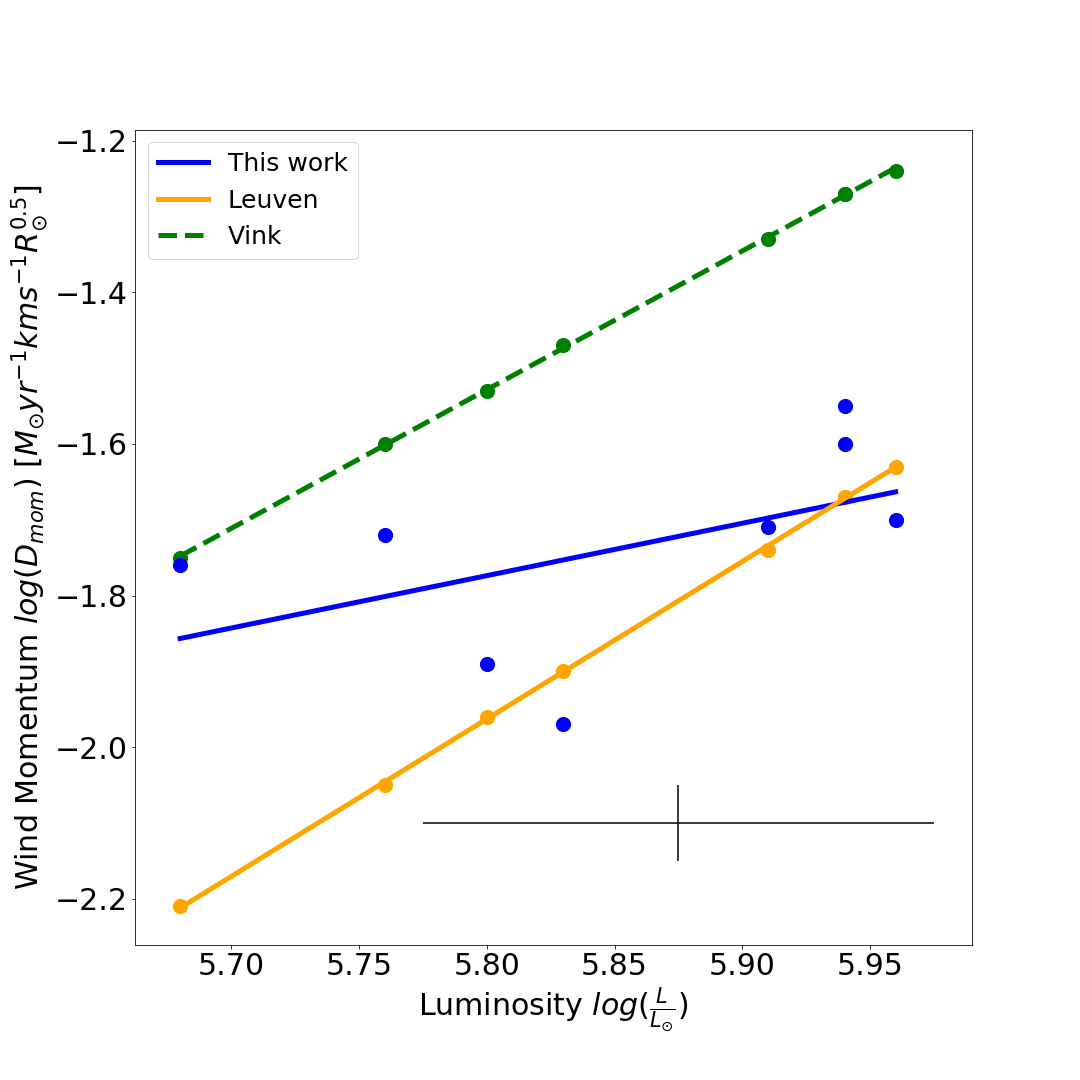}
    \caption{Mass-loss rates (upper panel) and modified wind-momentum (lower panel) found by the GA best fits, compared to the predictions made by Leuven (\citealp{Bjorklund2020}) and Vink (\citealp{Vink2000}) for each star in our sample. The GA mass-loss rates are those found from the best fit with optically thick clumping. Black cross in bottom right shows the minimum error on mass-loss rate and wind momentum $\pm0.05$, from this analysis, and error on luminosity $\pm0.1$ from BHL12.}
    \label{fig: Luminosity - mass-loss}
\end{figure}

We find, for example, for HD14947 that \cite{Vink2000} over-predicts the mass-loss rate by a factor of 3.9 compared to our results, which account for optically thick clumping, while the Bj\"{o}rklund et al. (2020) rates (hereafter the Leuven rates) predict values within 10\% of what is found by the GA. Such a reduction has been observed in a number of UV spectroscopic studies (\citealp{Crowther2002}, \citealp{Hillier2003}, \citealp{Bouret2004a}, \citealp{Fullerton2005}), and also in mass-loss rates determined from X-ray diagnostics \citep{Cohen2014}. Most of the mass-loss rates found with the GA sit slightly above the Leuven predictions, with the Leuven rates predicting a mass-loss rate 0.8 times of that observed, on average (Fig. \ref{fig: Luminosity - mass-loss}). The \cite{Vink2000} rates predict a mass loss that is on average a factor of 3.6 above the observed values. A similar trend can be seen in the modified wind-momentum (Fig. \ref{fig: Luminosity - mass-loss}). It is worth noting that clumping is not the source of discrepancy between these two theoretical mass loss predictions: the effect of clumping is prominent when comparing observational diagnostics to theoretical predictions, whereas the differences between these theoretical predictions come from the different techniques used to calculate the radiation force and solving the equation of motion (see discussions in \citealp{Sundqvist2019}; \citealp{Bjorklund2020}).

Throughout the sample we find clumping factors in the range of 20-32 for any wind prescription with an average of clumping factor of 25, while BHL12 finds values in the range 17-33 with an average clumping factor of 22. This is consistent with a radially average clumping factor found in the LDI simulations by \cite{Driessen2019} for an O-supergiant with $\zeta$ Puppis like input parameters. We note that these are maximum clumping factors. There is clear evidence that the clumping factor may decrease in the outer wind (\citealp{Puls2006}, \citealp{Najarro2011}); however, it is difficult to diagnose this with the spectral coverage used for this study. Therefore, we use a near-constant clumping law and leave a further investigation into the effect of clumping stratification on observational diagnostics to a future study with infrared spectral coverage. 

\subsection{Velocity filling factors}

For five stars in the sample the GA best fits have velocity filling factors in the range 0.2 - 0.6. These are the first systematic empirical constraints derived for this parameter. The exceptions are HD192639 with a lower velocity filling and HD16691/HD163758, which show uncharacteristically high velocity filling factors, bordering on recovering a velocity profile reminiscent of a smooth wind outflow. On average, we find a velocity filling factor of 0.44 for the full sample. If the three stars that show uncharacteristically high or low velocity filling factors are excluded the average is reduced slightly to 0.32. Theoretical constraints for the velocity filling factor can be made using LDI simulations from \cite{Driessen2019}. We calculate a clump velocity span and clump separation from a prototypical O-supergiant model using region cutoffs at $v/v_{\infty}$ = 0.05 or 1.1 $R_{*}$ and consider a clump to be any region with a density higher than $2<\rho>$. These limits predict a velocity filling factor $f_\mathrm{vel}=0.56$, which is overall consistent with the empirical constraints. Altering these limits will influence the velocity filling estimate. Therefore, within reasonable density limits $2-5<\rho>$ we find an estimate of $0.57 > f_\mathrm{vel} > 0.44$, as computed from the velocity span of the clumps. For the stars that do not agree with theoretical estimates we additionally attempt to fit the velocity filling parameters with fixed photospheric parameters, varying only the optically thick wind parameters. In these tests, it is possible to find fits of similar quality to the UV resonance lines with lower velocity filling but, to compensate, the interclump density converges to the upper limit of our parameter space. We are certainly limited by data quality, specifically S/N in this sample and so it is difficult to distinguish whether these stars do show uncharacteristically high velocity filling (for which the physical motivation is unclear) or if a higher S/N would allow fitting with lower velocity filling. 

Similar parameters have been explored in works including \cite{Oskinova2007} and \cite{Surlan2013} although the implementation of such a parameter is quite different; these authors use a clump separation parameter in physical space, with discrete clumps each providing independent contributions to line driving and absorption. Implementation in FASTWIND is a wind averaged velocity filling, so a direct comparison between parameters is difficult. In general our results are qualitatively consistent with these studies. We find that allowing optically thick clumping in the stellar wind has a significant impact on line profiles, and the resulting enhanced porosity reduces saturation in P-Cygni profiles. 

\subsection{Interclump density}

This parameter is hard to constrain empirically, as can be seen throughout the GA fits for this sample. We generally find flat distributions around a best $\chi^{2}$ with confidence intervals encompassing large portions of the parameter space. This parameter has a similar, although weaker, influence on the model as the velocity filling factor. Therefore, it may be the case that we are unable to significantly distinguish between values of interclump density in a global $\chi^{2}$ metric, mostly due to relatively low S/N in our UV observations and the fact that only a few lines are impacted, decreasing the impact of $f_\mathrm{ic}$ on the global $\chi^{2}$.

With this in mind, a loose constraint on the interclump density can be estimated by looking specifically at the fitness distribution for saturated P-Cygni line profiles. It is not possible for a line to saturate if the interclump medium is completely void and $f_\mathrm{vel}$ is significantly lower than unity, so these lines could act as a lower limit diagnostic \citep{Sundqvist2011a}. The \ion{C}{iv} $\lambda\lambda$1548-1550 line is mostly saturated in our sample, and this line provides confidence intervals around 0.15 - 0.3 on the $f_\mathrm{ic}$ value,  indicating a fairly high density in the interclump medium. Such a value is then higher than typically seen in 1D LDI models of clumping, which indicates that the lateral `filling in' of radially compressed gas, observed in 2D LDI simulations \citep{Sundqvist2018a}, likely plays an important role in setting the clump to interclump density contrast. While a global fit might point towards a more tenuous interclump medium, due to the degeneracy with the velocity filling factor, it is possible that actually a higher interclump density is preferred. Perhaps it will be possible to constrain this in other P-Cygni profiles with a higher S/N in the UV but for now saturated lines point towards a relatively high interclump density. \citet{Zsargo2008} also showed that a tenuous interclump medium is required in hot stars, in the context of reproducing the features of `super-ionised' species such as OVI in UV and X-ray observations.

\subsection{Beta and clumping onset}

\begin{figure}[t!]
    \centering
    \includegraphics[scale=0.25]{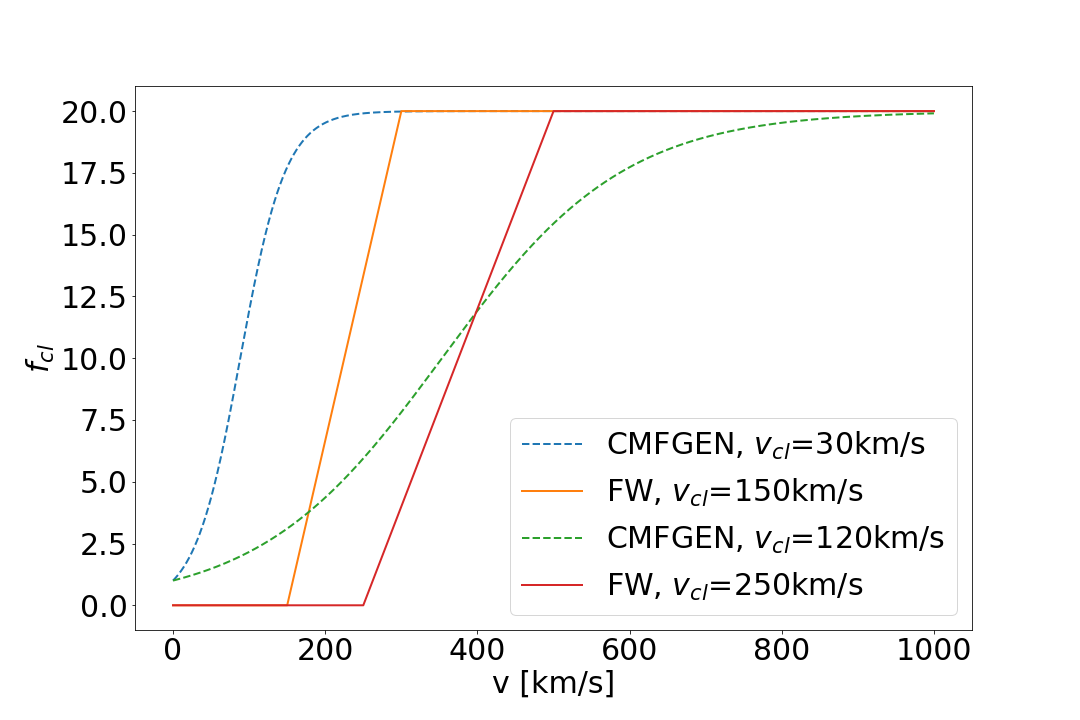}
    \caption{Clumping profiles for CMFGEN (BHL12) and GA fits, showing clumping profiles with the same maximum clumping factor but varying the onset velocity. The velocities shown correspond to best-fit onset velocities found for HD14947 (slow rotator - BHL12:vcl=30, GA:vcl=150) and HD210839 (fast rotator - BHL12:vcl=120, GA:vcl=250).} 
    \label{fig: Clumping Laws}
\end{figure}

We find high velocities for clumping onset relative to previous spectroscopic studies. BHL12 has an average value of 60 $\mathrm{km s^{-1}}$ for this sample, and our optically thick models give an average almost five times higher at 290 $\mathrm{km s^{-1}}$. However, the clumping profile is different between our study and BHL12. In BHL12, an exponential increase in clumping factor with velocity is used while we implement a linear increase. In practice, the profile is similar between both studies but the nature of the clumping profile in BHL12 means that the onset velocity is not a strict onset, but the velocity at which the clumping factor reaches a threshold value. Also, the clumping development is more gradual for higher onset velocities. An example of the clumping profiles used in CMFGEN and FASTWIND are shown in Fig. \ref{fig: Clumping Laws}. For most stars, BHL12 uses a clumping onset velocity of 30 $\mathrm{km s^{-1}}$ and we find higher onset velocities, meaning we still have quite low clumping factors in regions where BHL12 is approaching maximum clumping. For the fast rotators where BHL12 finds higher onset velocities our clumping profiles are very similar. 

For most lines, the clumping onset is unconstrained but it is clear that high onset velocities provide higher fitness in recombination lines throughout the sample. We note, however, that BHL12 had difficulty fitting the double peaked profiles that emerge in the emission components of the fast rotators in the sample. They were able to improve their fits by increasing the clumping onset velocity by around 100 $\mathrm{km s^{-1}}$. In this study we find that the fits can indeed be further improved if the onset velocity is increased by an even greater extent, around 200 $\mathrm{km s^{-1}}$. On the other hand, we point out that in this work we have convolved all lines with a rotational broadening profile assuming a fixed photospheric value for the projected rotation velocity $v \sin i$. But due to angular momentum conservation in the wind, this value should really decrease as one moves away from the stellar surface. \cite{Bouret2008} shows that if a depth-dependant rotation is applied to the models the observational profiles can be better reproduced. We have not investigated this, but our findings are inline with the predictions made by these studies and so we can suggest that depth-dependent effects on the rotational broadening profiles of these objects are likely to have a significant effect on some of the emergent wind profiles.

In some stars we observe very high clumping onsets and high $\beta$ values. In these cases the GA tries to fit the core and blue edge of emission profiles, to varying degrees of success, generally not very well. The high clumping onsets might not have a huge impact on photospheric diagnostics but the resulting high $\beta$ values may influence photospheric parameters. 

Other studies have also noted that a high $\beta$ mimics the effect of clumping as a higher $\beta$ leads to a denser wind near the wind onset region, having a similar effect as an increased clumping factor or mass-loss rate \citep{Petrov2014}. This certainly appears to hold true, at least if one were to consider recombination lines as shown for H$\alpha$ in Fig. \ref{fig: Beta Diagnostics}. It is clear from this study that an increased clumping onset velocity can act to compensate for this increased emission, leading to no clear systematic impact on clumping or mass loss. However, it is possible that such an offset could occur if the clumping onset was fixed to lower velocities. We note that in our fits with optically thin clumping the clumping onset velocity is fixed to $0.05v_{\infty}$, which leads to onset speeds around 100 $\mathrm{km s^{-1}}$. In these fits with clumping onset fixed to a lower velocity we do not find the same high $\beta$ values as derived with optically thick clumping, but values compatible with those from BHL12 in their analysis. However, in these stars we also find somewhat different mass-loss rates than obtained when fitting with optically thick clumping. An early onset of clumping has been found both from previous diagnostic studies (Puls et al. 2006; Cohen et al. 2011) and from theoretical LDI simulations (Sundqvist \& Owocki 2013). Moreover, since for Galactic O supergiants radiative acceleration might exceed gravity even in deep sub-surface layers (at the so-called `iron-opacity bump' at $T \approx 200$ kK), it is possible that this might trigger a very turbulent atmosphere \citep{Jiang2015}; resulting, not only in the observed large values of `macroturbulence', but also affecting clumping properties near the photosphere. Future multi-dimensional LDI simulations might examine such a potential connection by proper downward extension to those deep sub-surface layers. 

It is still unclear whether the high onsets seen in this sample are truly high clumping onset velocities or an observational proxy of another physical process. Regardless, it is clear from our current dataset that it is difficult to distinguish between slow wind acceleration (high $\beta$) and an earlier onset of clumping. As such, the overall result here of generally quite slow acceleration in near photospheric layers, accompanied by a quite high velocity-onset of clumping, should be re-investigated in future work. 

\subsection{Phosphorus}

A common problem in spectral analyses of stars in this regime is the phosphorus abundance. Using models with optically thin clumping, up until this point, it has only been possible to reproduce the P-Cygni absorption by reducing the phosphorus abundance to sub-solar levels. This was first noticed by \citet{Pauldrach1994a} and has been reobserved in a number of subsequent studies (see e.g. \citealp{Pauldrach2001}, \citealp{Crowther2002}, \citealp{Hillier2003}, \citealp{Bouret2004a}). The phosphorus abundance reduction is generally thought to be unphysical as phosphorus should not be processed to this extent throughout the stellar lifetime, this is discussed in detail in \citet{Fullerton2005}. To address this we exclude the phosphorus abundance as a parameter in our fitting, opting instead to keep phosphorus fixed to solar abundance. In a test run, we tried to fit the phosphorus abundance in the optically thin clumping case and indeed reproduced the problems discussed above. Therefore, we do not try to fit this line with optically thin clumping, we only include the phosphorus lines when modelling optically thick clumping to see if the added wind porosity in velocity space helps to reproduce the lines. Other studies (e.g. \citealp{Oskinova2007}, \citealp{Sundqvist2009}, \citealp{Sundqvist2011a}, \citealp{Surlan2013}) have been able to reproduce these lines with solar phosphorus abundance by including the effects of wind velocity porosity, and we obtain the same result here, albeit using a quantitative fitting approach. Throughout our sample we are able to consistently fit the UV phosphorus resonance doublet as a result of our inclusion of wind velocity porosity without lowering the phosphorus abundance. An example of this effect on the emergent line profile is shown in Fig. \ref{fig: Velocity Filling Diagnostics}. 

\section{Conclusions}

We performed automated spectroscopic fitting of stellar atmosphere models using a genetic algorithm for a sample of O supergiants, simultaneously across optical and UV wavelengths, to investigate the effect of optically thick clumping and wind porosity on their wind properties. Our results are:

- We are able to resolve the PV problem by including optically thick clumping. By including the effects of wind porosity in velocity space it is possible to reproduce the phosphorus line profiles with solar phosphorus abundance.

- If the PV lines are excluded from the analysis, optically thick clumping does not significantly affect the resulting mass loss measurements. Nonetheless, we stress the general importance of a multi-diagnostic approach to constraint this mass loss, where clumping must be considered in order to obtain consistent line-profile fits across the optical and UV ranges. 

- The resulting empirical mass-loss rates are well constrained by our GA fitting. Overall, they agree well with the new theoretical predictions by \citet{Bjorklund2020} and are on average a factor of 3.6 lower than the models by \citet{Vink2000}. Since the latter are often used as standard in evolutionary calculations, this strengthens earlier claims (e.g. \citealp{Sundqvist2011a}; \citealp{Najarro2011}; \citealp{Surlan2013}; \citealp{Cohen2014}; \citealp{Krticka2017}; \citealp{Keszthelyi2016}; \citealp{Sundqvist2019}; \citealp{Bjorklund2020}) that line-driven mass-loss rates included in current models of massive-star evolution should be reduced. Although the empirical study here needs to be extended to a larger sample of stars (and the theoretical models by Bj\"{o}rklund to a larger grid), our combined results suggest that simply scaling down the earlier rates by a factor of 3 is a quite reasonable first approximation for the considered spectral types. 

- From our systematic study, we are able to derive empirical constraints on wind structure parameters associated with the effects of optically thick clumping. On average, we find that clumps cover roughly half of the wind velocity field and that inter-clump densities are around 10-30 \% of the mean wind density. While such clump velocity filling factors agree well with 1D LDI simulations of clumping, the rather high interclump densities suggest that the lateral filling-in of radially compressed material seen in corresponding 2D models might be critical for setting the density scales of the rarefied interclump medium. Regarding the clumping factor, we derive mean maximum values on order 20, which again agrees well with the results of current 1D LDI simulations for Galactic O supergiants.  

- We notice our best fit models often have significantly higher clumping-onset speeds than what was found in previous studies; however, we stress both the importance of future studies accounting for a depth-dependent rotational broadening profile and current degeneracy-issues between the onset of clumping and the wind acceleration parameter $\beta$ in near-photospheric layers. 

We intend to extend this type of study to a wider range of spectral types across the upper Hertzsprung-Russell diagram and to lower metallicity environments. This method can also be applied within the framework of the ULLYSES and XShootU programmes which will provide large datasets of massive stars in the Magellanic Clouds.

\begin{acknowledgements}
This project has received funding from the  KU Leuven Research Council (grant C16/17/007: MAESTRO), the FWO through a FWO junior postdoctoral fellowship (No. 12ZY520N) as well as the European Space Agency (ESA) through the Belgian Federal Science Policy Office (BELSPO). The computational resources and services used in this work were provided by the VSC (Flemish Supercomputer Center), funded by the Research Foundation - Flanders (FWO) and the Flemish Government. This research has made use of the SIMBAD database, operated at CDS, Strasbourg, France.
\end{acknowledgements}

\bibliographystyle{aa} 
\bibliography{porosity.bib}

\begin{appendix}

\section{Star-by-star and optically thin versus thick clumping}

In this appendix we discuss the quality of the best fits found by the GA for each object in detail, highlighting specific shortcomings and discussing potential causes. We also introduce the initial round of fitting, for which we run the GA including the assumption that all clumps remain optically thin. 

In the fits with optically thin clumping we use the same clumping profile as we use for optically thick clumping, diverging from BHL12 in the clumping law as discussed in Sect. 7.4. We also opt to fix the clumping consistently throughout the optically thin clumping fits, highlighting issues that may arise if a general assumption for the clumping factor is used and this parameter is not optimised. In BHL12 a clumping factor of 20 is used for 5 objects and so we use this value. Again, this is not optimised to test the effect and in 3 objects (HD14947, HD190429A and HD16691) BHL12 adapts the clumping factor. In BHL12 the clumping onset velocity is fixed to 30 $\mathrm{km s^{-1}}$ (and increased for 3 objects: HD16691, HD66811 and HD210839), whereas we opt to fix our clumping onset speed to 0.05$v_{\infty}$, which results in an average clumping onset speed of 110 $\mathrm{km s^{-1}}$. HD210839 and HD163758 display some effects of a clumping structure that is not optimised. Here, the surface gravities are significantly lower than what would be determined using H$\gamma$ more or less alone; the low surface gravity comes as the GA attempts to fit H$\alpha$ and settles on a lower value as it cannot shape the profile with clumping. If one were to vary the clumping factor a more realistic surface gravity can be found. Another way to get around this is to increase the weight of the Balmer lines H$\gamma$ and $H\delta$ to constrain surface gravity. In a sense, these optically thin clumping `best fits' are not true best fits, but the best that can be achieved with the input assumptions and caveats. However, the optically thick clumping best fits allow more of these parameters to be free and result in more optimised fits. 

We have found temperature discrepancies larger than 1kK (which is more than twice our minimum error) compared to BHL12 for three stars in the sample with similar helium and metal line strengths. This may indicate a difference between CMFGEN and FASTWIND as seen in \citet{Massey2013}. However, it could also be that the GA converges on an alternate solution and other parameters are adjusted accordingly to compensate. It is difficult to make generalised statements across the sample when comparing to BHL12, a more complete picture is gained by looking at the fits to each object individually. We also notice that when fitting with optically thin clumping, we sometimes find significantly different surface abundances than found with optically thick clumping. This shows that the clumping assumption used can heavily influence abundance determination when fitting line profiles shaped by the winds.

\subsection{HD16691}

This is a difficult star to do a fit comparison to BHL12 as the GA best fit with optically thin clumping is particularly bad, largely due to difficulties reproducing H$\beta$ and H$\alpha$. The inclusion of optically thick clumping was also unable to reproduce $H\beta$ so we removed it from the global fit with optically thick clumping presented here. In this case, these lines drive the optically thin clumping fit to a $v \sin i$ = 250 $\mathrm{km s^{-1}}$ and $v \sin i$ = 85 $\mathrm{km s^{-1}}$. Such a high value is not supported by the metal absorption lines, which are all clearly overbroadened. The broadening parameters in the optically thin clumping fit are not consistent between UV and optical, it would be helpful to carry out a more extensive analysis of this star in the future. 

Despite the difficulties in fitting, when comparing optically thin wind clumping fits between the two studies we find consistent values for $T_{\rm{eff}}$, log $g$, $\beta$, $v_{\infty}$, and surface abundances. We find a significantly lower mass-loss rate than BHL12, by a margin of 0.2 dex (four times the minimum error). A difference was indeed expected as we use a higher clumping factor than BHL12. The cause of the difference could lie in the clumping prescription. Given similar onset values are found in both fits, the clumping profile used in BHL12 has a much more gradual development. With a more rapid onset of clumping, here we require a lower mass-loss rate to reproduce the spectrum. We obtain noticeably poorer fits to \ion{He}{i} $\lambda$4471 and \ion{He}{ii} $\lambda$4541 (our main temperature diagnostics), but this is due to the aforementioned overbroadening and helps to explain the discrepancy in helium abundance. 
We find significantly different best-fit parameters when allowing optically thick clumping. This is particularly notable as the maximum fitness found is significantly higher than that found with optically thin clumping, despite a peculiar velocity filling factor that borders on replicating a smooth wind. Our $T_{\rm{eff}}$ is 1.5kK lower than BHL12 (three times larger than minimum error) and log $g$ 0.2 higher (four times minimum error). Comparing the relevant diagnostics, both studies are able to reproduce optical HeI lines while BHL12 has a more accurate optical HeII strength. The optically thick clumping finds a particularly good fit to the strengths of $H\delta$ and $H\gamma$ compared to BHL12; however, the wing fits may be slightly better in BHL12, excluding the nitrogen line blend in H$\delta$, which BHL12 often has difficulty fitting. We note also that there is a sharp drop in fitness for both $H\delta$ and $H\gamma$ at $\beta$ lower than 1.6, while a $\beta$ consistent with BHL12 is preferred by wind emission lines $H\alpha$ and \ion{He}{ii} $\lambda$4686. The emission lines also drive a high clumping onset velocity, three times that found by BHL12, showcasing the issues discussed in Sect. 7.4.

We obtain a worryingly high velocity filling factor with a reasonable interclump density in our best fit. These parameters show no clear fitness distribution peaks in most lines except for \ion{N}{iv} $\lambda$1718, which clearly prefers a high velocity filling. In another GA run we fixed the photosphere properties to those found in the optically thin clumping run and vary only the wind parameters. In this run we see a more reasonable velocity filling and an overestimated interclump density. In this run the velocity filling converges on a more physically motivated value of 0.6; however, the fitness distributions per line do not change considerably, nor does the best-fit spectrum. It is possible that a lower velocity filling could be compensating for the fixed higher temperature and lower log $g$. As a result, the interclump density is increased to fit the UV P-Cygni troughs, highlighting the degeneracy between these wind parameters. 

In a further test run, we fix all photospheric parameters to those found by the GA optically thin fit and vary only wind parameters. In this test we allow for optically thick clumping and fit an even smaller subset of lines, only those heavily affected by the stellar wind. In this case our mass loss, $\beta$, and clumping onset velocity match BHL12, but $f_\mathrm{cl} = 10$. We also find a reasonable $f_\mathrm{vel} = 0.5$ but with an extremely high $f_\mathrm{ic} = 0.3$, which is only limited by our upper parameter boundary. This test has a very similar quality of fit as found by the GA with optically thick clumping. With all of this is mind, and the peculiar H$\beta$ profile, it is clear that this star has a unique wind profile worth further investigation.

\subsection{HD66811 - $\zeta$ Puppis}

HD66811 appears to have two solutions for $T_{\rm{eff}}$ and log $g$: after finding 41kK and 3.5 with optically thin clumping and 40kK and 3.5 with optically thick clumping, another GA run, which allows for optically thick clumping, was submitted that returned 41kK and 3.6. Comparing these to the 40kK and 3.6 found by BHL12, a safe conclusion is to say the solution lies within $T_{\rm{eff}}=$40-41kK and log$g=$3.5-3.6, this fits comfortably within the errors estimated by BHL12 but is beyond the margin provided by our perhaps too optimistic minimum errors. 

Most parameters agree between BHL12 and the GA with optically thick clumping, apart from the clumping onset velocity and some of the abundances. The onset velocity is again heavily restricted to higher values by $H\alpha$ and \ion{He}{ii} $\lambda$4686, this is discussed in Sect. 7.4. When looking at all the GA runs, each abundance found by BHL12 is found at least once. The carbon abundance is matched by the fit with optically thin clumping (run 1), the nitrogen matched by the run with optically thick clumping, which finds a low $T_{\rm{eff}}$ and log $g$ (run 3), and the oxygen abundance matches all but run 3. This is showing that there is an issue when not including enough lines in the GA to consistently constrain the abundances. For all GA runs we find a significantly higher helium abundance than BHL12; however, the run with optically thin clumping is close enough to cover the discrepancy within the error margin and this run has the best fit to \ion{He}{i} $\lambda$4471 so indeed an abundance close to BHL12 is the more realistic and likely solution. 

\subsection{HD190429A}

We find a significantly higher $T_{\rm{eff}}$ than BHL12 when fitting with optically thick clumping, with a consistent log $g$. Even with a large lower boundary margin on our best fit the temperature is still 3kK higher. The high temperature is being constrained by the optical hydrogen lines. Generally, the GA is able to reproduce these lines more accurately than BHL12; however, the \ion{He}{i} $\lambda$4471 profile is too weak in the GA best fit and it is clear that if we were to focus on this line the temperature would be lowered. In our fit with optically thin clumping we see a slightly better fit to this profile with $T_{\rm{eff}}$ only 1kK higher than BHL12. 

When fitting with optically thick clumping we match BHL12 for all other parameters except clumping factor and onset velocity. We find a difference of a factor of 1.2 in clumping factor and an onset velocity 50 $\mathrm{km s^{-1}}$ higher than BHL12. Although, the differences in these parameters between studies are not so large considering the difference in clumping laws used.

The GA fit with optically thin clumping is clearly a poorer fit than that with optically thick clumping for all lines except for HeI lines. However, with the lower $T_{\rm{eff}}$ found with optically thin clumping also comes a low surface gravity, 0.2 dex lower than the other best fits. Another issue in the optically thin clumping fit for this star is the abundance measurements. As mentioned previously the subset of lines used here result in incorrect abundances when a limited prescription for the wind physics is used as, in the UV, the strength of the lines are generally dominated by the stellar wind. Therefore, when attempting to fit without a good enough wind implementation the abundances vary unrealistically in order to reproduce these profiles.

\subsection{HD15570}

In the best fit with optically thick clumping we match BHL12 for nearly all parameters within our error margins. There are also no clear differences in the best-fit spectra for the line list we use between the two studies, except the strength of \ion{N}{iv} $\lambda$4058 appears to be overestimated in BHL12. As for parameter discrepancies, our $\beta$ is larger by 0.1 but our clumping onset velocity is 200 $\mathrm{km s^{-1}}$ compared to 30 $\mathrm{km s^{-1}}$ in BHL12, again highlighting the issue discussed in Sect 7.4. Our macroturbulence is higher than BHL12 but this is to be expected given the difference between the isotropic and radial-tangential prescriptions. The discrepancy of 400 $\mathrm{km s^{-1}}$ in terminal wind speed is interesting as both the GA and BHL12 achieve roughly the same quality of fit to the bluest edge of the P-Cygni trough in \ion{C}{iv} $\lambda\lambda$1548-1550.

Most of the best-fit parameters returned by the GA with optically thin clumping are not significantly different than with optically thick clumping, with the exception of the surface gravity and abundances. These issues have also been discussed in other objects. A low surface gravity can be limited by an attempt to fit emission features while using a fixed clumping prescription. Also, abundances can vary in an attempt to fit UV metal lines shaped by the winds when an insufficient wind prescription is used in the models. A way out of this dilemma would be to allow clumping to vary, and either allow optically thick clumping or add more photospheric metal lines into the fit.

\subsection{HD14947}

We find one of the largest discrepancies in temperature between the GA and BHL12 in this star, our best fit with optically thick clumping settling at 40.5kK. This is 3kK higher than BHL12, which is especially odd as the quality of fits of the main $T_{\rm{eff}}$ diagnostics between the two models are remarkably similar. This discrepancy is certainly aided by the offset in surface gravity; we find a log $g$ almost 0.2 dex higher than BHL12 and again a by-eye comparison of the models in each study reveals little as both best fits replicate the wings of Balmer profiles very well. Abundances, broadening, mass loss, and $\beta$ are all in good agreement. We find a significantly lower clumping factor, a difference of a factor of 1.5, but again the best-fit models do not appear to be significantly different. BHL12 finds a terminal wind speed 300 $\mathrm{km s^{-1}}$ higher than the GA but the GA best fit does appear to fall short of the P-Cygni edge so the higher velocity is the more realistic result.

The best fit with optically thin clumping agrees well with the other best fits. The optically thin clumping best fit finds some middle ground in $T_{\rm{eff}}$ and log $g$ with a surface gravity to match BHL12 and a temperature 1.5kK higher than BHL12. We notice again an issue in surface abundances with our optically thin clumping method, but in this case the differences are not too concerning. 

\subsection{HD210839 - $\lambda$ Cep}

Comparing $T_{\rm{eff}}$ and log $g$ between BHL12 and the GA with optically thick clumping, we notice discrepancies slightly larger than can be accounted for by our errors; however, the two solutions are not too far apart, the $T_{\rm{eff}}$ certainly lies between 36-37kK and log $g$ appears to be slightly weighted to lower values by optical recombination emission lines, so a higher value around 3.55 found by BHL12 is more likely the better solution. All other parameters agree quite well, our helium abundance is slightly higher and strongly weighted by the Balmer lines, and the GA does replicate the strengths of the Balmer series quite well here. A noteworthy feature of this spectrum is the prominent double peak shown in the $H\alpha$ and \ion{He}{ii} $\lambda$4686 lines, this leads the GA to settle on a high clumping onset velocity as discussed in Sect. 7.4.

Our optically thin fit once again suffers some discrepancies due to our clumping prescription, a log $g$ lower by 0.2 dex is found as well as a $\beta$ value close to 0.6 with these lower values being preferred by the double peak optical recombination profiles, especially \ion{He}{ii} $\lambda$4686. The abundances found in our optically thin run are significantly lower than those found in BHL12, a common problem in this study with limited line list but again the problem is alleviated when including optically thick clumping.

The interclump density for this object is an intermediate value of 0.16, which is fairly common in this sample. The velocity filling is quite low but reasonable, driven by the PV lines, which have a weaker absorption depth than some of the other objects in this sample. The physical cause of the reduction in PV absorption depth is unclear. 

\subsection{HD163758}

Bestfit temperatures agree comfortably within error regions for all fits, comparing the fit with optically thick clumping to BHL12 we have a slightly lower $T_{\rm{eff}}$. Our fit with optically thick clumping finds a log $g$ consistent with BHL12, the fit quality is very good in both studies, both have some trouble fitting H$\delta$ due to a strong blend. We inspect the fitness for just the H$\gamma$ line, and it is clear that a slightly lower surface gravity is allowed, so the true value is somewhere between the bounds provided by both fits, therefore log $g$ $\approx$ 3.4-3.55. We find a mass-loss rate consistent with BHL12 with a clumping factor 1.5 times higher, although a difference of 1.5 times in clumping factor only accounts for a 0.05 dex shift in mass-loss rate. From the models one can see that, in the GA, H$\alpha$ is possibly underestimated in an attempt to fit the double-peak profile while \ion{He}{ii} $\lambda$4686 is replicated well. However, in BHL12 the strength of the synthetic H$\alpha$ is closer to the observed spectrum but \ion{He}{ii} $\lambda$4686 is overestimated. We also find one of the highest clumping onset velocities in this star of 320 $\mathrm{km s^{-1}}$, while BHL12 used 30 $\mathrm{km s^{-1}}$ in their fitting. Our high onset causes $\beta$ to converge to a high value of 2.3. This problem is discussed in Sect 7.4 and certainly can influence the mass loss. 

Rotation rates and macroturbulence are fairly consistent in both fits. We find nitrogen and oxygen abundances that match BHL12 but our carbon abundance is lower. BHL12 also takes care to fit optical carbon lines so the carbon abundance of 8.6 estimated by BHL12 is likely more accurate. We also find a helium abundance significantly higher than BHL12 of 0.28, this is mainly driven by \ion{He}{ii} $\lambda4686$. Difficulty fitting this line likely has a negative impact on our helium abundance. Evidence for this comes from the fact that this line drives a high $\beta$ value in our fit with optically thick clumping, while in our fit with optically thin clumping fit $\beta$ remains low. In our fit with optically thin clumping the helium abundance matches BHL12 and \ion{He}{ii} $\lambda4686$ is poorly reproduced. 

Our fit with optically thin clumping is very similar to that found in BHL12, with most best-fit parameters in agreement.

\subsection{HD192639}

For this object, we find a similar $T_{\rm{eff}}$, 0.3 dex higher log $g$, 0.07 dex difference in mass-loss rate and a very high $\beta$ and clumping onset velocity. Rotation rates are similar while we find a significantly higher macroturbulence. Carbon, nitrogen and oxygen abundances are in agreement and we find a slightly higher helium abundance. An additional problem present in our fitting of this object is that we find a much too high terminal wind speed, 800 $\mathrm{km s^{-1}}$ higher than BHL12. The blue edge of the P-Cygni trough is not so well defined in the spectrum due to the iron forest. If we were to fit the spectrum by eye, we would aim to fit the point of bluest complete absorption as opposed to the point at which the edge meets the continuum, while the GA appears to fulfil the latter criteria. It is clear from our fit to \ion{Si}{iv} $\lambda\lambda1393-1402$ that our terminal wind speed is too high, this line reaches a maximum fitness at $v_{\infty}$ $\simeq 2100 \mathrm{km s^{-1}}$, which is a much more consistent value for the full UV spectrum. 

\onecolumn

\section{Best fits}

In this appendix we present models from our fits with optically thick clumping, showing only the line profiles used in our GA analysis. The observed spectrum is shown by the black points, the solid red line is our best fitting model, and green lines represent any models generated during the GA iterations that lie within the error regions.

\begin{figure}[ht]
    \centering
    \includegraphics[scale=0.45]{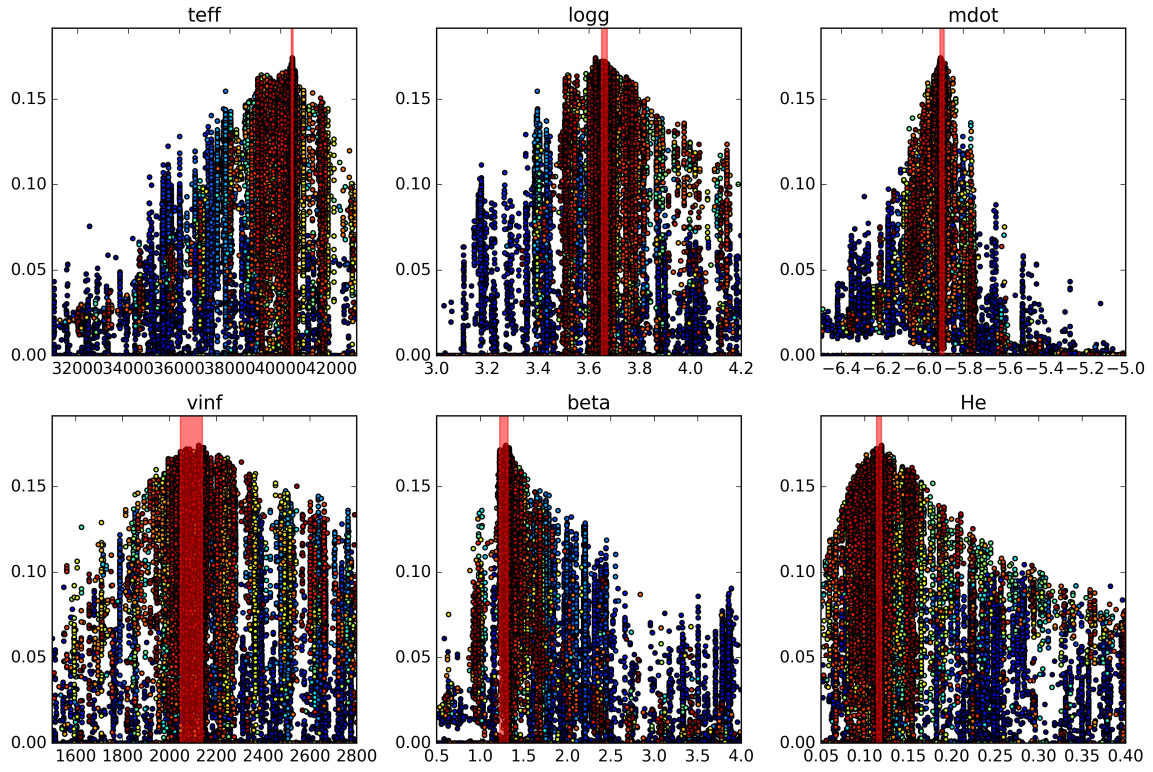}
    \caption{Fitness distribution for HD14947 fit with optically thick clumping. The colour scheme of the points goes from blue to red, corresponding to the earliest and latest models in the exploration, respectively. The region highlighted in red shows the confidence interval. This figure is included to demonstrate the good agreement between an assessment of fit quality by the fitness metric F as defined in Eq. 4 and $\chi^{2}$ as shown in Fig. \ref{fig: Chi2 distrib HD14947}.} 
\end{figure}

\begin{figure}[ht]
    \ContinuedFloat
    \centering
    \includegraphics[scale=0.35]{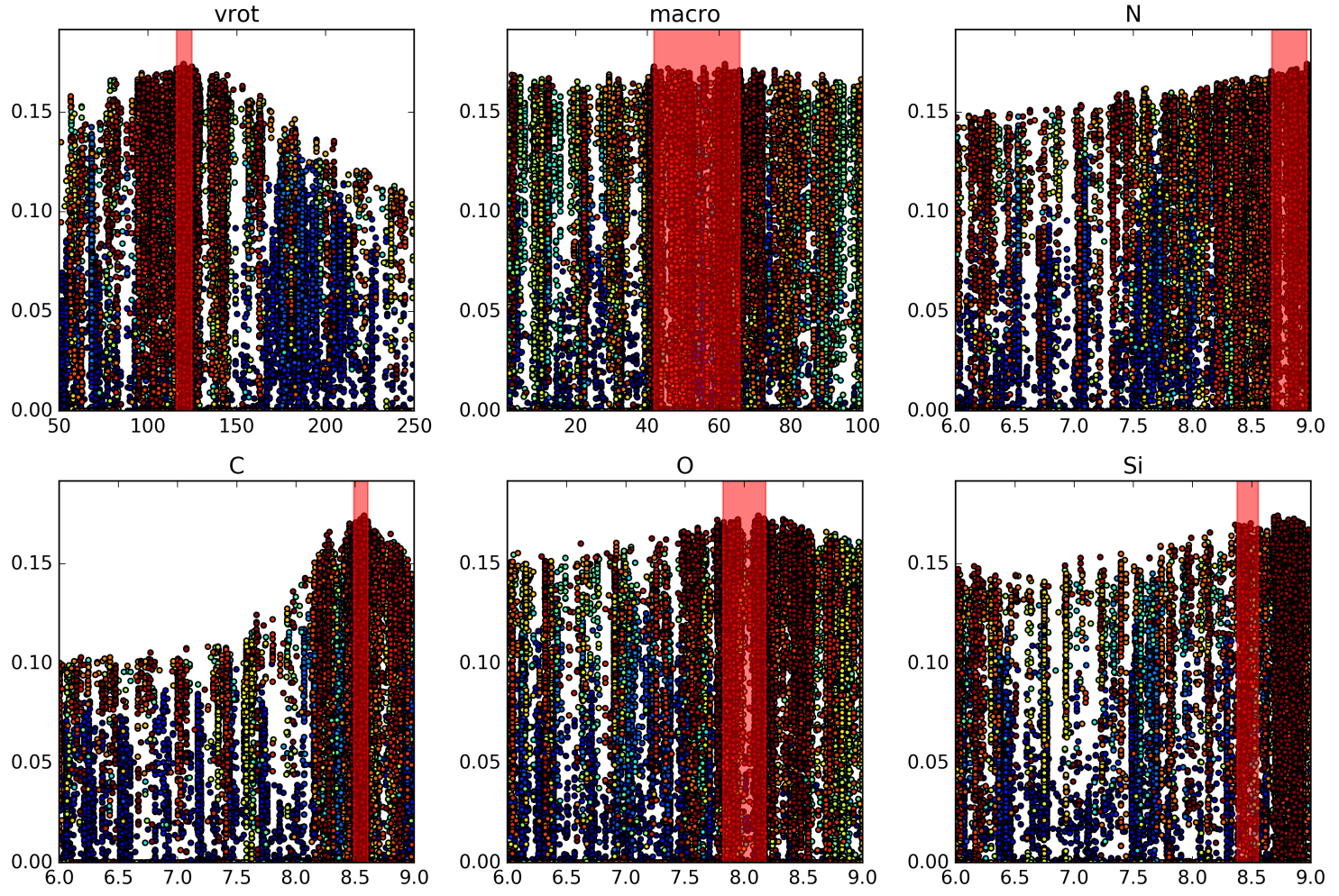}
    \includegraphics[scale=0.35]{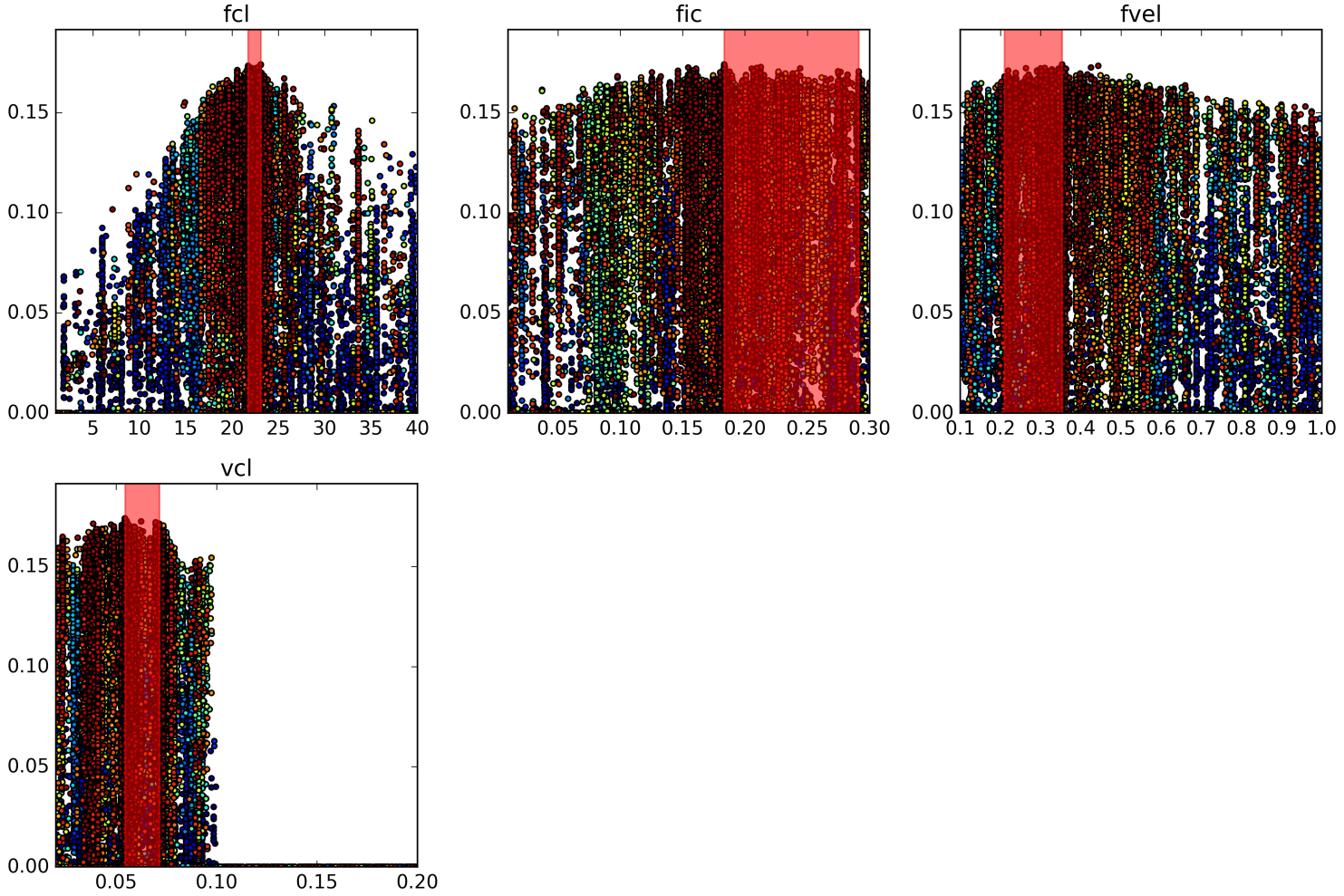}
    \caption{Continued.} 
    \label{fig: Fitness distrib HD14947}
\end{figure}

\onecolumn

\thispagestyle{plain}

\begin{landscape}

\begin{table}

\caption{Photospheric Parameter best fits. The first row, for each object, shows the best-fit model parameters from BHL12 with optically thin clumping, the second row shows the GA run with optically thin clumping and the third row shows the GA run that allows for optically thick clumping.}

\vspace{2mm}

\label{tbl: Parameters Thin}
\begin{tabular}{lllllllllllllll}
\hline
Star & Teff & logg & $\dot{M}$ & $v_{\infty}$ & beta & fcl  & vsini & vmac & He/H & C & N & O\cr
Sp Type & [K] & & & [$km s^{-1}$] & &  & [$km s^{-1}$] & [$km s^{-1}$] & \\
\hline
HD16691	&	$41.0\pm1.0$ & $3.66\pm0.1$ &$-5.52\pm0.03$& $2300\pm 100$ &	1.2	&	17&	135	&	37	&	0.15&	$6.5\pm0.2$ 	&$9.0\pm0.2$&	$7.8\pm0.3$	\\
O4 If	& $41.5 \pm 0.5$ & $3.74_{-0.11}^{+0.20}$ & $-5.72\pm 0.05 $&$	2100_{-100	}^{+147	}$ & $	1.1	\pm 0.1 $ & 	20	&	$247	_{-	37	}^{+	10	}$ & $	86	_{-	84	}^{+	13	}$ & $	0.12\pm 0.02$ & $	6.6	_{-	0.2	}^{+	0.6	}$ & $	8.8	_{-	0.4	}^{+	0.2	}$ & $	8.1	_{-	0.3	}^{+	0.3	}$ \\
& $39.5 \pm 0.5$ & $	3.85_{-	0.05}^{+0.14}$&$	-5.60	\pm 0.05 $&$	2300	\pm 100 $&$	*1.6	\pm 0.1 $&$	20	_{-	2	}^{+	2	}$&$	188	_{-	25	}^{+	42	}$&$	68	_{-	21	}^{+	25	}$&$	0.13	\pm 0.02 $&$	6.6	_{-	0.1	}^{+	0.6	}$&$	9.0	\pm 0.1 $&$ 8.4	_{-	0.4	}^{+	0.3	}$ \\
	&	&	&	&	&	&	&	&	&	&	&	&	&	\\																									
HD66811	&	$40.0 \pm 1.0$					&	$3.64 \pm 0.1$					&	$-5.7	_{-	0.05	}^{+	0.04	}$&	$2300\pm 100$					&	0.9					&	20					&	210					&	90					&	0.16					&	$7.6\pm0.3$					&	$9.1\pm0.2$					&	$8.1\pm0.3$		\\

O4 I(n)fp	& $	41.5 \pm 0.5$ & $	3.52\pm 0.05 $ & $	-5.7\pm 0.05 $ & $	2200	\pm 100 $ & $	0.6	\pm 0.1 $&	20	&	$234	_{-	20	}^{+	10	}$ & $36	_{-	10	}^{+	56	}$ & $	0.21	_{-	0.04	}^{+	0.05	}$&$8.0	_{-	1.0	}^{+	0.3	}$&$	7.7	\pm 0.02 $&$	7.8	\pm 0.2 $\\
$\zeta$ Pup &$	40.0\pm 0.5 $&$	*3.52\pm 0.05 $&$	-5.7	\pm 0.05 $&$	2300	\pm 100 $&$	0.8	_{-	0.2	}^{+	0.1	}$&$	23	_{-	1	}^{+	8	}$&$	203	_{-	10	}^{+	40	}$&$	96	_{-	65	}^{+	10	}$&$	*0.25	_{-	0.03	}^{+	0.02	}$&$	*8.3	_{-	0.3	}^{+	0.1	}$&$	9.0	_{-	0.1	}^{+	0.2	}$&$	*6.9	_{-	0.7	}^{+	0.1	}$\\
	&	&	&	&	&	&	&	&	&	&	&	&	&	\\																																																																	
HD190429A	&	$39.0 \pm 1.0$					&	$3.62 \pm 0.1$					&	$-5.68\pm0.04$					&	$2300\pm 100$					&	1.0					&	25					&	150					&	57					&	0.15					&	$7.1\pm0.2$					&	$8.9\pm0.2$					&	$7.8\pm0.3$		\\
O4 If	&$	40.0 \pm 0.5$&$	3.38	\pm 0.05 $&$	-5.59	\pm 0.05 $&$	2400	\pm 100 $&$	0.8	\pm 0.1 $&	20	&	$172	\pm 10 $&$	72	_{-	10	}^{+	15	}$&$	0.19\pm 0.02 $&$	6.9	_{-	0.5	}^{+	0.1	}$&$	8.2		\pm 0.1 $&$	8.2	_{-	0.4	}^{+	0.1	}$\\
&$	42.5	_{-	0.7	}^{+	0.5	}$&$	3.62	\pm 0.05 $&$	-5.69	\pm 0.05 $&$	2400	\pm 100 $&$	0.9	\pm 0.1 $&$	30	_{-	1	}^{+	3	}$&$	192	_{-	54	}^{+	10	}$&$	44	_{-	10	}^{+	56	}$&$	0.16	\pm 0.02 $&$	7.7	_{-	0.3	}^{+	0.1	}$&$	9.3	_{-	0.4	}^{+	0.1	}$&$	8.0	_{-	0.2	}^{+	0.4	}$\\
	&	&	&	&	&	&	&	&	&	&	&	&	&	\\																																																																	
HD15570	&	$38.0 \pm 1.0$					&	$3.51 \pm 0.1$					&	$-5.66\pm0.04$					&	$2200\pm 100$					&	1.1					&	20					&	97					&	40					&	0.10					&	$7.5\pm0.2$					&	$8.6\pm0.2$					&	$8.3\pm0.2$		\\

O4 If	&$	38.5 \pm 0.5$&$	3.37	\pm 0.05 $&$	-5.69	\pm 0.05 $&$	2600	\pm 100 $&$	1.1		\pm 0.1 $& 20	& $	122	\pm 10$ & $	74	_{-	14	}^{+	16	}$&$	0.08	\pm 0.02 $&$	8.3	_{-	0.6	}^{+	0.1	}$&$	7.6	_{-	0.2	}^{+	0.1	}$&$	8.2	_{-	0.6	}^{+	0.1	}$\\
&$	38.0 \pm 0.5 $&$	3.45	_{-	0.06	}^{+	0.05	}$&$	-5.63	\pm 0.05 $&$	2700	\pm 100 $&$	*1.6	\pm 0.1 $&$	20	_{-	1	}^{+	3	}$&$	103	_{-	10	}^{+	11	}$&$	98	\pm 10 $&$	0.09	\pm 0.02 $&$	7.9	_{-	0.2	}^{+	0.2	}$&$	8.7	_{-	0.5	}^{+	0.5	}$&$	8.3	_{-	0.4	}^{+	0.1	}$\\
	&	&	&	&	&	&	&	&	&	&	&	&	&	\\																																																																	
HD14947	&	$37.0 \pm 1.0$					&	$3.52 \pm 0.1$					&	$-5.85	_{-	0.07	}^{+	0.06	}$&	$2300\pm 100$					&	1.3					&	33					&	130					&	36					&	0.12					&	$8.3\pm0.2$					&	$8.8\pm0.1$					&	$8.1\pm0.2$		\\
O4.5 If	&$	38.5 \pm 0.5$&$	3.49	\pm 0.05 $&$	-5.86	\pm 0.05 $&$	2200	\pm 100 $&$	1.3	\pm 0.1 $&	20	&	$139	\pm 10 $&$	53	\pm 10 $&$	0.11	\pm 0.02 $&$	8.0	\pm 0.01 $&$	8.4	\pm 0.01 $&$	8.0	\pm 0.01 $\\
&$	40.5 \pm 0.5 $&$	3.67	\pm 0.05 $&$	-5.89	\pm 0.05 $&$	2000	\pm 100 $&$	1.2	\pm 0.1 $&$	22	\pm 1 $&$	116	\pm 10 $&$	66	_{-	24	}^{+	10	}$&$	0.11	\pm 0.02 $&$	8.5	_{-	0.1	}^{+	0.1	}$&$	8.8	_{-	0.2	}^{+	0.1	}$&$	8.1	_{-	0.3	}^{+	0.1	}$\\
	&	&	&	&	&	&	&	&	&	&	&	&	&	\\																																																																	
HD210839	&	$36.0 \pm 1.0$					&	$3.54 \pm 0.1$					&	$-5.85	_{-	0.07	}^{+	0.06	}$&	$2100\pm 100$					&	1.0					&	20					&	210					&	80					&	0.12					&	$8.2\pm0.2$					&	$8.7\pm0.2$					&	$8.5\pm0.1$		\\
O6 I(n)fp	&$	36.0 \pm 0.5$ & $	3.28	\pm 0.05 $&$	-5.72	_{-	0.09	}^{+	0.05	}$&$	2100	_{-	118	}^{+	100	}$&$	0.6	\pm 0.01 $&	20		&	$212	_{-	10	}^{+	20	}$&$	96	_{-	21	}^{+	10	}$&$	0.16	\pm 0.02 $&$	7.7	_{-	0.1	}^{+	0.2	}$&$	8.2	_{-	0.2	}^{+	0.1	}$&$	8.3	_{-	0.1	}^{+	0.2	}$\\
$\lambda$ Cep&$	37.0	\pm 0.5 $&$	3.47	_{-	0.09	}^{+	0.05	}$&$	-5.86	\pm 0.05 $&$	2100	_{-	128	}^{+	100	}$&$	0.9	\pm 0.1 $&$	23	_{-	1	}^{+	2	}$&$	214	_{-	10	}^{+	22	}$&$	62	_{-	12	}^{+	12	}$&$	0.17 \pm 0.02 $&$	8.0	\pm 0.1 $&$	8.9	_{-	0.6	}^{+	0.1	}$&$	8.7	_{-	0.6	}^{+	0.2	}$\\
	&	&	&	&	&	&	&	&	&	&	&	&	&	\\																																																																	
HD163758	&	$34.5 \pm 1.0$					&	$3.41 \pm 0.1$					&	$-5.8	_{-	0.06	}^{+	0.05	}$&	$2100\pm 100$					&	1.1					&	20					&	94					&	34					&	0.15					&	$8.6\pm0.2$					&	$8.8\pm0.2$					&	$8.4\pm0.2$		\\
O6.5 If	&$	35.0 \pm 0.5$&$	3.27	\pm 0.05 $&$	-5.85	\pm 0.05 $&$	2200	\pm 100 $&$	1.2	\pm 0.1 $& 	20	&	$95	\pm 10 $&$	96	_{-	21	}^{+	10	}$&$	0.16	\pm 0.02 $&$	8.9	\pm 0.1 $&$	9.0	\pm 0.1 $&$	8.3	\pm 0.1 $\\
&$	34.0	\pm 0.5 $&$	3.45\pm 0.05$&$	-5.81	\pm 0.05 $&$	2600	\pm 100 $&$	*2.3	\pm 0.1 $&$	30 \pm 1$&$	116	\pm 10 $&$	34_{-	20	}^{+	10	}$&$	0.28	\pm 0.02 $&$	7.9	\pm 0.1$&$	9.3	\pm 0.1 $&$	8.6	\pm 0.1 $ \\
	&	&	&	&	&	&	&	&	&	&	&	&	&	\\																																																																	
HD192639	&	$33.5 \pm 1.0$	&	$3.42\pm0.1$	&	$-5.92	_{-	0.08	}^{+	0.07	}$&	$1900\pm 100$	&	1.3	&	20	&	90	&	43	&	0.15	&	$8.2\pm0.2$	&	$8.8\pm0.2$	&	$8.6\pm0.2$	\\
O7.5 Iabf	&$	34.5 \pm 0.5$&$	3.48	\pm 0.05 $&$	-5.84	\pm 0.05 $&$	2400	\pm 100 $&$	1.4	\pm 0.1 $&	20	&	$109	_{-	11	}^{+	10	}$&$	84	_{-	10	}^{+	11	}$&$	0.10	\pm 0.02 $&$	7.8 	\pm 0.01 $&$	8.4 	\pm 0.05 $&$	8.4	_{-	0.1	}^{+	0.2	}$\\
&$	33.5	\pm 0.5 $&$	3.71	\pm 0.05 $&$	-5.85	\pm 0.05 $&$	*2700	\pm 100 $&$	*2.5_{-	0.1	}^{+	0.3	} $&$	32 \pm 2 $&$	103	\pm 14 $&$	100	\pm 19 $&$	0.19	\pm 0.02 $&$	7.8 \pm 0.1 $&$	8.7 \pm 0.1 $&$	8.6 \pm 0.1 $\\
\hline
\\              
\end{tabular}
\end{table}

\end{landscape}

\onecolumn

\begin{figure}[htb]
\begin{tikzpicture}
  \node (img1)  {\includegraphics[scale=0.28]{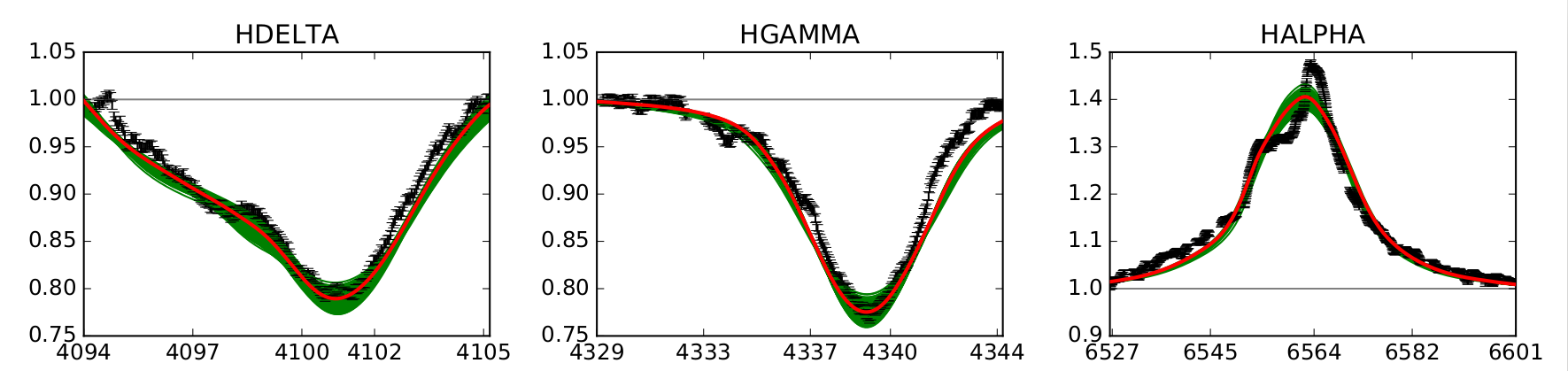}};
  \node[below=of img1, node distance=0cm, yshift=1cm] (img2) {\includegraphics[scale=0.28]{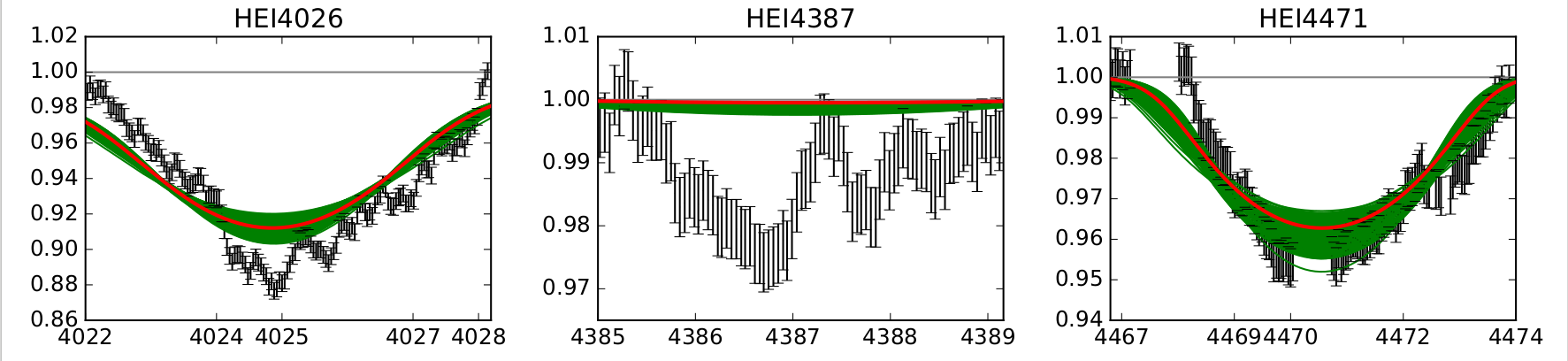}};
  \node[below=of img2, node distance=0cm, yshift=1cm] (img3) {\includegraphics[scale=0.28]{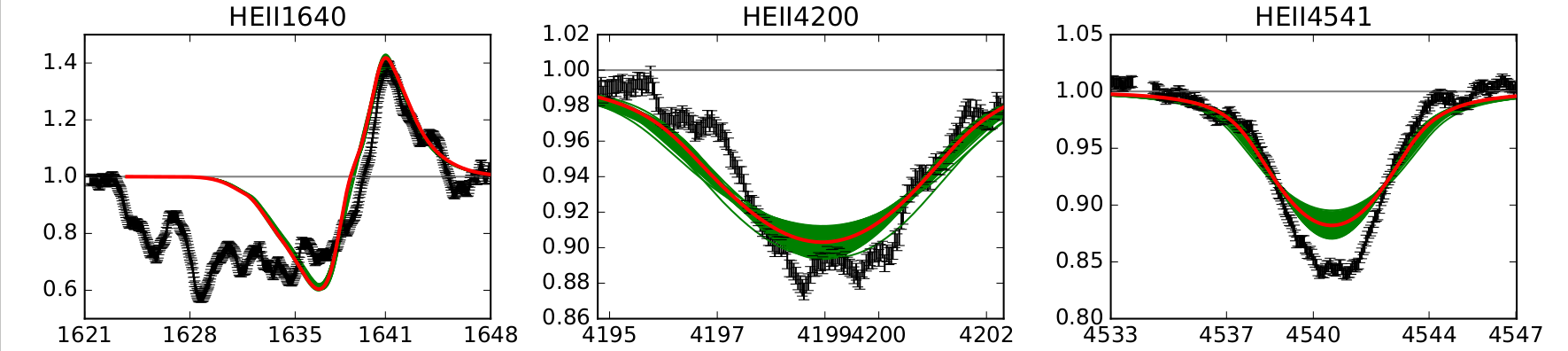}};
  \node[below=of img3, node distance=0cm, yshift=1cm] (img4) {\includegraphics[scale=0.28]{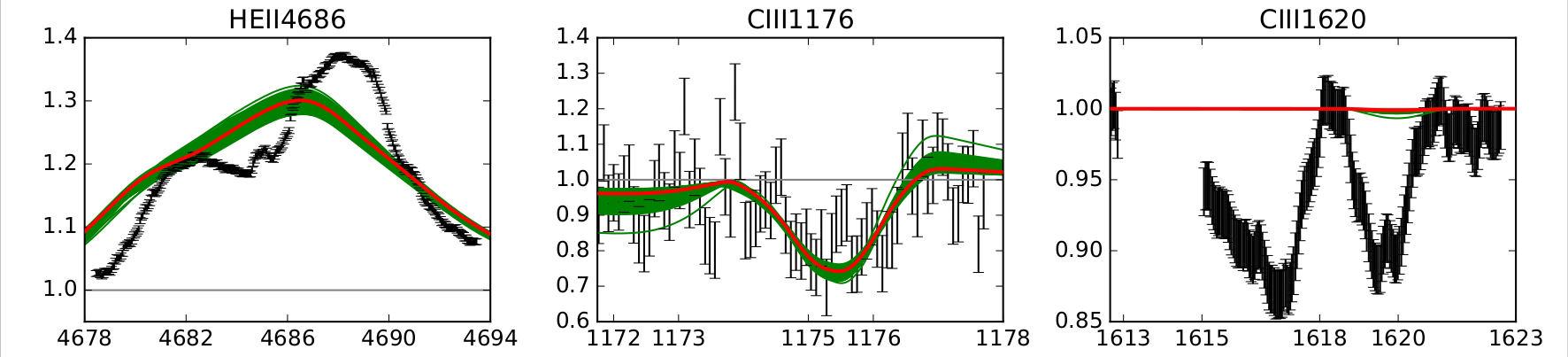}};
  \node[below=of img4, node distance=0cm, yshift=1cm] (img5) {\includegraphics[scale=0.28]{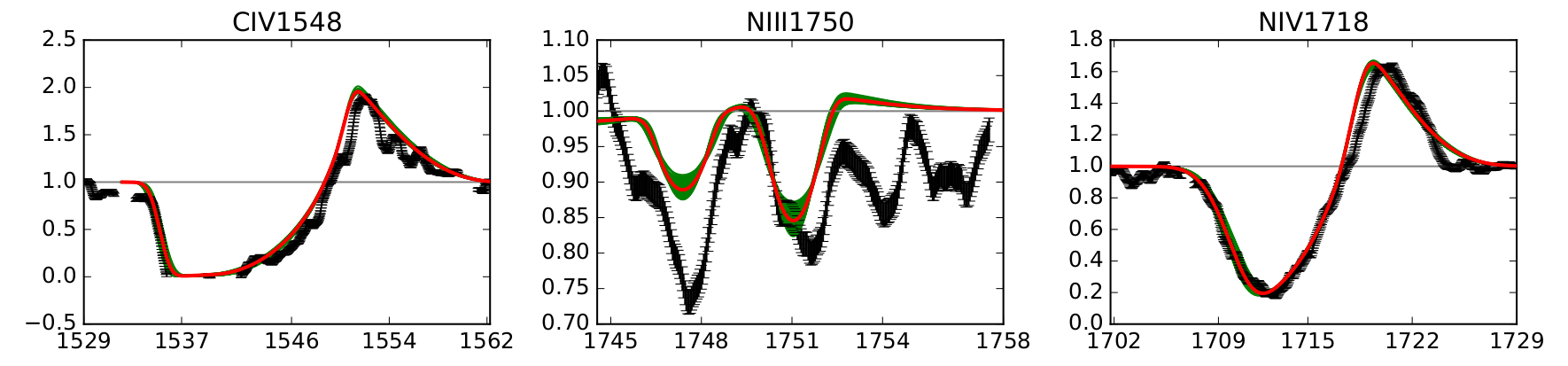}};
  \node[below=of img5, node distance=0cm, yshift=1cm] {Wavelength};
  \node[left=of img3, node distance=0cm, rotate=90, anchor=center,yshift=-0.7cm] {Normalised Flux};
\end{tikzpicture}
\caption{Best fit for HD16691 from GA with optically thick clumping.} 
\end{figure}

\begin{figure}[htb]
\ContinuedFloat
\begin{tikzpicture}
  \node (img1)  {\includegraphics[scale=0.28]{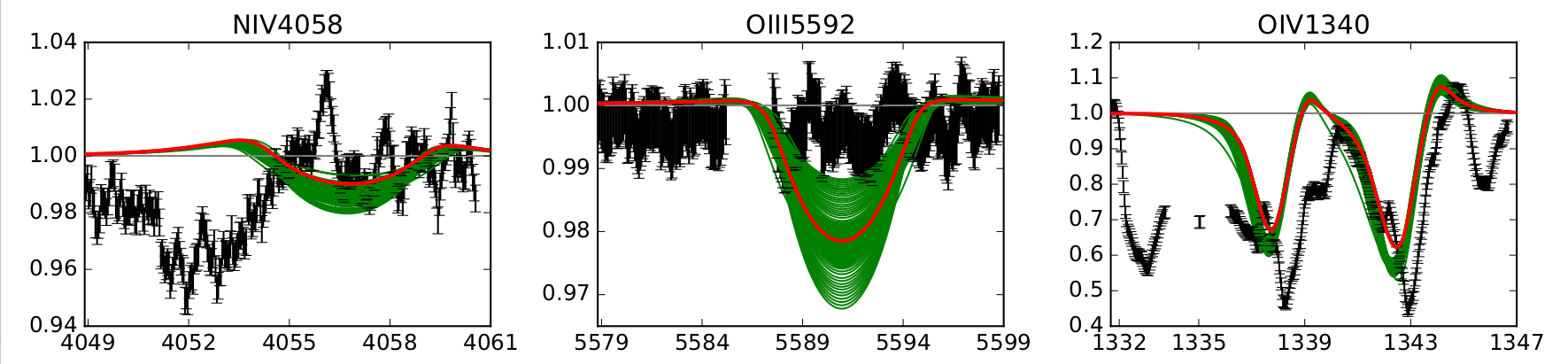}}; 
  \node[below=of img1, node distance=0cm, yshift=1cm] (img2) {\includegraphics[scale=0.28]{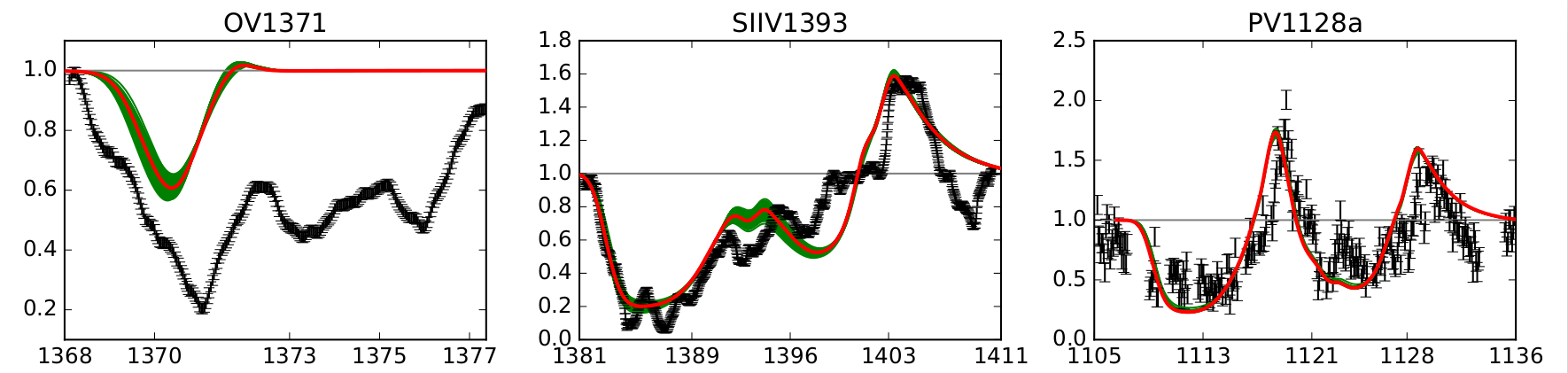}};
  \node[below=of img2, node distance=0cm, yshift=1cm] {Wavelength};
  \node[left=of img1, node distance=0cm, rotate=90, anchor=center,yshift=-0.7cm] {Normalised Flux};
\end{tikzpicture}
\caption{Continued.} 
\label{fig: HD16691 best-fit}
\end{figure}

\begin{figure}[htb]
\begin{tikzpicture}
  \node (img1)  {\includegraphics[scale=0.28]{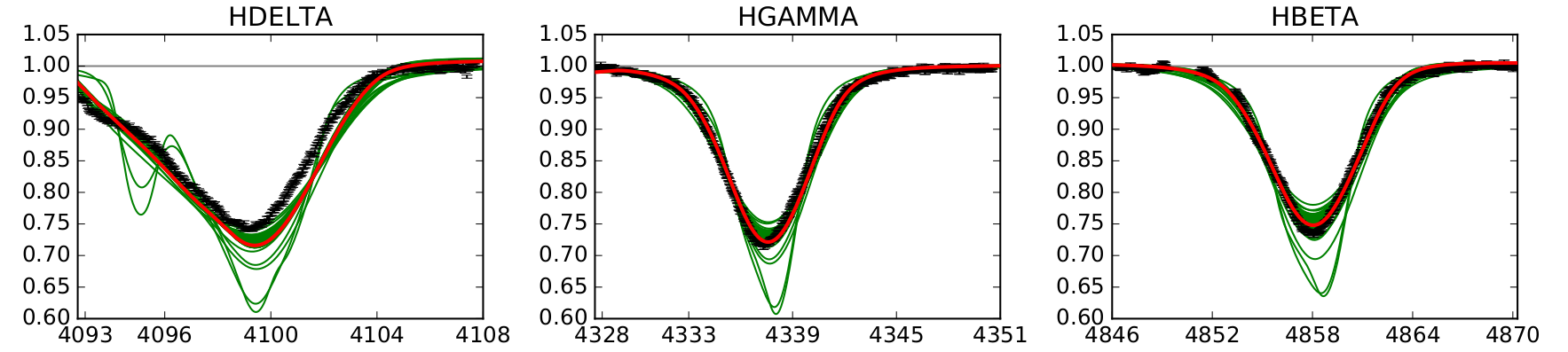}}; 
  \node[below=of img1, node distance=0cm, yshift=1cm] (img2) {\includegraphics[scale=0.28]{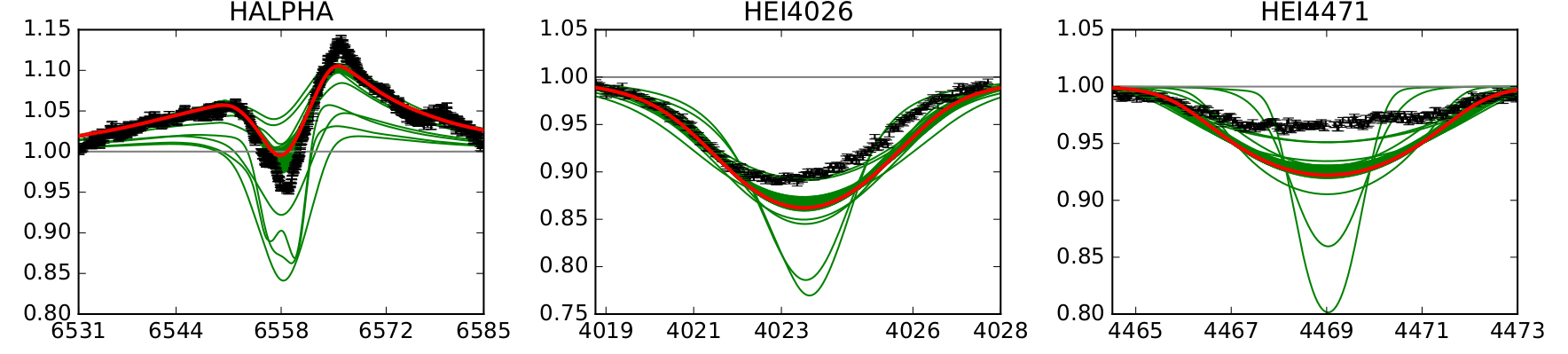}};
  \node[below=of img2, node distance=0cm, yshift=1cm] (img3) {\includegraphics[scale=0.28]{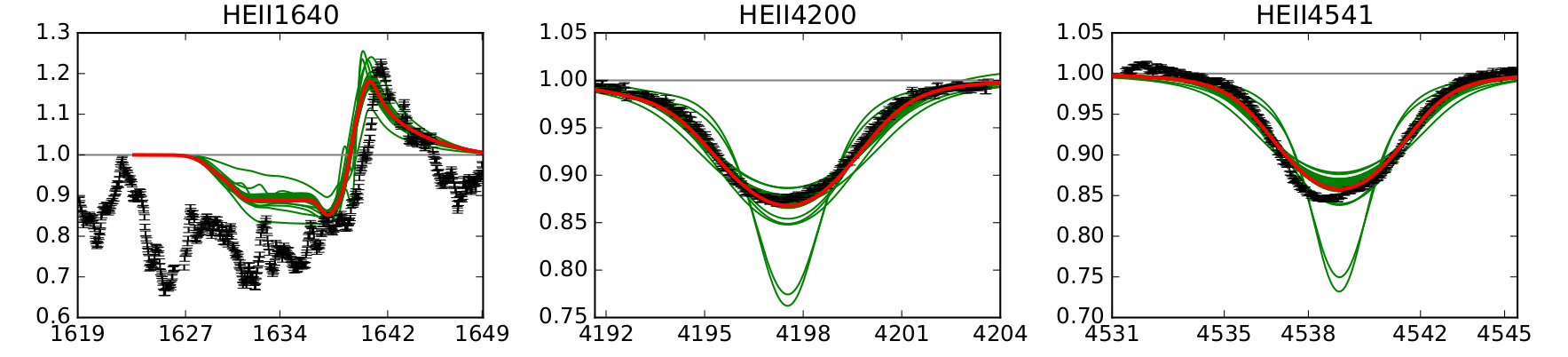}};
  \node[below=of img3, node distance=0cm, yshift=1cm] {Wavelength};
  \node[left=of img2, node distance=0cm, rotate=90, anchor=center,yshift=-0.7cm] {Normalised Flux};
\end{tikzpicture}
\caption{Best fit for HD66811 from GA with optically thick clumping.} 
\end{figure}

\begin{figure}[htb]
\ContinuedFloat
\begin{tikzpicture}
  \node (img1)  {\includegraphics[scale=0.28]{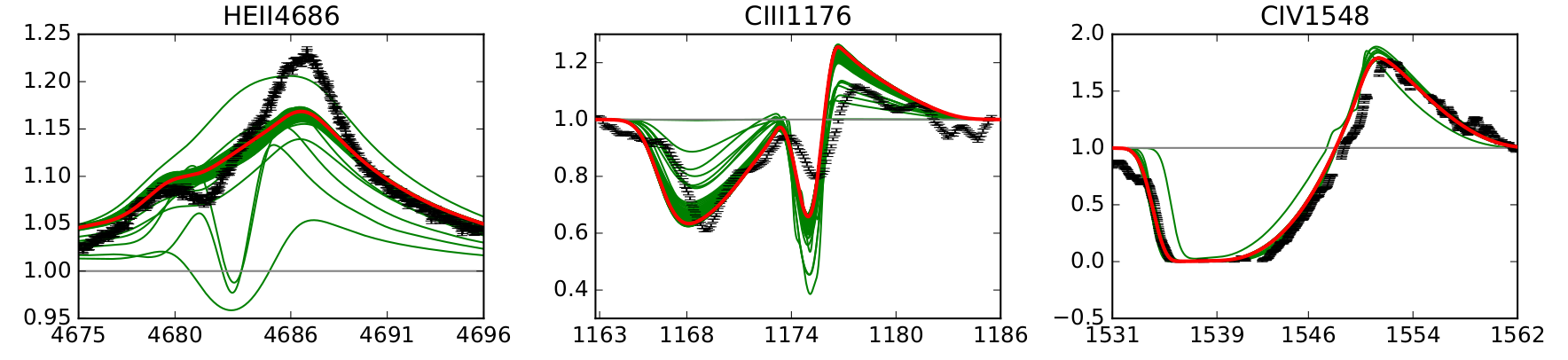}}; 
  \node[below=of img1, node distance=0cm, yshift=1cm] (img2) {\includegraphics[scale=0.28]{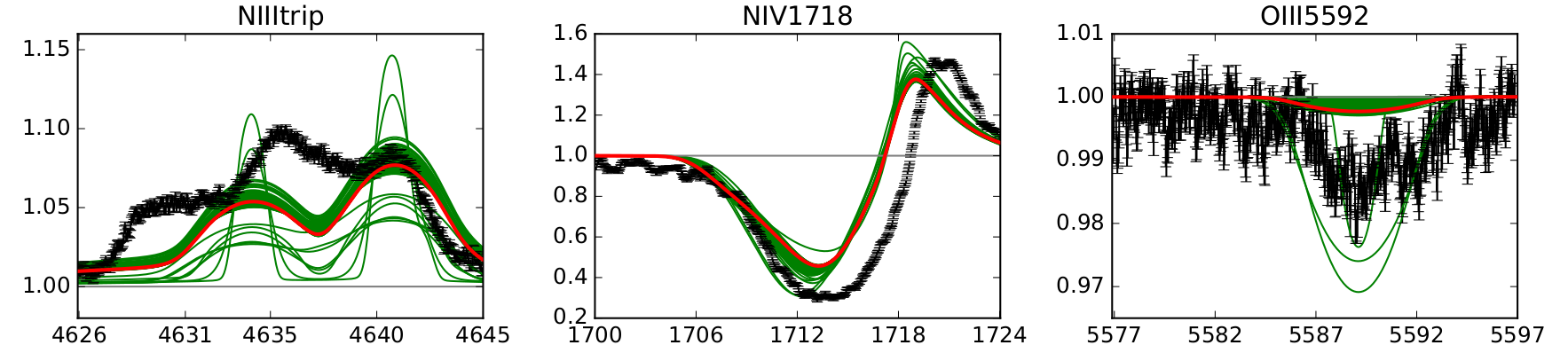}};
  \node[below=of img2, node distance=0cm, yshift=1cm] (img3) {\includegraphics[scale=0.28]{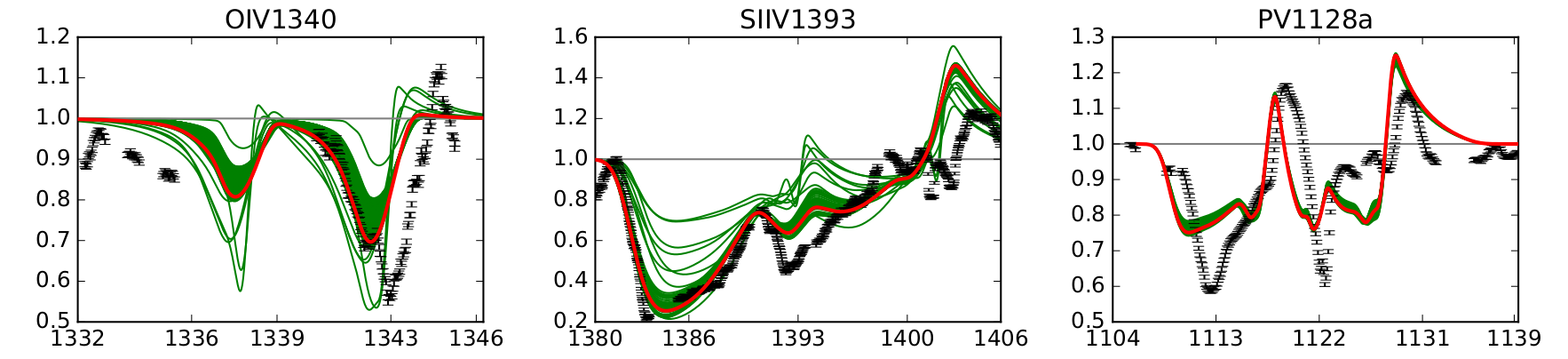}};
  \node[below=of img3, node distance=0cm, yshift=1cm] {Wavelength};
  \node[left=of img2, node distance=0cm, rotate=90, anchor=center,yshift=-0.7cm] {Normalised Flux};
\end{tikzpicture}
\caption{Continued.} 
\label{fig: HD66811 best-fit}
\end{figure}

\begin{figure}[htb]
\begin{tikzpicture}
  \node (img1)  {\includegraphics[scale=0.28]{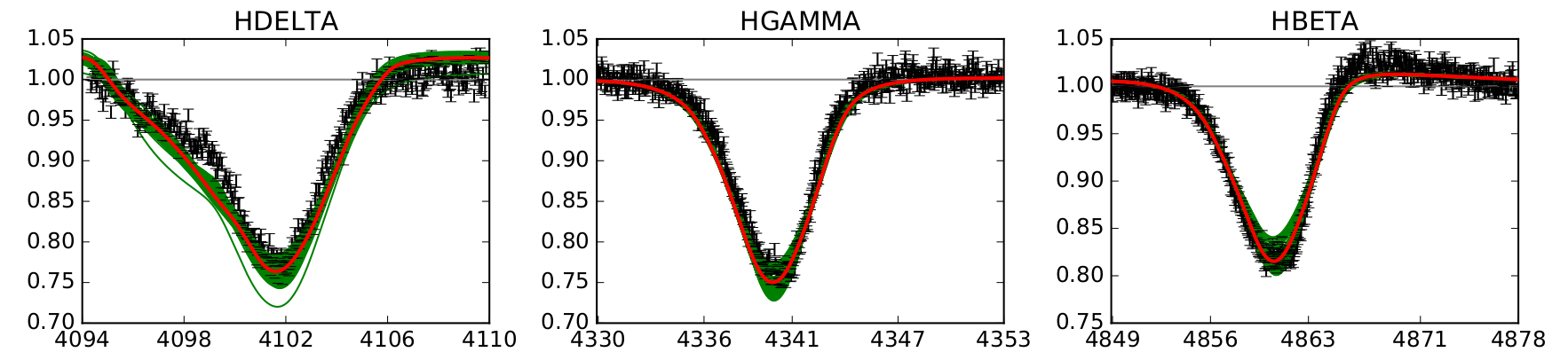}};
  \node[below=of img1, node distance=0cm, yshift=1cm] (img2) {\includegraphics[scale=0.28]{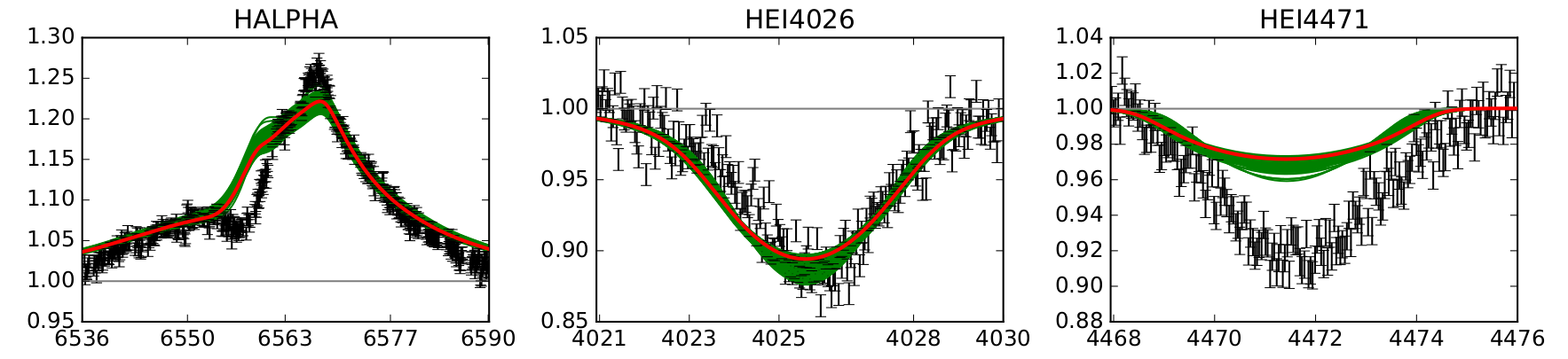}};
  \node[below=of img2, node distance=0cm, yshift=1cm] {Wavelength};
  \node[left=of img1, node distance=0cm, rotate=90, anchor=center,yshift=-0.7cm] {Normalised Flux};
\end{tikzpicture}
\caption{Best fit for HD190429A from GA with optically thick clumping.} 
\end{figure}

\begin{figure}[htb]
\ContinuedFloat
\begin{tikzpicture}
  \node (img1)  {\includegraphics[scale=0.28]{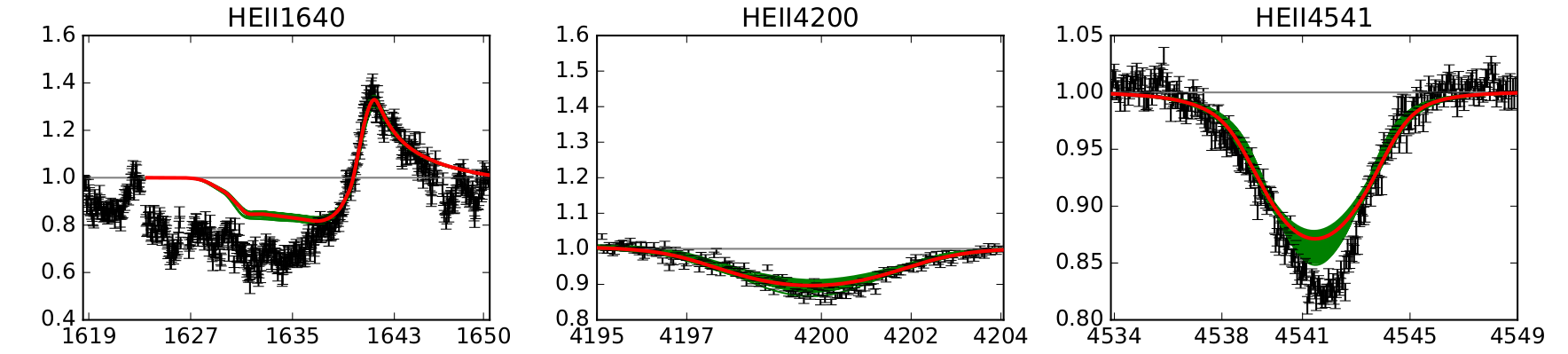}};
  \node[below=of img1, node distance=0cm, yshift=1cm] (img2) {\includegraphics[scale=0.28]{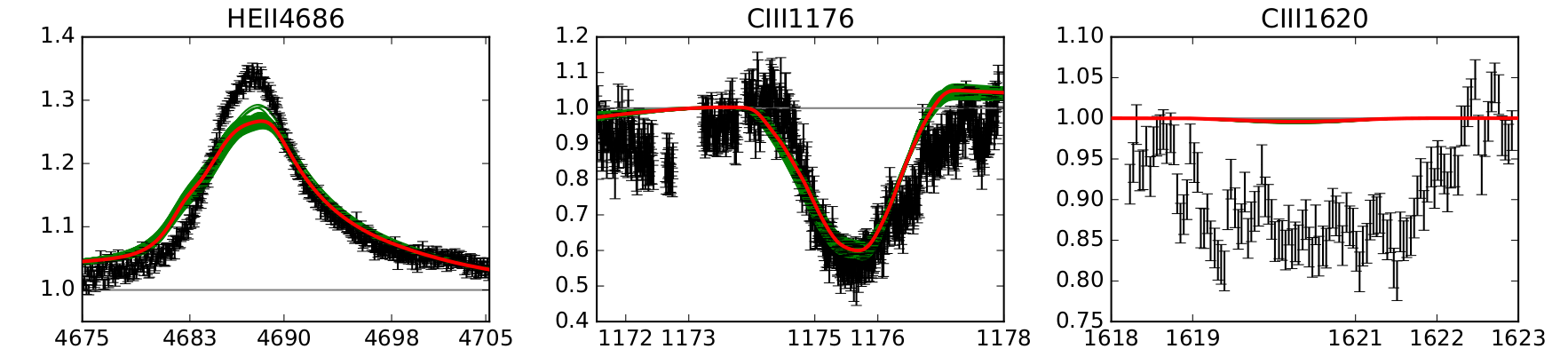}};
  \node[below=of img2, node distance=0cm, yshift=1cm] (img3) {\includegraphics[scale=0.28]{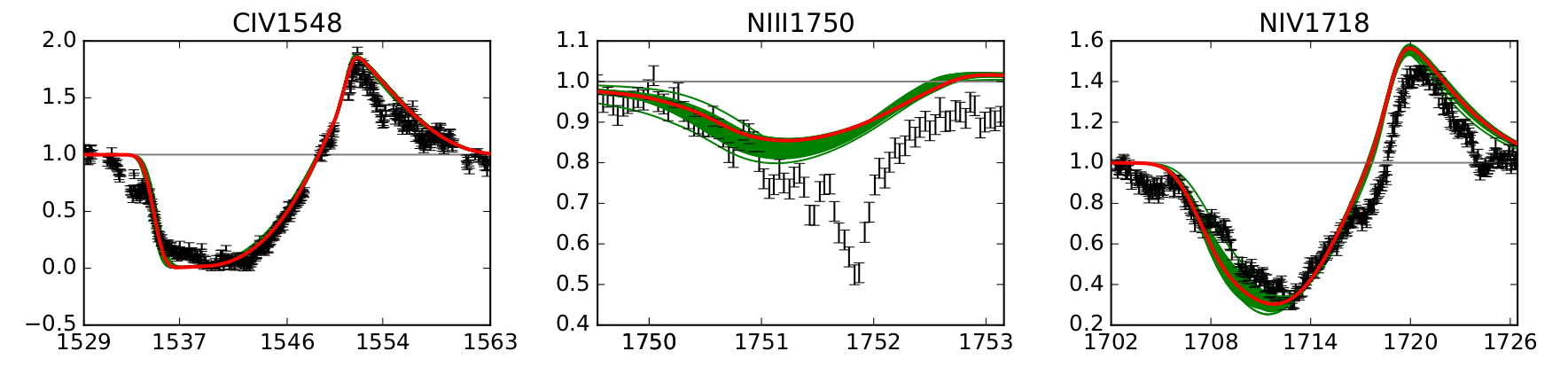}};
  \node[below=of img3, node distance=0cm, yshift=1cm] (img4) {\includegraphics[scale=0.28]{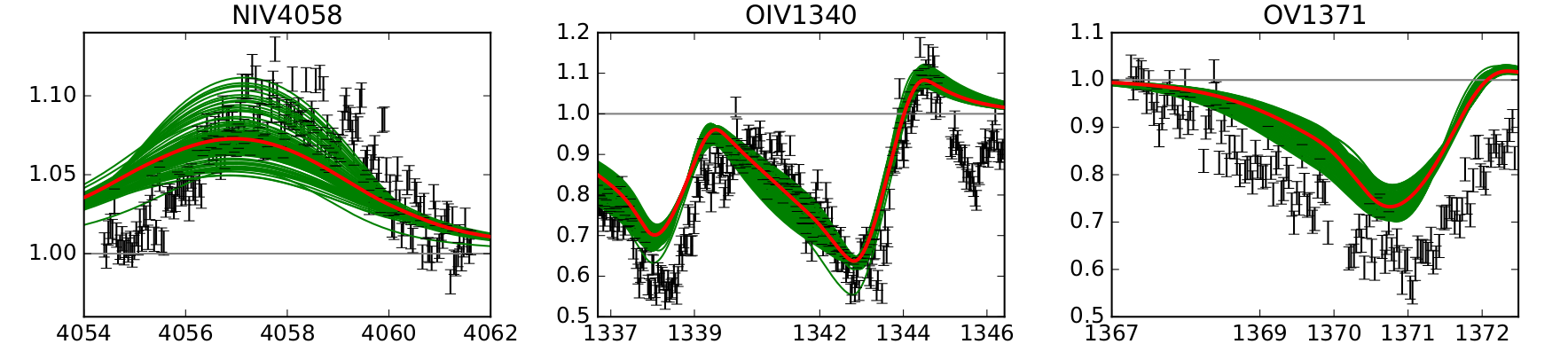}};
  \node[below=of img4, node distance=0cm, yshift=1cm] (img5) {\includegraphics[scale=0.28]{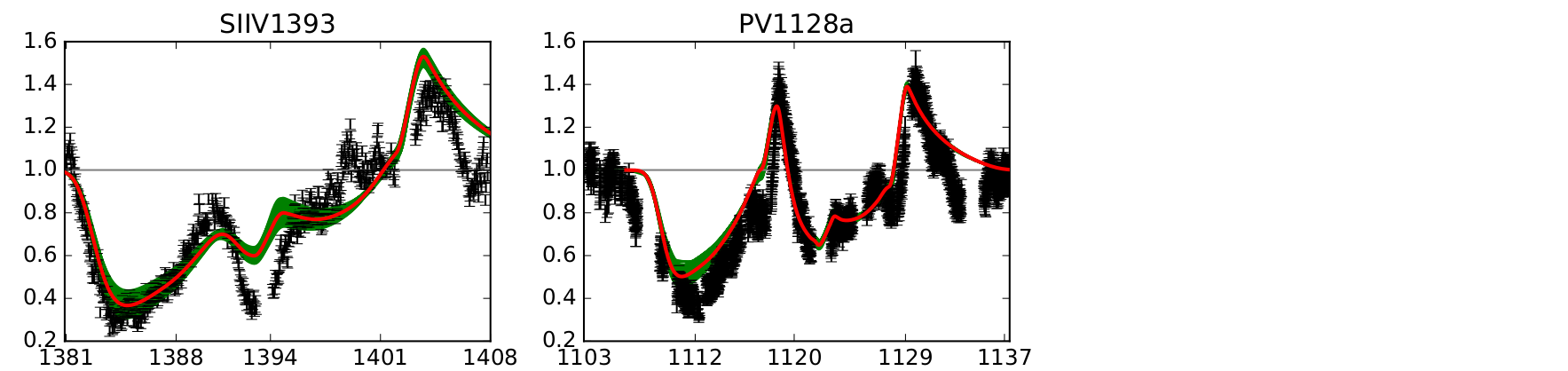}};
  \node[below=of img5, node distance=0cm, yshift=1cm] {Wavelength};
  \node[left=of img3, node distance=0cm, rotate=90, anchor=center,yshift=-0.7cm] {Normalised Flux};
\end{tikzpicture}
\caption{Continued.} 
\label{fig: HD190429A best-fit}
\end{figure}

\begin{figure}[htb]
\begin{tikzpicture}
  \node (img1)  {\includegraphics[scale=0.28]{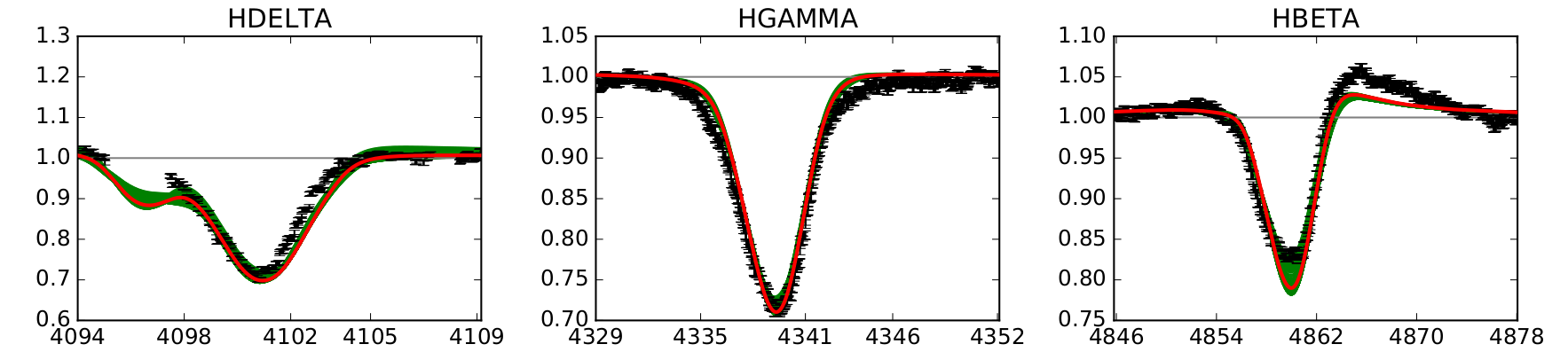}}; 
  \node[below=of img1, node distance=0cm, yshift=1cm] (img2) {\includegraphics[scale=0.28]{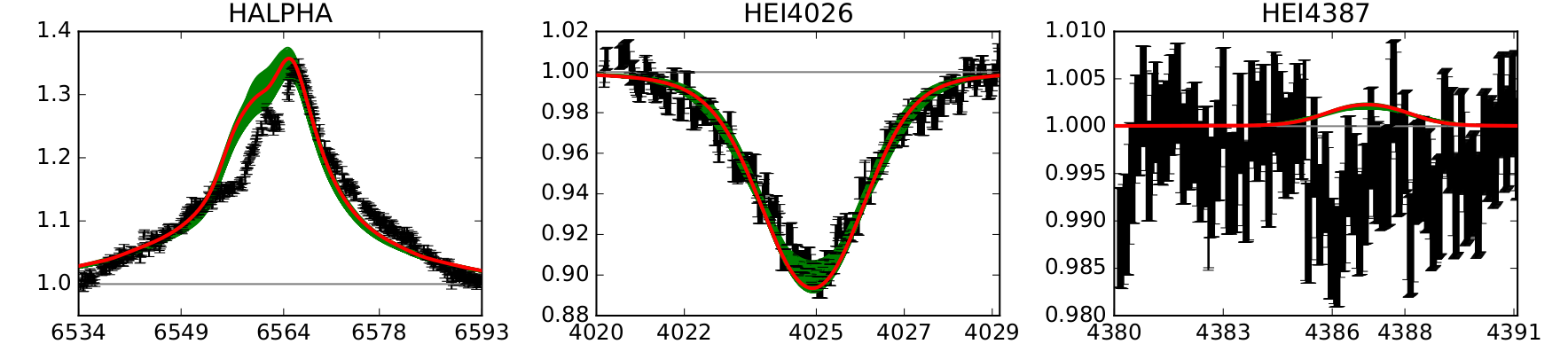}};
  \node[below=of img2, node distance=0cm, yshift=1cm] (img3) {\includegraphics[scale=0.28]{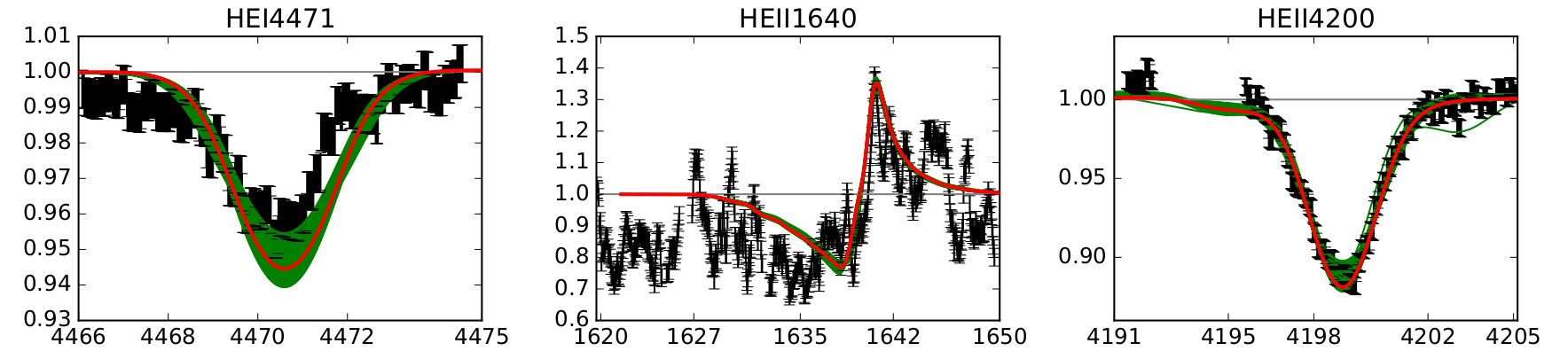}};
  \node[below=of img3, node distance=0cm, yshift=1cm] (img4) {\includegraphics[scale=0.28]{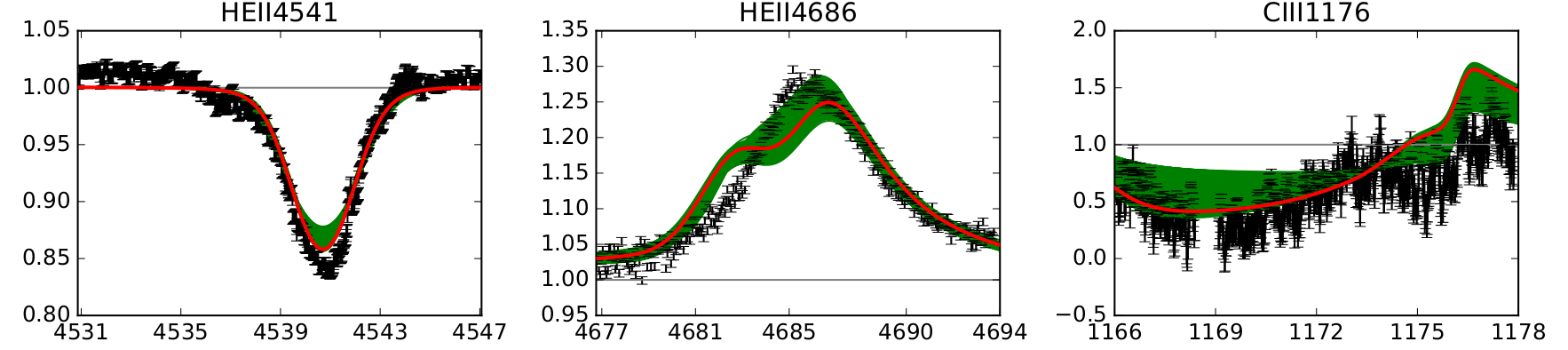}};
  \node[below=of img4, node distance=0cm, yshift=1cm] (img5) {\includegraphics[scale=0.28]{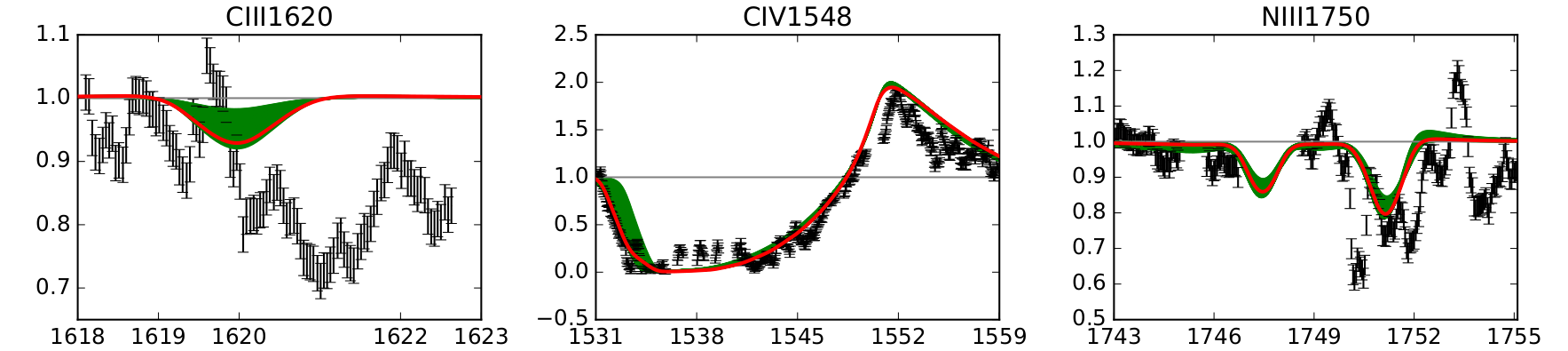}};
  \node[below=of img5, node distance=0cm, yshift=1cm] {Wavelength};
  \node[left=of img3, node distance=0cm, rotate=90, anchor=center,yshift=-0.7cm] {Normalised Flux};
\end{tikzpicture}
\caption{Best fit for HD15570 from GA with optically thick clumping.} 
\end{figure}

\begin{figure}[htb]
\ContinuedFloat
\begin{tikzpicture}
  \node (img1)  {\includegraphics[scale=0.28]{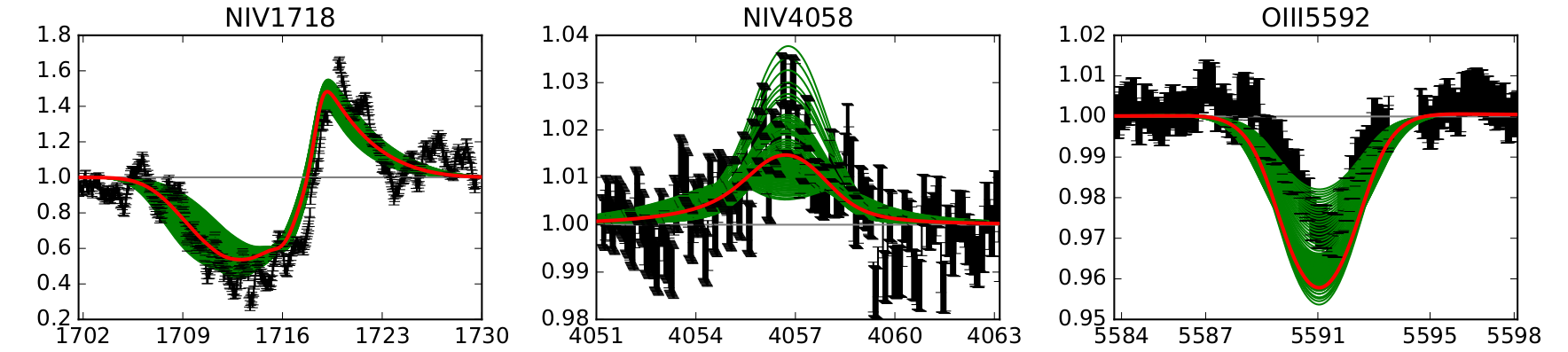}}; 
  \node[below=of img1, node distance=0cm, yshift=1cm] (img2) {\includegraphics[scale=0.28]{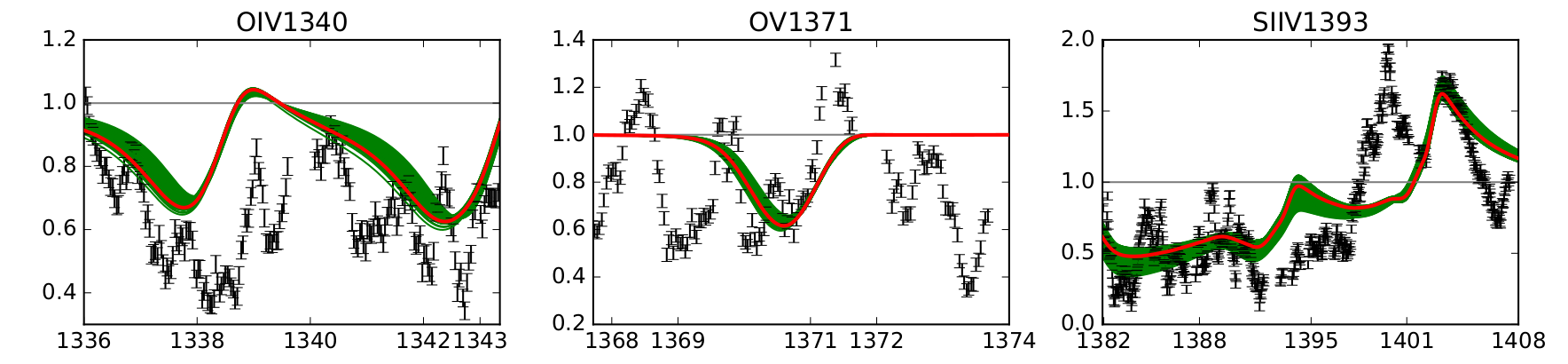}};
  \node[below=of img2, node distance=0cm, yshift=1cm] (img3) {\includegraphics[scale=0.28]{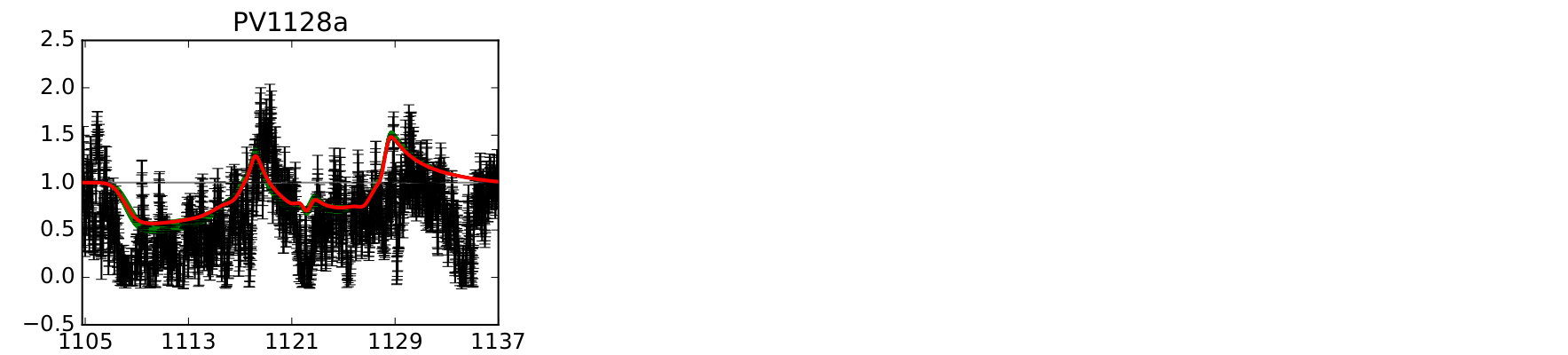}};
  \node[below=of img3, node distance=0cm, yshift=1cm] {Wavelength};
  \node[left=of img2, node distance=0cm, rotate=90, anchor=center,yshift=-0.7cm] {Normalised Flux};
\end{tikzpicture}
\caption{Continued.} 
\label{fig: HD15570 best-fit}
\end{figure}

\begin{figure}[htb]
\begin{tikzpicture}
  \node (img1)  {\includegraphics[scale=0.28]{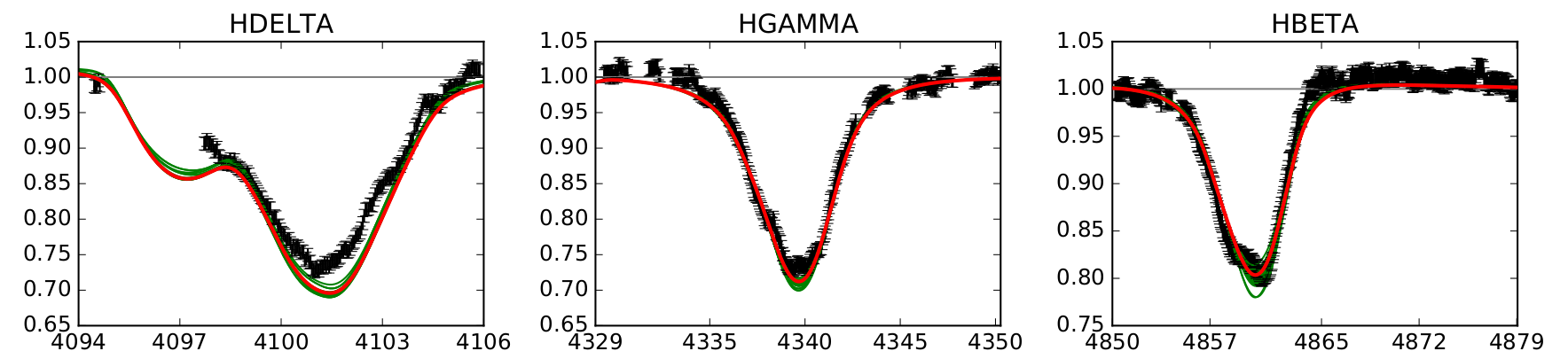}};
  \node[below=of img1, node distance=0cm, yshift=1cm] (img2) {\includegraphics[scale=0.28]{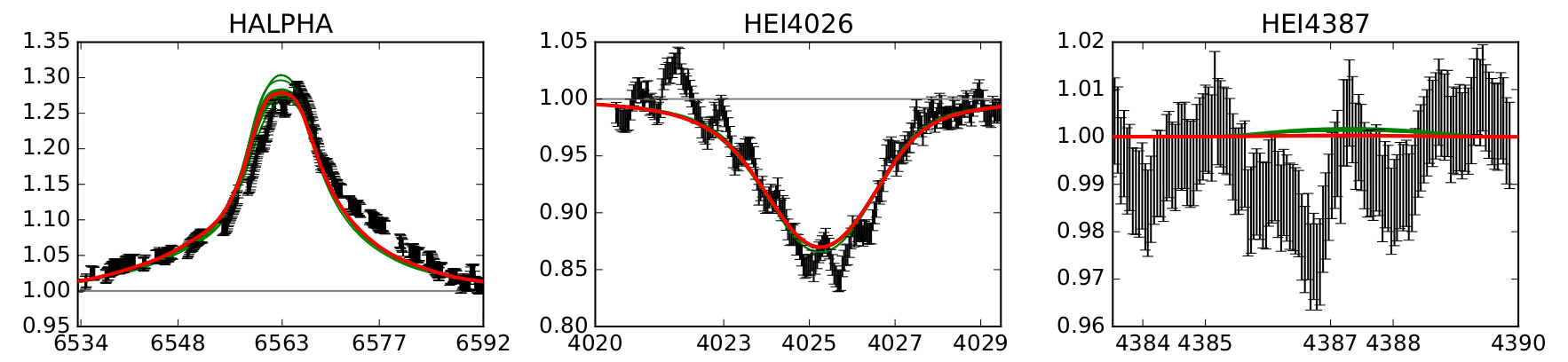}};
  \node[below=of img2, node distance=0cm, yshift=1cm] {Wavelength};
  \node[left=of img1, node distance=0cm, rotate=90, anchor=center,yshift=-0.7cm] {Normalised Flux};
\end{tikzpicture}
\caption{Best fit for HD14947 from GA with optically thick clumping.}
\end{figure}

\begin{figure}[htb]
\ContinuedFloat
\begin{tikzpicture}
  \node (img1)  {\includegraphics[scale=0.28]{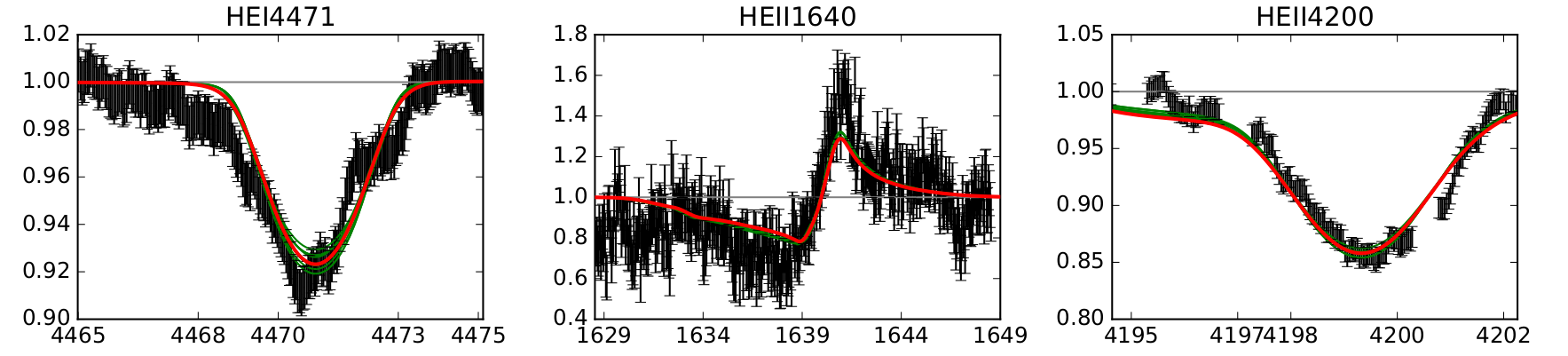}};
  \node[below=of img1, node distance=0cm, yshift=1cm] (img2) {\includegraphics[scale=0.28]{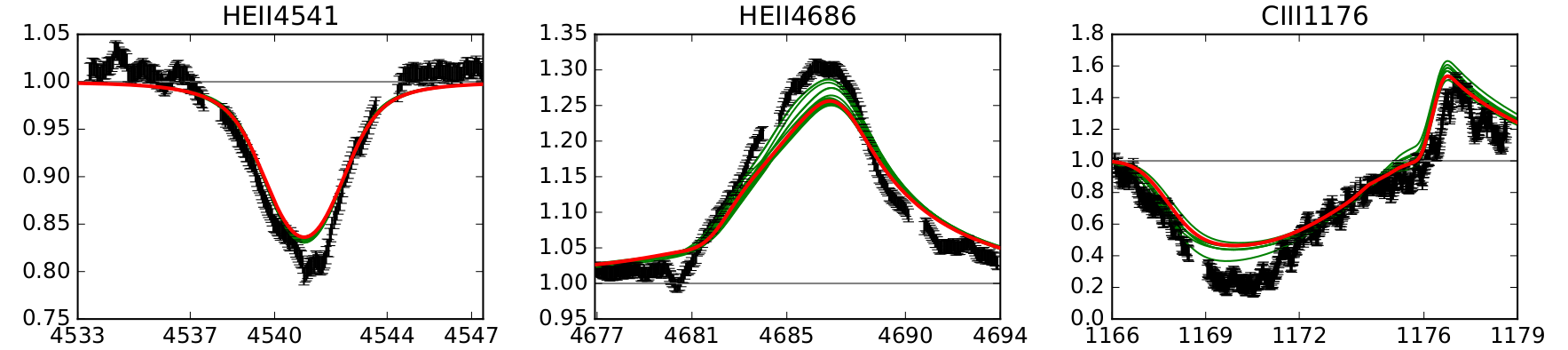}};
  \node[below=of img2, node distance=0cm, yshift=1cm] (img3) {\includegraphics[scale=0.28]{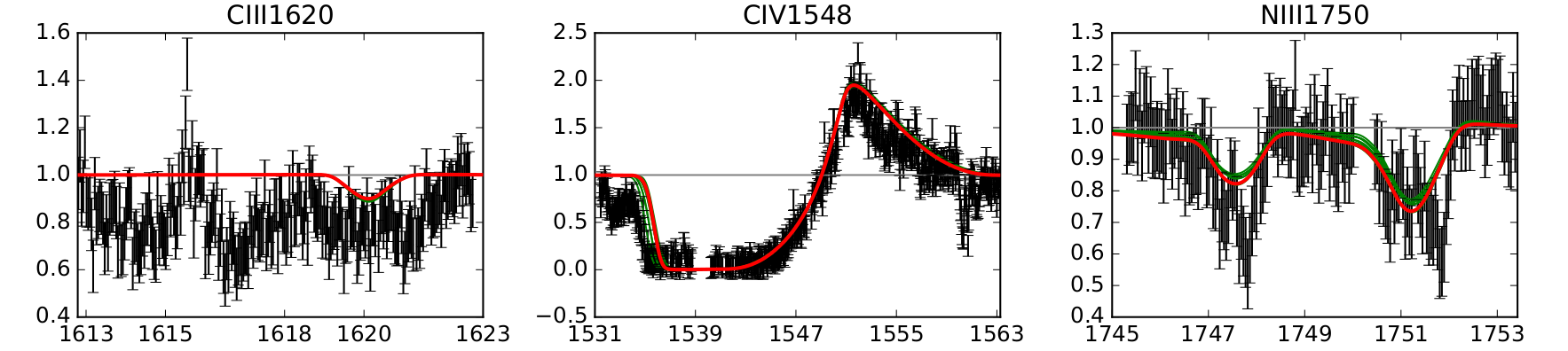}};
  \node[below=of img3, node distance=0cm, yshift=1cm] (img4) {\includegraphics[scale=0.28]{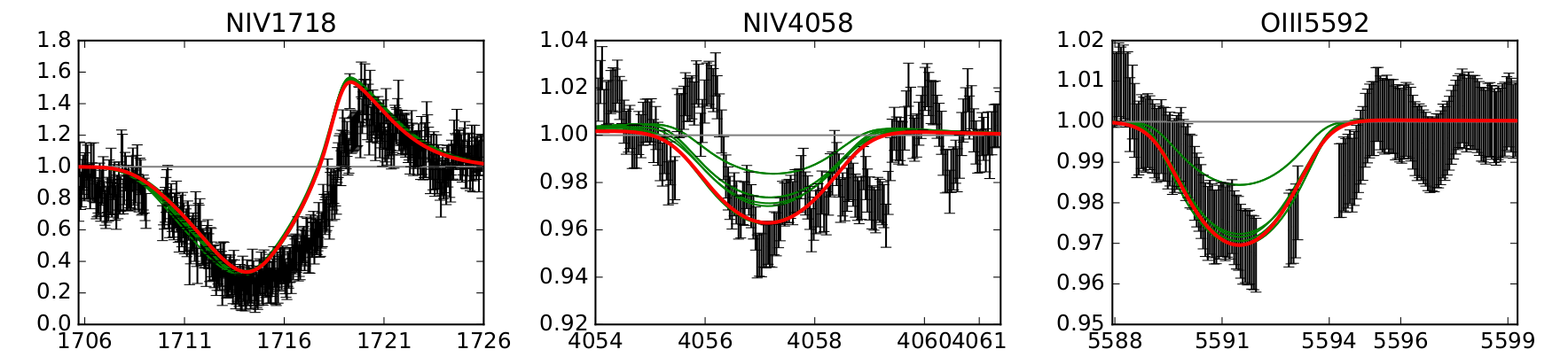}};
  \node[below=of img4, node distance=0cm, yshift=1cm] (img5) {\includegraphics[scale=0.28]{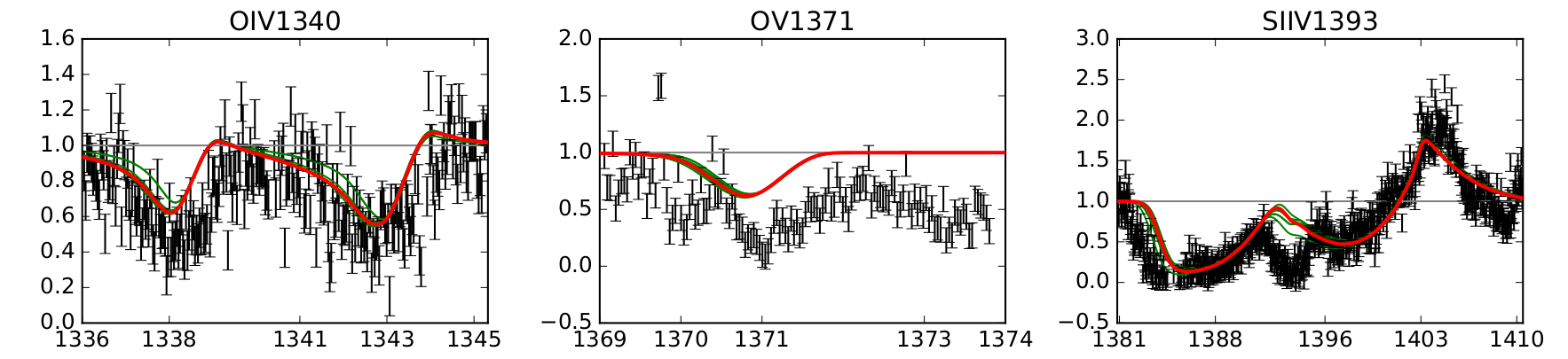}};
  \node[below=of img5, node distance=0cm, yshift=1cm] {Wavelength};
  \node[left=of img3, node distance=0cm, rotate=90, anchor=center,yshift=-0.7cm] {Normalised Flux};
\end{tikzpicture}
\caption{Continued.}
\end{figure}

\begin{figure}[htb]
\ContinuedFloat
\begin{tikzpicture}
  \node (img1)  {\includegraphics[scale=0.28]{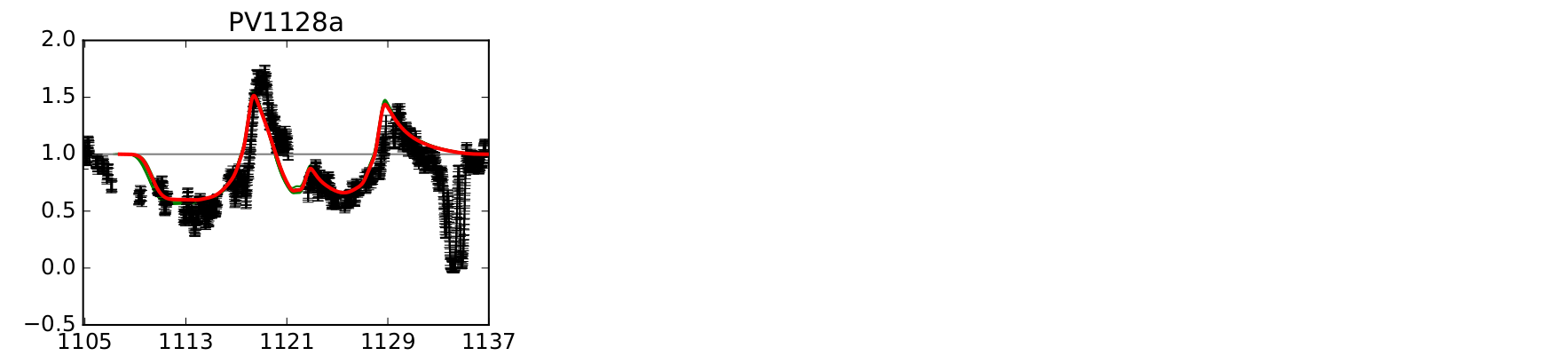}};
  \node[below=of img1, node distance=0cm, yshift=1cm] {Wavelength};
  \node[left=of img1, node distance=0cm, rotate=90, anchor=center,yshift=-0.7cm] {Normalised Flux};
\end{tikzpicture}
\caption{Continued.}
\label{fig: HD14947 best-fit}
\end{figure}

\begin{figure}[htb]
\begin{tikzpicture}
  \node (img1)  {\includegraphics[scale=0.28]{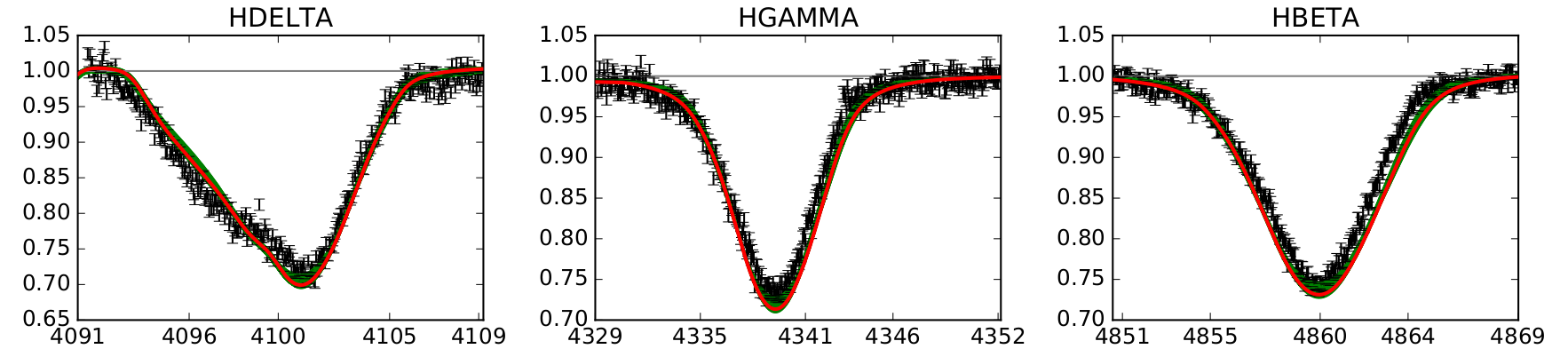}}; 
  \node[below=of img1, node distance=0cm, yshift=1cm] (img2) {\includegraphics[scale=0.28]{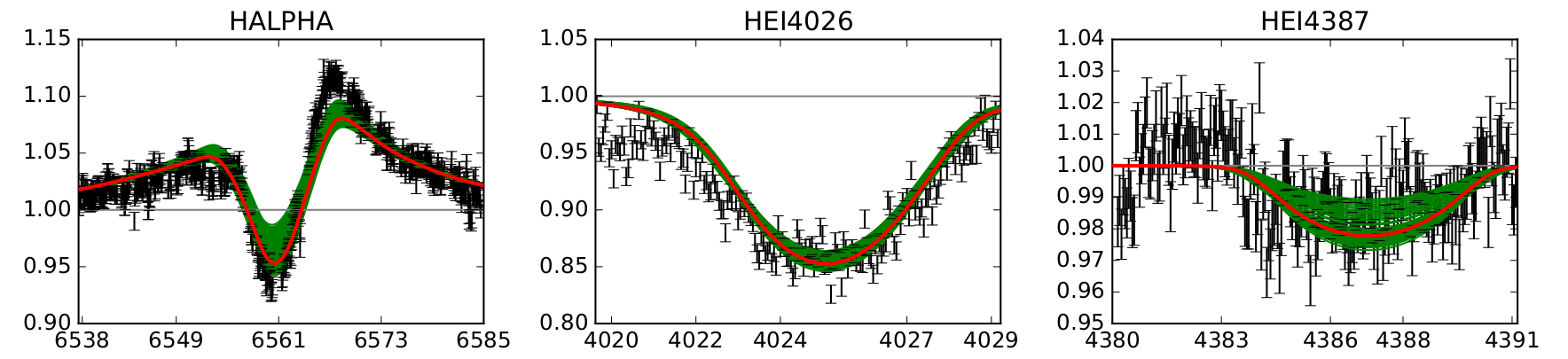}};
  \node[below=of img2, node distance=0cm, yshift=1cm] (img3) {\includegraphics[scale=0.28]{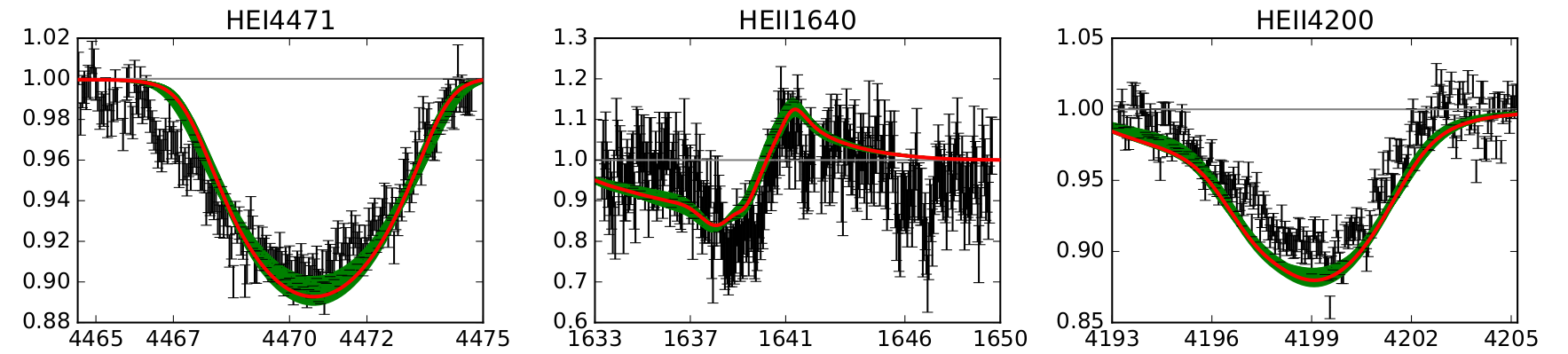}};
  \node[below=of img3, node distance=0cm, yshift=1cm] (img4) {\includegraphics[scale=0.28]{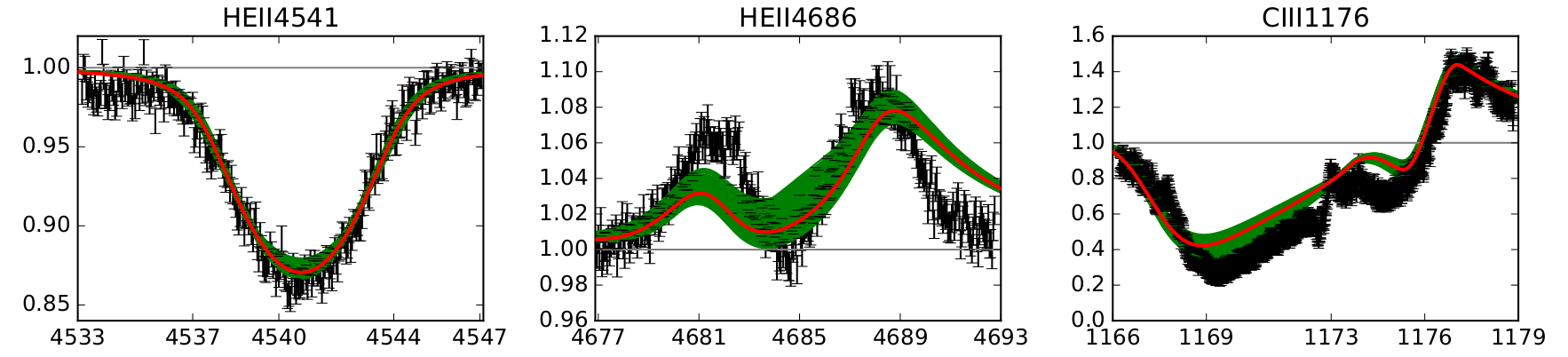}};
  \node[below=of img4, node distance=0cm, yshift=1cm] {Wavelength};
  \node[left=of img2, node distance=0cm, rotate=90, anchor=center,yshift=-0.7cm] {Normalised Flux};
\end{tikzpicture}
\caption{Best fit for HD210839 from GA with optically thick clumping.} 
\end{figure}

\begin{figure}[htb]
\ContinuedFloat
\begin{tikzpicture}
  \node (img1)  {\includegraphics[scale=0.28]{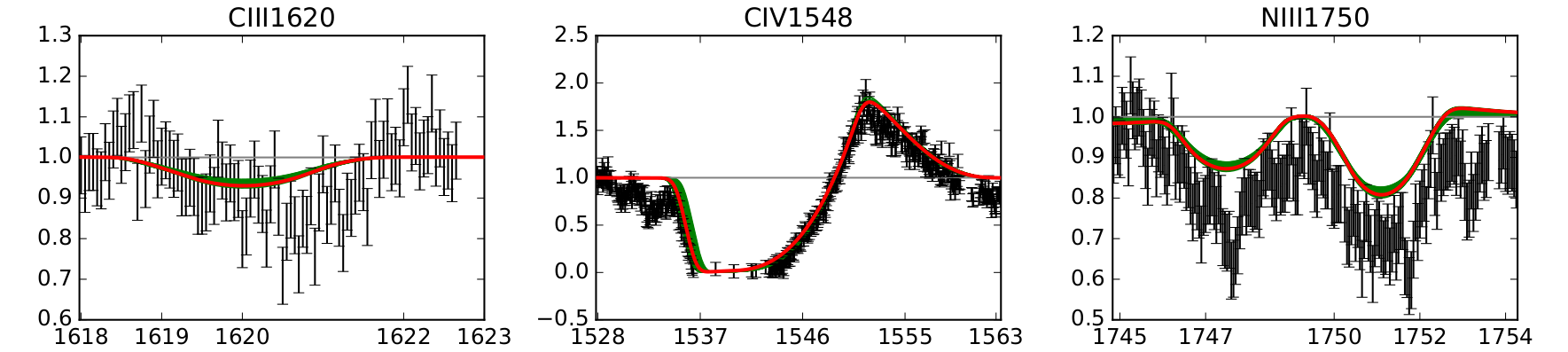}}; 
  \node[below=of img1, node distance=0cm, yshift=1cm] (img2) {\includegraphics[scale=0.28]{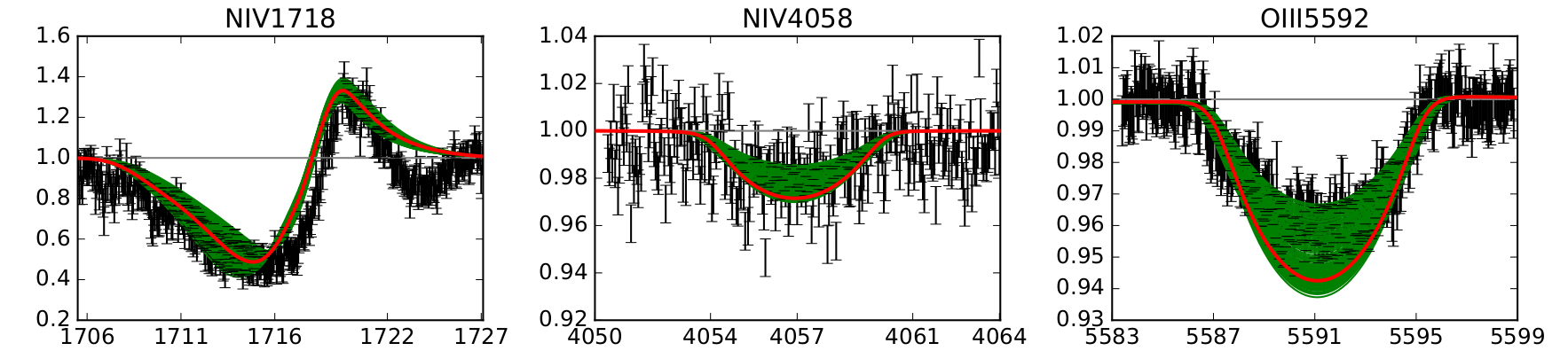}};
  \node[below=of img2, node distance=0cm, yshift=1cm] (img3) {\includegraphics[scale=0.28]{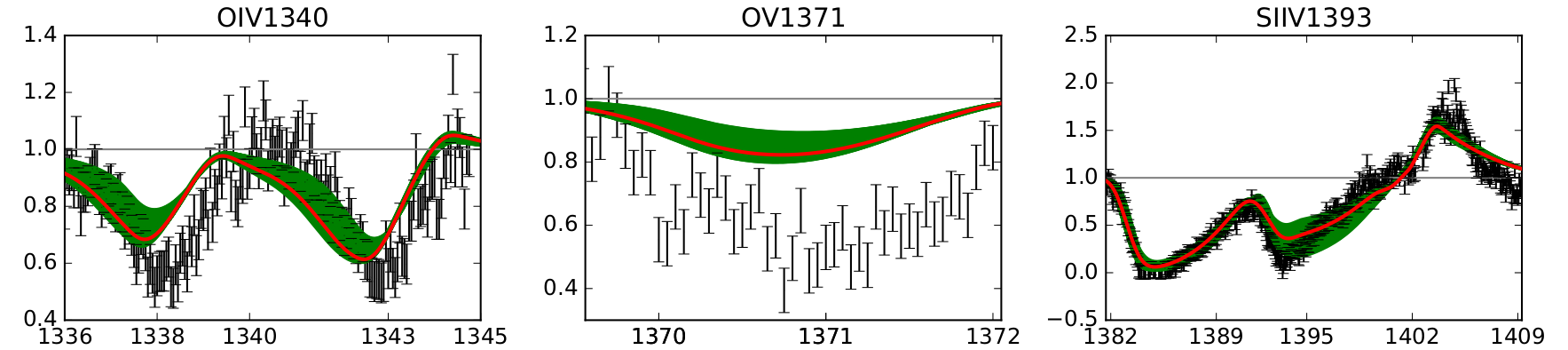}};
  \node[below=of img3, node distance=0cm, yshift=1cm] (img4) {\includegraphics[scale=0.28]{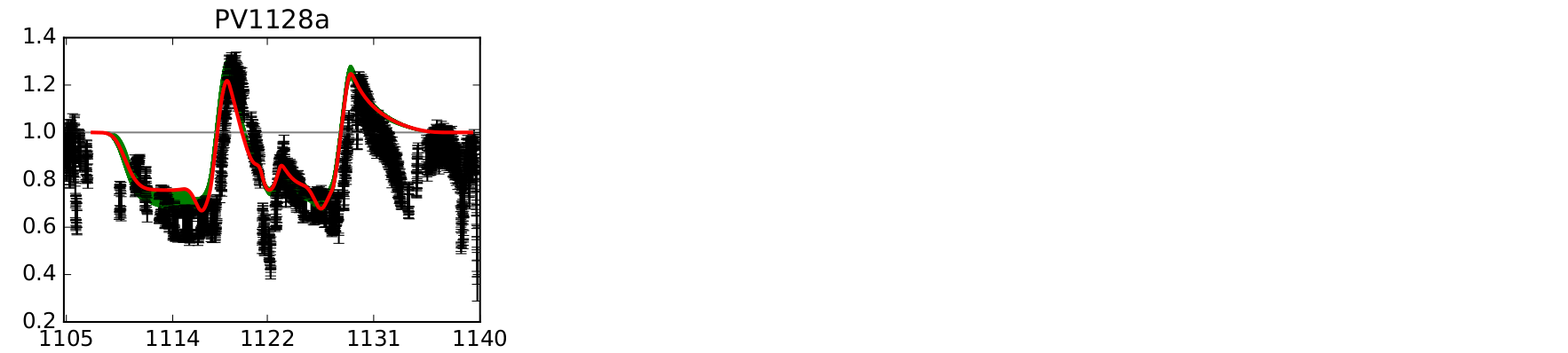}};
  \node[below=of img3, node distance=0cm, yshift=1cm] {Wavelength};
  \node[left=of img2, node distance=0cm, rotate=90, anchor=center,yshift=-0.7cm] {Normalised Flux};
\end{tikzpicture}
\caption{Continued.}
\label{fig: HD210839 best-fit}
\end{figure}

\begin{figure}[htb]
\begin{tikzpicture}
  \node (img1)  {\includegraphics[scale=0.28]{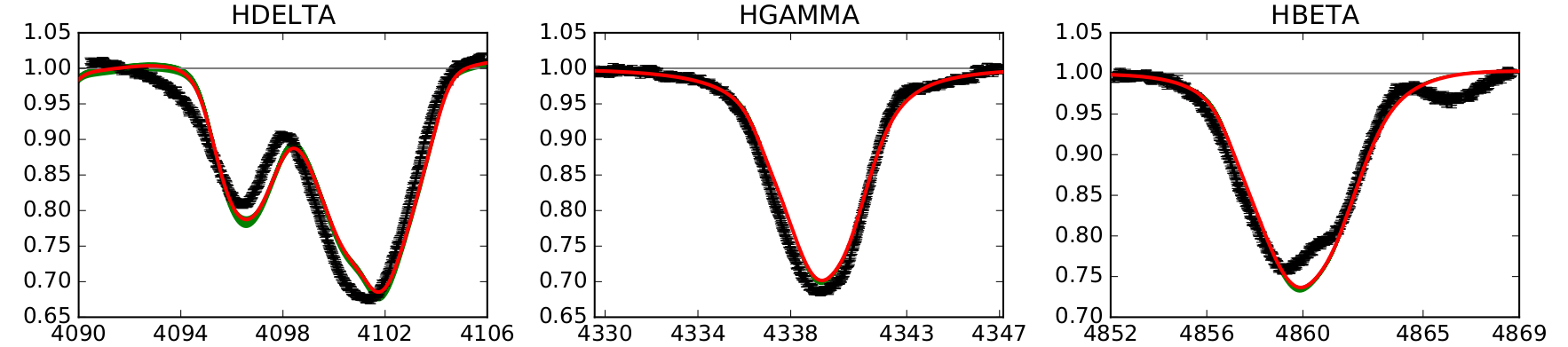}}; 
  \node[below=of img1, node distance=0cm, yshift=1cm] {Wavelength};
  \node[left=of img1, node distance=0cm, rotate=90, anchor=center,yshift=-0.7cm] {Normalised Flux};
\end{tikzpicture}
\caption{Best fit for HD163758 from GA with optically thick clumping.} 
\end{figure}

\begin{figure}[htb]
\ContinuedFloat
\begin{tikzpicture}
  \node (img1)  {\includegraphics[scale=0.28]{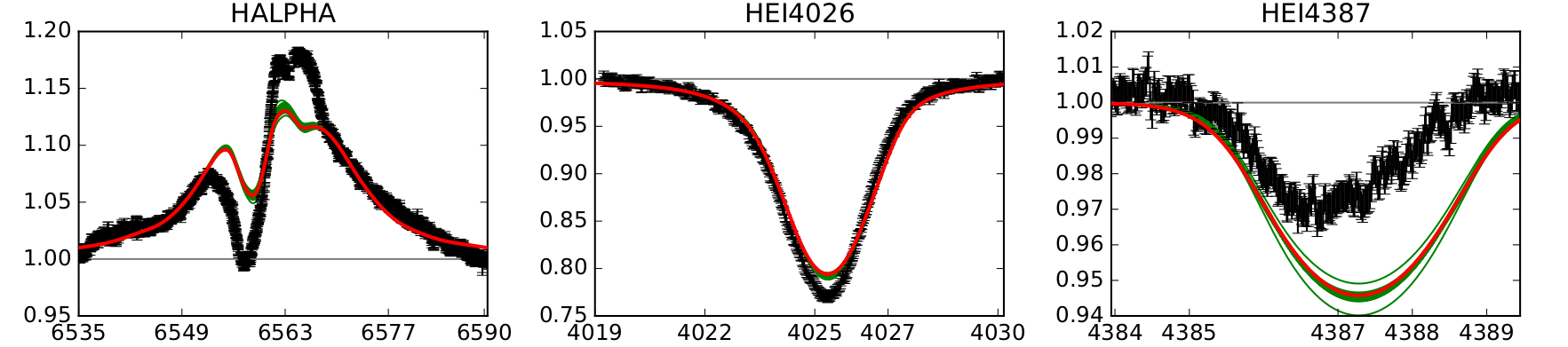}}; 
  \node[below=of img1, node distance=0cm, yshift=1cm] (img2) {\includegraphics[scale=0.28]{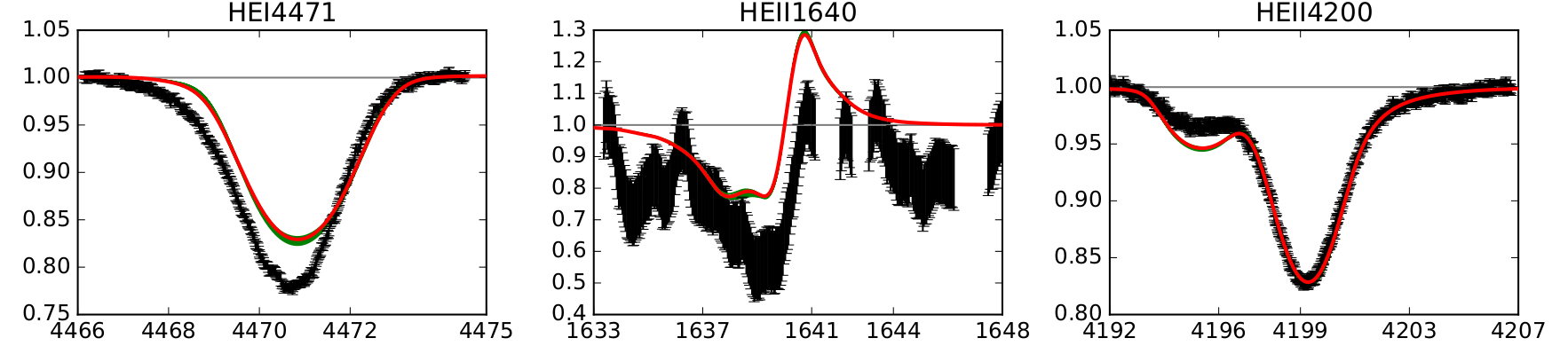}};
  \node[below=of img2, node distance=0cm, yshift=1cm] (img3) {\includegraphics[scale=0.28]{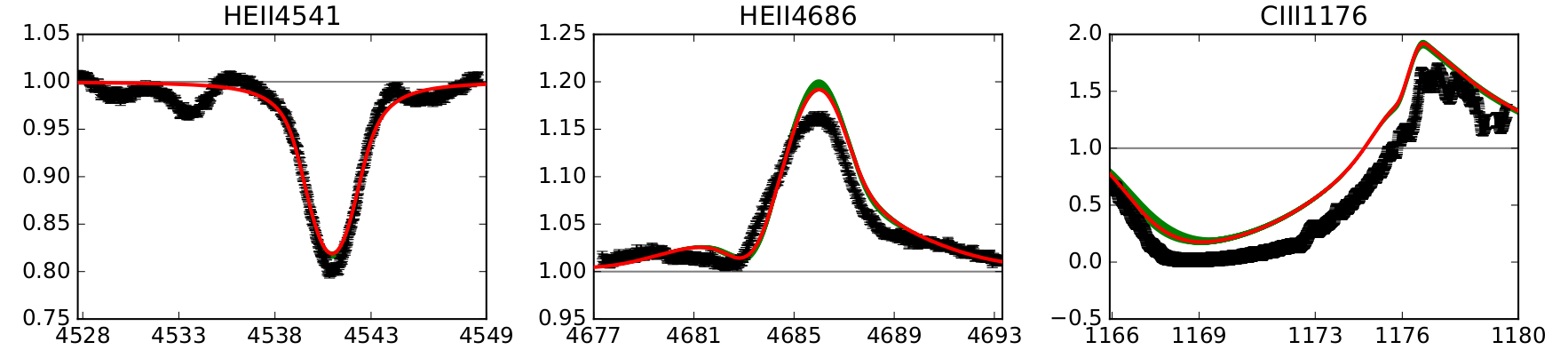}};
  \node[below=of img3, node distance=0cm, yshift=1cm] (img4) {\includegraphics[scale=0.28]{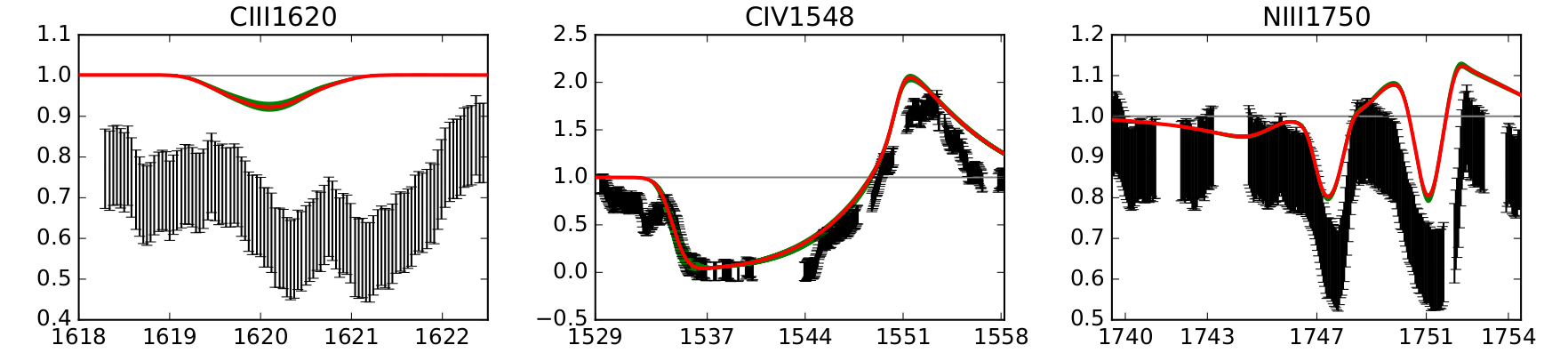}};
  \node[below=of img4, node distance=0cm, yshift=1cm] (img5) {\includegraphics[scale=0.28]{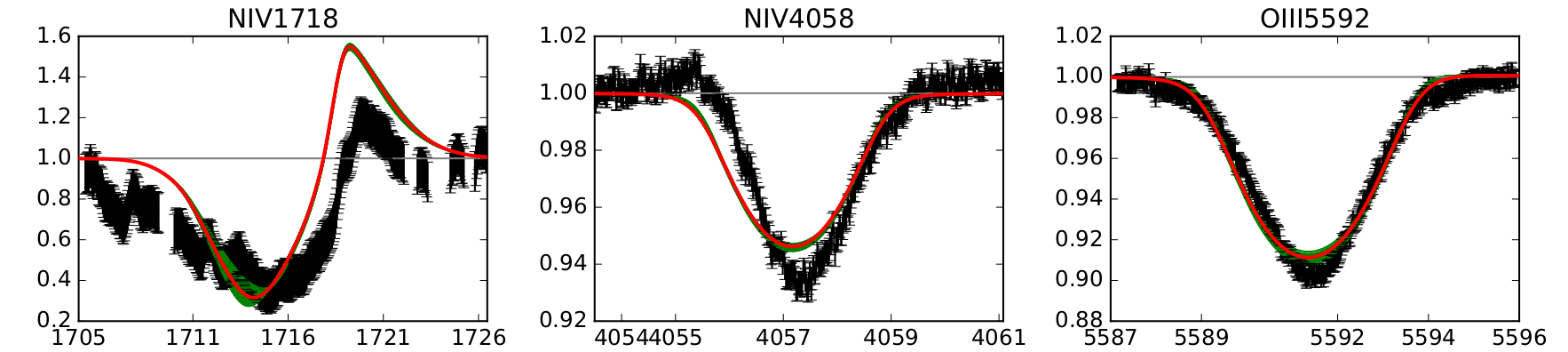}};
  \node[below=of img5, node distance=0cm, yshift=1cm] {Wavelength};
  \node[left=of img3, node distance=0cm, rotate=90, anchor=center,yshift=-0.7cm] {Normalised Flux};
\end{tikzpicture}
\caption{Continued.}
\end{figure}

\begin{figure}[htb]
\ContinuedFloat
\begin{tikzpicture}
  \node (img1)  {\includegraphics[scale=0.28]{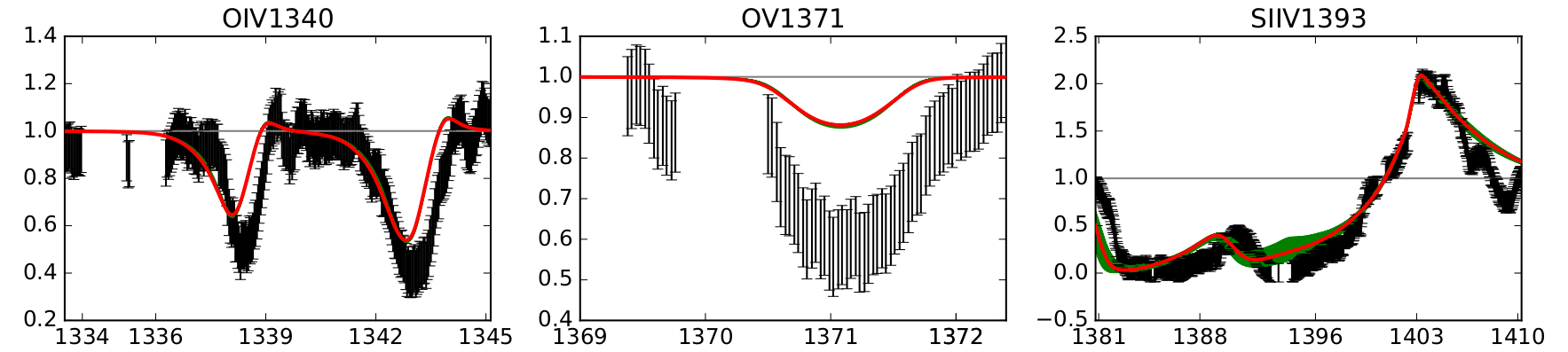}}; 
  \node[below=of img1, node distance=0cm, yshift=1cm] (img2) {\includegraphics[scale=0.28]{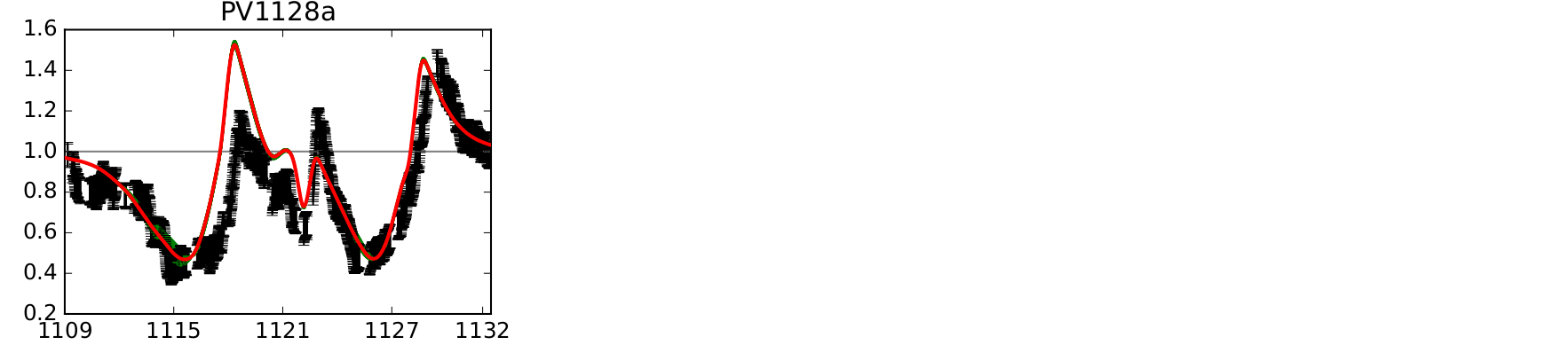}};
  \node[below=of img2, node distance=0cm, yshift=1cm] {Wavelength};
  \node[left=of img1, node distance=0cm, rotate=90, anchor=center,yshift=-0.7cm] {Normalised Flux};
\end{tikzpicture}
\caption{Continued.}
\label{fig: HD163758 best-fit}
\end{figure}

\begin{figure}[htb]
\begin{tikzpicture}
  \node (img1)  {\includegraphics[scale=0.28]{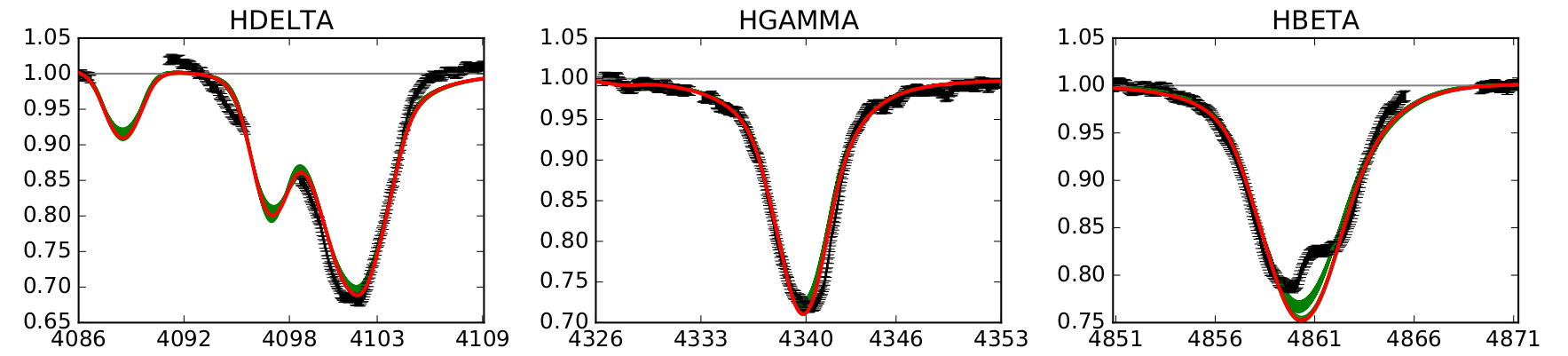}}; 
  \node[below=of img1, node distance=0cm, yshift=1cm] (img2) {\includegraphics[scale=0.28]{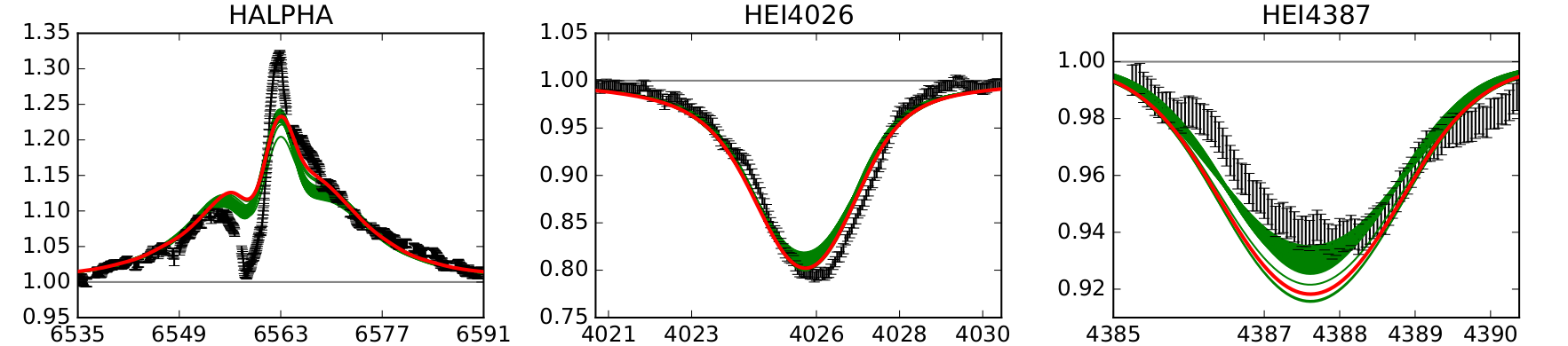}};
  \node[below=of img2, node distance=0cm, yshift=1cm] (img3) {\includegraphics[scale=0.28]{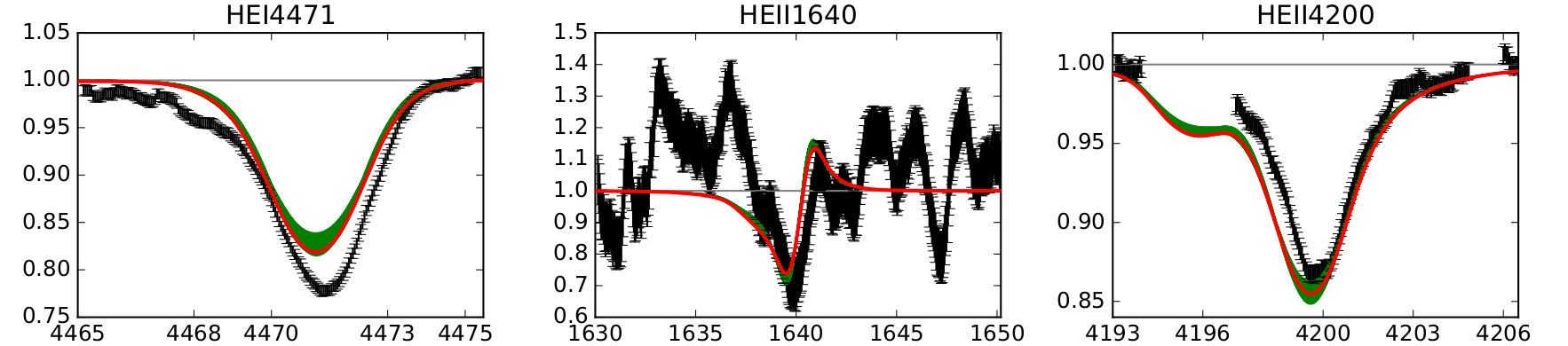}};
  \node[below=of img3, node distance=0cm, yshift=1cm] {Wavelength};
  \node[left=of img2, node distance=0cm, rotate=90, anchor=center,yshift=-0.7cm] {Normalised Flux};
\end{tikzpicture}
\caption{Best fit for HD192639 from GA with optically thick clumping.}
\end{figure}

\begin{figure}[htb]
\ContinuedFloat
\begin{tikzpicture}
  \node (img1)  {\includegraphics[scale=0.28]{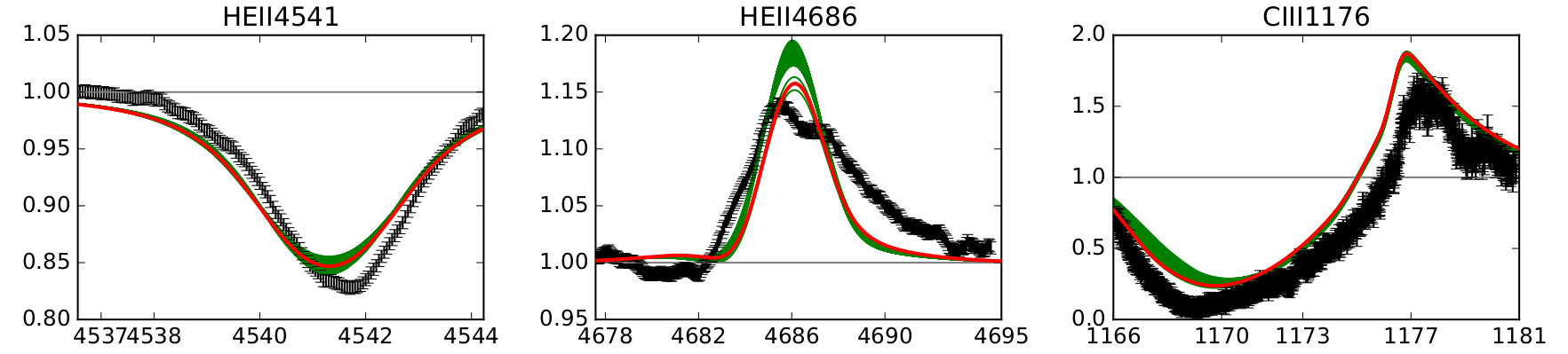}}; 
  \node[below=of img1, node distance=0cm, yshift=1cm] (img2) {\includegraphics[scale=0.28]{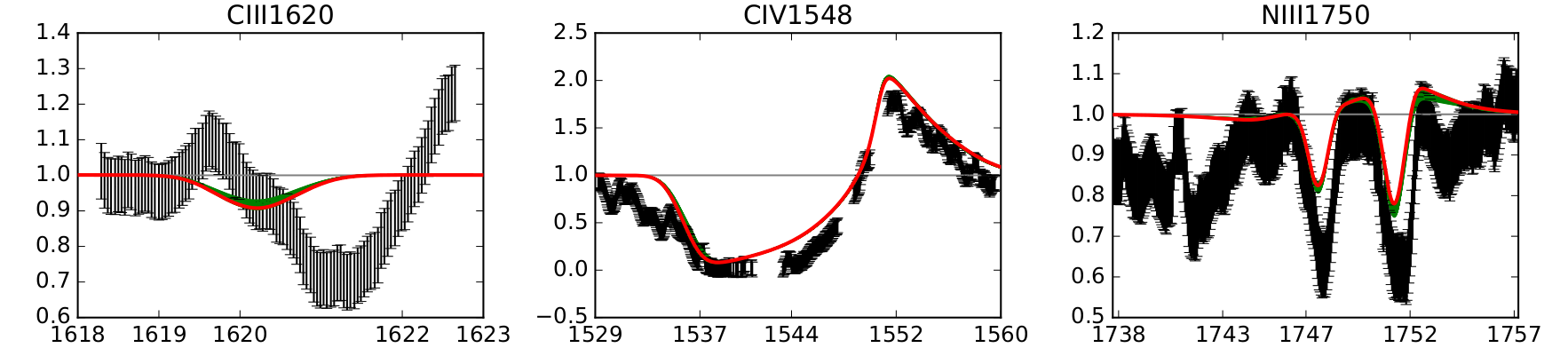}};
  \node[below=of img2, node distance=0cm, yshift=1cm] (img3) {\includegraphics[scale=0.28]{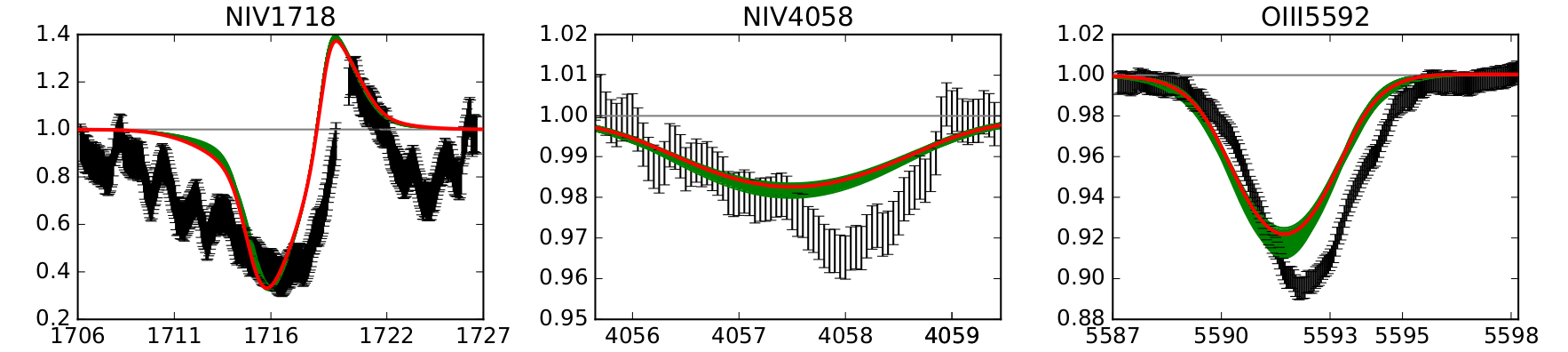}};
  \node[below=of img3, node distance=0cm, yshift=1cm] (img4) {\includegraphics[scale=0.28]{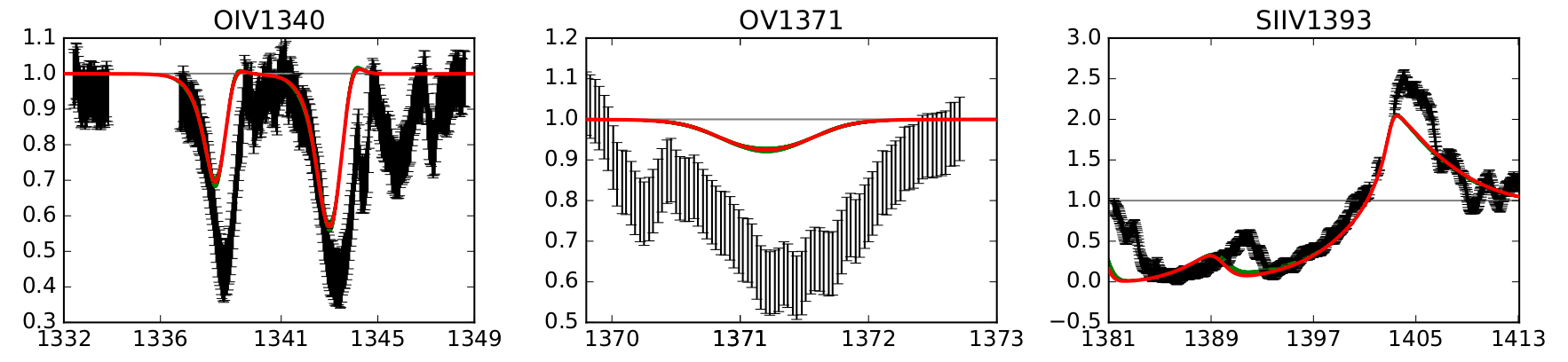}};
  \node[below=of img4, node distance=0cm, yshift=1cm] (img5) {\includegraphics[scale=0.28]{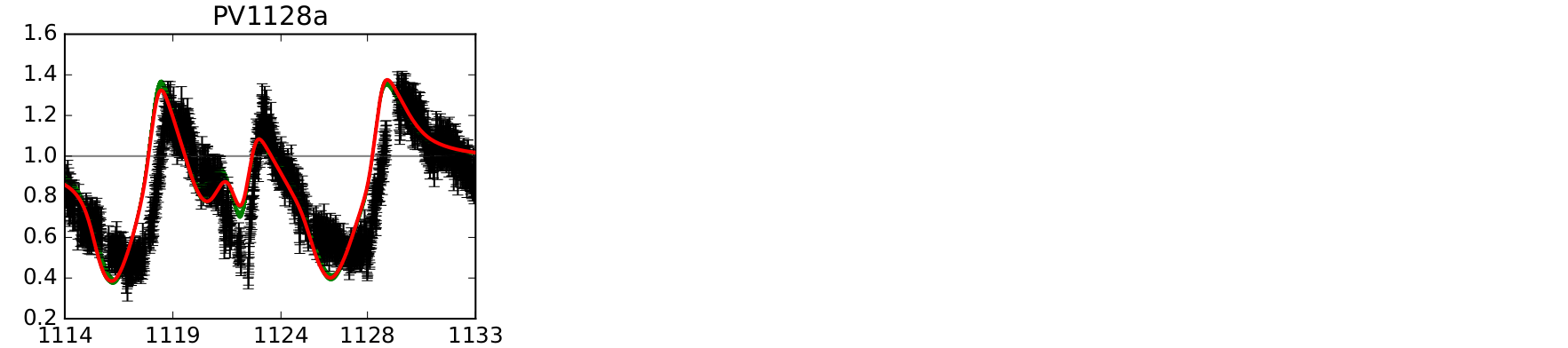}};
  \node[below=of img5, node distance=0cm, yshift=1cm] {Wavelength};
  \node[left=of img3, node distance=0cm, rotate=90, anchor=center,yshift=-0.7cm] {Normalised Flux};
\end{tikzpicture}
\caption{Continued.}
\label{fig: HD192639 best-fit}
\end{figure}

\end{appendix}

\end{document}